\documentclass[11pt]{article}

\usepackage{lineno}
\usepackage{tabularx}

\modulolinenumbers[1]

\usepackage{pstricks}

\usepackage{amsmath}
\usepackage{amssymb}
\usepackage{color}
\usepackage{soul}
\usepackage{graphicx}

\usepackage{setspace}

\begin{document}

\begin{center}
{\LARGE {\bf A multi-phase thermo-mechanical model for rock-ice avalanche}}
\\[8mm]
\end{center}
{Shiva P. Pudasaini$^{\,1,2}$} \\[5mm]
$^{1}$Technical University of Munich, School of Engineering and Design\\
\mbox{~\,}Chair of Landslide Research, Arcisstrasse 21, D-80333, Munich, Germany.
\\[2mm]
$^{2}$Kathmandu Institute of Complex Flows, Kageshwori Manohara - 3, Bhadrabas, Kathmandu, Nepal.
\\[2mm]
\mbox{~\,}{E-mail: shiva.pudasaini@tum.de}\\[6mm]
{\bf Abstract:}
  We propose a novel physically-based multi-phase thermo-mechanical model for rock-ice avalanche. It considers rock, ice and fluid; includes a rigorously derived ice melting rate, the melting efficiency dependent general fluid production rate, and a general temperature equation. It explains advection-diffusion of heat including the heat exchange across the rock-ice avalanche, basal heat conduction, production and loss of heat due to frictional shearing and changing temperature, and the enhancement of temperature due to basal entrainment. The temperature equation couples the rate of change of thermal conductivity and temperature. Ice melt intensity determines these rates as mixture conductivity evolves, characterizing distinctive thermo-mechanical processes. The model includes internal mass and momentum exchanges between the phases and mass and momentum productions due to entrainment. The latter significantly changes the state of temperature; yet, the former exclusively characterizes the rock-ice avalanche. A strong coupling exists between phase mass and momentum balances and the temperature equation. The new model offers the first-ever complete dynamical solution for simulating rock-ice avalanche with changing temperature and ice melting. We also develop a thermo-mechanical advection-diffusion-decay-source model and its exact-analytical solutions providing fundamentally novel understanding of the temperature evolution. The 2021 Chamoli event simulations with r.avaflow illustrate the functionality of thermo-mechanical rock-ice avalanche model. Four scenarios are considered: variations in the ice-melt-efficiency, fraction of ice, and ice and rock frictions, focusing on how these govern the process of melting, flow transformation, propagation, spreading and mobility. The ice melting inherently characterizes the motion and explains the mobility of rock-ice avalanche: a phenomenal thermo-mechanical play. Essentially different controls of the ice and rock frictions on the state of flow and mobility are great new revelations, explaining seemingly plausible counter-intuitive, complex thermo-mechanical processes. This provides a useful method for practitioners and mountain engineers in better solving the problems associated with the hazard mitigation and planning.

\section{Introduction}

Glaciarized or permafrost-affected high mountain slopes often witness high-magnitude, low-frequency rock-ice avalanches. The 2023 Fluchthorn (Zhuang et al., 2025) and the 2025  Blatten (Islam et al., 2025) rock-ice avalanches are the very recent huge catastrophic instances.
As rock-ice avalanche volumes can exceed hundreds of millions of cubic meter and can rush down the slope at speeds as high as 100s of ms$^{-1}$,  sometimes traveling 10s of kilometers, these events usually pose serious risks and cause catastrophic damages when they reach populated regions (Evans et al., 2009; Huggel et al., 2009). People and infrastructure in these regions are increasingly threatened by rock-ice avalanches as their frequencies are increasing in recent decades (Schneider et al., 2011a,1011b; Sosio et al., 2012).
 \\[3mm]
  Permafrost degradation, thermal perturbations, heat conduction and advection caused by climatic and hydrological causes result in the initiation of the rock detachment and rockfalls in high-Alpine steep rockwalls (Gruber and Haeberli, 2007; Huggel, 2009; Ravanel and Deline, 2011). One example is the huge rock-ice masses that detached in Huascaran, Peru in 1962 and 1970 that caused thousands of casualties (Evans et al., 2009; Schneider et al., 2011a,2011b; Mergili et al., 2018). The extremely mobile 2002 rock-ice avalanche in Kolka glacier, Russian Caucasus, is another recent example of such catastrophic event (Haeberli et al., 2004; Huggel et al., 2005; Evans et al., 2009; Schneider et al., 2011a). Particularly, these two events highlight a mechanical and a flow dynamical complexity of rock-ice avalanches. The main point is that, these events consisted of rock, ice, snow and water. Crucially, high ice content and melting of ice can quickly transform the initially solid-type rock-ice avalanches into highly viscous debris flows, which, however, depends on the fragmented fine particles and the colloids (Pudasaini and Krautblatter, 2014; Mergili et al., 2018; Pudasaini and Mergili, 2019). Consequently, such flows can propagate exceptionally long travel distances with very high velocities. The rock-ice avalanches are characterized by these effects as they are coupled with the relatively smooth sliding path consisting of the rock beds, glacial deposits, glaciers and vegetation (Schneider et al., 2011a).
  So, due to the two main reasons, rock-ice avalanche events are more challenging than the other types of mass flows (Schneider et al., 2011a; Pudasaini and Krautblatter, 2014): (i) They consist of rock, ice, snow and fluid with different physical and rheological properties. (ii) During propagation, they show changing flow behavior as the ice fragments and ice and snow melt by frictional heating and changing temperature producing substantial amount of fluid that transforms the initial solid-type rock-ice avalanche into a multi-phase rock-ice-fluid debris flow.   
\\[3mm]
Considerable attention has been paid in the past to investigate the physical modeling of flows and their numerical simulations, related to rock-ice avalanches (Schneider et al., 2010; Sosio et al., 2012; Pudasaini and Krautblatter, 2014; Sansone et al., 2021).
Pudasaini and Krautblatter (2014) presented the first-ever two-phase model for rock-ice avalanches capable of performing dynamic strength weakening due to internal fluidization and basal lubrication, and internal mass and momentum exchanges between the phases. In this model, effective basal and internal friction angles are variable and correspond to evolving effective solid volume fraction, friction factors, volume fraction of the ice, true
friction coefficients, and lubrication and fluidization factors. Their benchmark numerical simulations demonstrate
that the two-phase model can explain some basic, dynamically changing frictional properties of rock-ice avalanches
that occur internally and along the flow path. 
Internal mass and momentum exchanges between the phases, and the associated
internal and basal strength weakening provide a more
realistic simulation and explain the
exceptionally long runout distances and dynamically changing high mobilities of rock-ice avalanches.
\\[3mm]
However, field and laboratory observations on rock-ice mass movements clearly reveal changing flow properties from solid-like to fluid-like and thus largely showing the multi-phase nature of the flow as a mixture of rock, ice and fluid as  the rock-ice mass propagates downslope (Schneider et al., 2011b; Mergili et al., 2018; Shugar et al., 2021). 
 Thus, the more realistic simulations can be performed with a real multi-phase (or, three-phase) mass flow simulation model. Moreover, it often entrains a large amount of basal material. Due to the shear heating and changing state of temperature of the rock-ice avalanche, the ice melts. So, changing temperature, melting ice and entrainments of the basal material are some fundamental characteristic features of naturally multi-phase rock-ice avalanche motion. However, the existing rock-ice avalanche model by Pudasaini and Krautblatter (2014) is only for two-phase flows, it does not include the basal entrainment, and the temperature evolution utilizing the ice melting is not considered. But, these aspects play important role in rock-ice avalanche. For these reasons, here, by including all these important features, we aim at fundamentally advancing our knowledge by proposing a physically-based, multi-phase mechanical model for rock-ice avalanches.
\\[3mm]
There are five major aspects of the proposed multi-phase mechanical model for rock-ice avalanches. (i) It is based on the multi-phase mass flow model (Pudasaini and Mergili, 2019), considering the rock, ice and fluid as three constituents in the avalanching mixture of the rock-ice mass. 
(ii) Extending the two-phase mechanical rock-ice avalanche model (Pudasaini and Krautblatter, 2014), it utilizes the principle of ice melting, internal mass and momentum exchanges between the solid (rock and ice) and fluid, basal lubrication and internal fluidization, and the resulting mechanical strength weakening of the avalanching mass. (iii) The unified multi-phase entrainment model (Pudasaini, 2025) is incorporated. (iv) A novel and general temperature evolution equation for rock-ice avalanche is developed here, that includes dynamically evolving mixture properties: density, heat capacity, thermal conductivity, heat exchange across the boundary of the avalanche, heat conduction at the sliding base, 
 frictional shear heat production, internal heat loss due to the ice melting for which a general model is formally derived, and enhancement of the temperature due to the entrainment of the basal material. (v) Whereas (i)-(iii) are already existing separately, (iv) is fundamentally new for the motion of rock-ice avalanches. Yet, there are two main tasks here. First, developing a general temperature equation for the rock-ice avalanche and a unified ice melt rate. And, second, a description on how to combine all components (i)-(iv), creating a general, operational multi-phase thermo-mechanical model for rock-ice avalanche. The resulting model is truly governed by the dynamics of the phases with different physical properties and mechanical responses, interfacial momentum exchanges between the phases, material strength weakening, phase change with the melting of ice and the internal mass and momentum exchanges, entrainment of the basal material by the rock-ice avalanche, and the complex evolution of its temperature. Consequently, this constitutes a comprehensive rock-ice avalanche dynamical simulation model, the first of this kind in the mass flow simulation.  
 \\[3mm]
 Moreover, we develop a simple thermo-mechanical advection-diffusion-decay-source model for the evolving temperature and its exact-analytical solutions, revealing some important thermo-mechanical and dynamical features of the rock-ice avalanche propagation.
 \\[3mm]
With r.avaflow computational tool, simulation results are presented for the 2021 Chamoli event explaining the essential functionality of the thermo-mechanical rock-ice avalanche model. This involves four scenarios: variations in the ice-melt-efficiency, the fraction of ice, and the ice and rock quality. Discussions are provided on how these aspects govern the process of melting, flow transformation, mass propagation and mobility. There appears a complex counter-play between the frictional weakening and the melt-rate weakening in controlling the net flow mobility with decreased ice friction. However, a positive interplay exists between these processes with decreased rock friction. The most important revelation is that the ice melting consistently describes the physics of the flow transformation and the mobility of the rock-ice avalanche.

\section{The model construction}

In order to understand the material response to thermal changes of the rock-ice mass, we need the knowledge of its thermal properties. To our purpose, these properties include the thermal conductivity, heat capacity, thermal diffusivity and the latent heat (Andersland and Ladanyi, 2003; Heinze, 2024). Often, these parameters (properties) change with the type of material under consideration, its density, temperature, ice and water content and the saturation degree. These thermal parameters are defined as follows.  Thermal conductivity refers to the heat conduction in the rock-ice mass that involves a transfer of kinetic energy from a worm part to a cooler part. The amount of heat required to raise the temperature of a unit of the rock-ice mass by one degree Celsius is called its heat capacity. Then, the specific heat of the rock-ice mass is defined as the heat capacity per unit mass. The thermal conductivity characterizes the rate at which the heat is transferred in the rock-ice mass. The produced heat causes the rise in temperature that varies inversely with the heat capacity and the bulk density of the rock-ice mass. The ratio of the thermal conductivity and the product of these quantities is called the thermal diffusivity of the rock-ice mass.  The amount of heat energy absorbed when a unit of ice mass is converted into a liquid at the melting point is defined as its latent heat of fusion. The same amount of heat is released when the water is converted into ice with no change in temperature. For more on it, we refer to Robertson (1988), Andersland and Ladanyi (2003) and Heinze (2024). With these descriptions of the thermal quantities, we now proceed to the model development.

\subsection{The basic temperature equation}

First, we define some basic and induced (or composite) variables and parameters involved in the derivation of the temperature equation for rock-ice avalanches.
Let, $t$ be time, $x, y, z$ be the coordinates along and across the slope and perpendicular to the sliding surface, and $g^x, g^y, g^z$ be the components of gravitational acceleration along the coordinate directions, respectively.  
 Let, $k = r, i, f$ indicate the rock, ice and fluid phases in the mixture. Then,
$\alpha_k$, ${\bf u}_k = (u_k, v_k, w_k)$, $\rho_k$, $c_k$ and $\kappa_k$ are the volume fractions, velocities, densities, specific heat capacities and thermal conductivities of the phases. Next, we introduce the bulk properties of the mixture of rock, ice and fluid.
The following bulk properties are involved (Clauser, 2009; Heinze, 2024): temperature $T$, density $\rho = \sum{\alpha_k\rho_k}$, velocity ${\bf u} = \sum{\alpha_k {\bf u}_k}$ (${\bf u} = u, v, w$), specific heat capacity $c = \sum{\alpha_k c_k}$, thermal conductivity $\kappa = \sum{\alpha_k\kappa_k}$, and 
 heat source and sink $Q^+, Q^-$, respectively. 
\\[3mm]
As the first major task, here, we develop a dynamically evolving general temperature equation for the rock-ice avalanche. For this, we begin with the basic temperature equation.
The temperature evolution plays a crucial role in the thermo-mechanical response and the dynamics of the rock-ice avalanche. This is so, because, the ice-melting and the temperature evolution are tightly related. 
\\[3mm]
Considering the low internal shear rate, as often considered in mass flow simulation (Pudasaini and Hutter, 2007), we assume that the internal viscous heating can be ignored. Then, the temperature equation takes the form (Clauser, 2009; Dall'Amico et al., 2011; Heinze, 2021): 
\begin{eqnarray}
\begin{array}{lll}
\displaystyle{\frac{\partial }{\partial t} \left( \rho c\, T\right)
+ \frac{\partial}{\partial x}\left (\rho c\, uT\right) 
+ \frac{\partial}{\partial y}\left (\rho c\, vT\right)  
+ \frac{\partial}{\partial z}\left (\rho c\, wT\right)
-\frac{\partial}{\partial x}\left (\! \kappa \frac{\partial T}{\partial x}\right)
-\frac{\partial}{\partial y}\left (\! \kappa \frac{\partial T}{\partial y}\right)
-\frac{\partial}{\partial z}\left (\! \kappa \frac{\partial T}{\partial z}\right)
\!= Q^+ - Q^-
}.
\label{Eqn_2}
\end{array}    
\end{eqnarray} 
This is an advection-diffusion equation for the evolution of the temperature of a body with source terms for the heat production and loss.
This is the basic temperature equation for the bulk mixture of rock, ice and fluid.

\subsection{The depth-averaging and thermo-mechanical closures}

Functional landslide, avalanche and debris flow models are depth-averaged (Hungr and McDougall, 2004; Pitman and Le, 2005; Pudasaini and Hutter, 2007;  Christen et al., 2010; Pudasaini, 2012; Pudasaini and Mergili, 2019). 
Yet, none of them include the important physical properties and dynamical evolution of the temperature of the sliding mass. 
We need to depth-average the temperature equation (\ref{Eqn_2}) in order to combine it with the existing balance equations for mass and momentum such that the resulting set of equations describe the overall picture of the rock-ice mass in motion.  
For this, we integrate (\ref{Eqn_2}) through the depth $h \,\,(h = s - b)$ from the basal surface $b$ to the free-surface $s$ of the avalanche (Pudasaini and Hutter, 2007; Pudasaini, 2012). 
Depth-averaging of a quantity ${*}$ is denoted by $\overline{*}$, and is defined as: 
 \begin{eqnarray}
\begin{array}{lll}
\displaystyle{\overline{*} = \frac{1}{h}\int_b^s * \,dz
}.
\label{Eqn_3}
\end{array}    
\end{eqnarray}
Depth-averaging requires the implementation of the kinematic boundary conditions at the free- and the basal- surfaces, which, respectively, are:
\begin{eqnarray}
\begin{array}{lll}
\displaystyle{
\frac{\partial s}{\partial t} 
+ u^s\frac{\partial s}{\partial x}
+ v^s\frac{\partial s}{\partial y}
-w^s = 0},
\label{Eqn_4}
\end{array}
\end{eqnarray}
\begin{eqnarray}
\begin{array}{lll}
\displaystyle{
\frac{\partial b}{\partial t} 
+ u^b\frac{\partial b}{\partial x}
+ v^b\frac{\partial b}{\partial y}
-w^b = - E_t},
\label{Eqn_5}
\end{array}
\end{eqnarray}
where $E_t$ is the flow-induced total erosion rate at the basal surface (Pudasaini, 2025), and the basal material is entrained into the rock-ice avalanche with this rate. The quantities with superscripts $s$ and $b$ indicate that these quantities are evaluated at the free-surface of the flow and the basal surface, respectively. Here, we assume that the mean of the product is the product of the mean (Pudasaini, 2012), which is equivalent to assume that the variations of the bulk properties (e.g., $\rho$ and $c$) through the avalanche depth is negligible.

\subsubsection{Advection}

Now, depth-averaging the terms associated with 
 the time variation and advection of temperature
 on the left hand side of (\ref{Eqn_2}), following the Leibniz rule of integrating the differentiation (Pudasaini and Hutter, 2007), and applying the kinematic boundary conditions (\ref{Eqn_4})-(\ref{Eqn_5}), we obtain:
{\small
\begin{eqnarray}
\begin{array}{lll}
\displaystyle{\int_b^s\left[ \frac{\partial }{\partial t} \left( \rho c\, T\right)
+ \frac{\partial}{\partial x}\left (\rho c\, uT\right) 
+ \frac{\partial}{\partial y}\left (\rho c\, vT\right)  
+ \frac{\partial}{\partial z}\left (\rho c\, wT\right) \right]dz
= \frac{\partial}{\partial t} \left( h \overline{\rho c \,T}\right) 
+ \frac{\partial}{\partial x}\left ( h \overline{\rho c \,uT}\right) 
+ \frac{\partial}{\partial y}\left ( h \overline{\rho c \,vT}\right)  
- \rho^b c^b\,T^bE_t 
},
\label{Eqn_6}
\end{array}    
\end{eqnarray}
}
where $\rho^b c^b$ is the thermal capacity (Clauser, 2009) and $T^b$ is the temperature 
 of the entrainable bed material.

\subsubsection{Diffusion, heat transfer across the boundary}

Depth-averaging diffusion terms is mathematically and thermo-mechanically much more demanding and challenging than
 the time rate of change and 
 advection terms.
Repeatedly depth-averaging the first diffusion term on the left hand side of (\ref{Eqn_2}) yields:
{\small
\begin{eqnarray}
\begin{array}{lll}
\displaystyle{\int_b^s{\frac{\partial}{\partial x}\left ( \kappa \frac{\partial T}{\partial x}\right) dz}
= \frac{\partial}{\partial x}\left[\int_b^s{ \left ( \kappa \frac{\partial T}{\partial x}\right )} dz\right] 
-\left[ \kappa \frac{\partial T}{\partial x}\frac{\partial z}{\partial x} \right]_b^s
= \frac{\partial}{\partial x}\left[\overline{\kappa}\left\{ \frac{\partial}{\partial x}\left(\int^s_b T dz \right)
- \left ( T\frac{\partial z}{\partial x}\right)_b^s\right\} \right]
-\left[ \kappa \frac{\partial T}{\partial x}\frac{\partial z}{\partial x} \right]_b^s
}.
\label{Eqn_7}
\end{array}    
\end{eqnarray}
\\[-1mm]
}
We assume that the depth variation of $\kappa$, $T$ and the lateral gradients of $T$ are negligible. Within the framework of the depth-averaged modeling, these approximations can be perceived for the quantities attached to the landslide body (Pudasaini, 2012). Then, (\ref{Eqn_7}) reduces to 
{\small
\begin{eqnarray}
\begin{array}{lll}
\displaystyle{\int_b^s{\frac{\partial}{\partial x}\left ( \kappa \frac{\partial T}{\partial x}\right) dz}
= \frac{\partial}{\partial x}\left[ \overline{\kappa} \frac{\partial}{\partial x}\left(h \overline{T} \right)\right]
-\frac{\partial}{\partial x}\left[ \overline{\kappa} \overline{T}\frac{\partial h}{\partial x}\right]
-\overline{\kappa}\frac{\partial \overline{T}}{\partial x}\frac{\partial h}{\partial x}
}.
\label{Eqn_8}
\end{array}    
\end{eqnarray}
}
Repeatedly applying the product rule of differentiation to (\ref{Eqn_8}), and simplifying the outcome results in:
\begin{eqnarray}
\begin{array}{lll}
\displaystyle{\int_b^s{\frac{\partial}{\partial x}\left ( \kappa \frac{\partial T}{\partial x}\right) dz}
= h \left[ \overline{\kappa} \frac{\partial^2 \overline{T}}{\partial x^2} 
+ \frac{\partial \overline{\kappa}}{\partial x}\frac{\partial \overline{T}}{\partial x}\right]
}.
\label{Eqn_9}
\end{array}    
\end{eqnarray}
Analogously, repeatedly depth-averaging the second diffusion term on the left hand side of (\ref{Eqn_2}) and applying the product rule of differentiation to it yields:
\begin{eqnarray}
\begin{array}{lll}
\displaystyle{\int_b^s{\frac{\partial}{\partial y}\left ( \kappa \frac{\partial T}{\partial y}\right) dz}
= h \left[ \overline{\kappa} \frac{\partial^2 \overline{T}}{\partial y^2} 
+ \frac{\partial \overline{\kappa}}{\partial y}\frac{\partial \overline{T}}{\partial y}\right]
}.
\label{Eqn_10}
\end{array}    
\end{eqnarray}
Moreover, depth-averaging the third diffusion term on the left hand side of (\ref{Eqn_2}) turns it into:
\begin{eqnarray}
\begin{array}{lll}
\displaystyle{\int_b^s{\frac{\partial}{\partial z}\left ( \kappa \frac{\partial T}{\partial z}\right) dz}
= \left[ \kappa \frac{\partial T}{\partial z}\right]^s_b 
= \kappa^s \left(\frac{\partial T}{\partial z}\right)^s 
 -\kappa^b \left(\frac{\partial T}{\partial z}\right)^b
\approx -\frac{\kappa^s}{h_s}\left (T - T^s \right )
  +\frac{\kappa^b}{h_b}\left (T^b - T \right )
},
\label{Eqn_11}
\end{array}    
\end{eqnarray}
where, $h_s$ and $h_b$ are some fractions of the flow depths ($1/h_s$ and $1/h_b$ are called the specific heat exchange areas, area per unit volume; Heinze, 2021, 2024) that are significantly influenced by the heat exchange across the free-surface and the basal substrate. And, $T^s$, $T^b$ are the ambient atmospheric and ground temperatures, respectively. 
The structure (\ref{Eqn_11}) is the formal description of heat fluxes across the avalanche boundary.
Nevertheless, for rapid motion of the rock-ice avalanche, conductive heat transfer may not be the primary process of heat transfer between the body and its surrounding.
\\[3mm]
In practice, however, the heat fluxes across the boundary are empirically expressed differently.
For simplicity, we can apply the Newton's law of heat exchange (transfer) across the free-surface of the avalanche and the basal sliding surface. Then, (\ref{Eqn_11}) can be written in alternative form as:
\begin{eqnarray}
\begin{array}{lll}
\displaystyle{\int_b^s{\frac{\partial}{\partial z}\left ( \kappa \frac{\partial T}{\partial z}\right) dz}
\approx -\Lambda^s (1+\lambda_v) A^s\left ( T - T^s\right)
  +\Lambda^b (1+\lambda_v) A^b\left ( T^b - T\right) 
},
\label{Eqn_12}
\end{array}    
\end{eqnarray}
where, $\Lambda^s, \Lambda^b$ [Wm$^{-2}$K$^{-1}$] are the free-surface and basal heat exchange coefficients; $A^s \,[-], A^b \,[-]$ are some fractions of the area exposed to the heat exchange at the free-surface and at the base; and $\lambda_v$ is a non-dimensional number with the magnitude resembling that of the landslide velocity, $|{\bf u}|$. As introduced here, $\lambda_v$ emerges legitimately, because, the convective heat exchange can also depend on the landslide velocity when it moves sufficiently fast, similar to the fluid flow in porous media (Heinze, 2021).
\\[3mm]
 Equations (\ref{Eqn_11}) and (\ref{Eqn_12}) represent two fundamentally different processes and two different ways of modelling heat exchanges between the solid surface and the ambient fluid (at the free-surface), and between two solid surfaces (along the bed). As the derivation shows, (\ref{Eqn_11}) is the formal description of heat fluxes across the avalanche boundary, in principle, it is perceived physically and mathematically better than (\ref{Eqn_12}). 
  As $\kappa^s$ and $\kappa^b$ are known from the phase fractions and the thermal conductivities of the phases at the free-surface and at the basal surface, 
   with (\ref{Eqn_11}),
  the heat fluxes (heat transfers) across the avalanche boundary is fully explained. This is an advantage over the model (\ref{Eqn_12}) as the heat transfer coefficient in (\ref{Eqn_12}) are difficult to measure and determine (Heinze, 2021).
\\[3mm] 
A crucial point in fast moving rock-ice avalanche is that, at the free-surface the heat exchange can be predominantly due to convective heat transfer. However, at the bottom, both the conductive and convective processes may contribute significantly and jointly. So, a better way to include both of these processes at the bed is by considering their linear combination with ${\mathcal P_b} \in [0, 1]$. Yet, in slow moving rock-ice avalanche, at the free-surface the heat exchange can be primarily due to conductive heat transfer.  
 This demands for the inclusion of both of the conductive and convective heat transfer at the free-surface by considering their linear combination, say, with ${\mathcal P_s} \in [0, 1]$. These considerations allow us combining (\ref{Eqn_11}) and (\ref{Eqn_12}) to develop a general model for the heat transfer across the boundary as:
{\small
\begin{eqnarray}
\begin{array}{lll}
\displaystyle{\int_b^s{\frac{\partial}{\partial z}\left ( \kappa \frac{\partial T}{\partial z}\right) dz}
= -\left [{\mathcal P_s}\Lambda^s (1+\lambda_v) A^s + \left (1 - {\mathcal P_s}\right)\frac{\kappa^s}{h_s} \right]\left ( T - T^s\right)
  +\left [{\mathcal P_b}\Lambda^b (1+\lambda_v) A^b + \left (1 - {\mathcal P_b} \right) \frac{\kappa^b}{h_b}\right]\left ( T^b - T\right) }.
\label{Eqn_12_total}
\end{array}    
\end{eqnarray}
\\[-1mm]
}
Depending on the situation, the basal heat transfer may be dominated either by the conductive $\left ({\mathcal P_b} \to 0\right )$ or, by the convective $\left ({\mathcal P_b} \to 1\right )$ process. However, for simplicity, one may consider the mean value with ${\mathcal P_b} = 0.5$.
 In principle, this also applies for the heat transfer at the free-surface of the avalanche.
\\[3mm]
The specific surface area can be calculated as $\displaystyle{\frac{1}{h_b} = 6(1-\alpha_{f})\frac{1}{\lambda_n d_{ri}}}$, where $d_{ri}$ is a typical grain diameter of the rock-ice mixture, which also applies to the cubic element of size $d_{ri}$, and $\lambda_n$ indicates that the effective heat exchange depth is a multiple of the grain size, typically, say 5 to 10. In general, for larger particle size $\lambda_n$ can be small, and for smaller particle size $\lambda_n$ can take larger values. More complex expressions for the specific area for flows in porous media are provided in Heinze (2024).
 For more on these and other parameters appearing in the above expressions, we refer to Robertson (1988), Crisp and Baloga (1990), Andersland and Ladanyi (2003), Costa and Macedonio (2005) and Clauser (2009).
 
\subsubsection{Frictional shear heating}

In avalanching motion, frictional shearing produces heat (Sosio et al., 2012; Pudasaini and Krautblatter, 2014).
We assume that, dominant shearing takes place in the $xz$ and $yz$ planes (in the flow depth direction) perpendicular to the sliding surface $xy$.
 Then, the heat produced by shear ($Q^+$) per unit volume, per unit time, is given by (Leloup et al., 1999; Young et al., 2022):
 \begin{eqnarray}
\begin{array}{lll}
\displaystyle{Q^+ = \tau_{xz}\frac{\partial u}{\partial z} + \tau_{yz}\frac{\partial v}{\partial z}},
\label{Eqn_13a}
\end{array}    
\end{eqnarray}
where $\tau_{xz}$ and $\tau_{yz}$ are the shear stresses in the $xz$ and $yz$ planes, respectively.
In (\ref{Eqn_13a}), $\tau_{xz} =\displaystyle{\rho_{ri} g^z (s-z)\mu_{ri} {\alpha_{ri}}} \frac{u}{|{\bf u}|}$ 
and
 $\tau_{yz} =\displaystyle{\rho_{ri} g^z (s-z)\mu_{ri} {\alpha_{ri}}} \frac{v}{|{\bf u}|}$ 
are the shear stresses with the slope normal load $\rho_{ri} g^z (s-z){\alpha_{ri}}$ of the avalanche, $\alpha_{ri} ={\left(\alpha_r+\alpha_i\right)}$ is the fraction, 
 $\rho_{ri} = (\rho_r\alpha_r + \rho_i\alpha_i)/(\alpha_r + \alpha_i)$ is the density, and  $\mu_{ri} = (\alpha_r\mu_r + \alpha_i\mu_i)/(\alpha_r + \alpha_i)$ is the basal friction coefficient for the rock and ice combined.
 Now, depth-averaging the heat source term $Q^+$ in (\ref{Eqn_13a}), we obtain:
\begin{eqnarray}
\begin{array}{lll}
\displaystyle{\overline{Q^+} = \l_s\,\frac{1}{2}\overline{\rho_{ri}}g^z{h}\overline{\alpha_{ri}}\,\overline{\mu_{ri}} \left [\frac{\overline{u}}{|{\overline{\bf u}|}}\left(u^s-u^b \right) + \frac{\overline{v}}{|{\overline{\bf u}|}}\left(v^s-v^b \right)\right]},
\label{Eqn_13}
\end{array}    
\end{eqnarray}
 where, 
 $\l_s$ is a number indicating that only a fraction of the shear heat is transferred to the avalanche causing the ice melt (Sosio et al., 2012), typically $\l_s\in(0.1, 0.5)$, but can even be lower for the rock-ice avalanches, and $|{\overline{\bf u}|}= \sqrt{{\overline{u}^2 + \overline{v}^2}}$. The velocity differences through the depth $\left(u^s-u^b \right)$ and $\left(v^s-v^b \right)$ can be approximated by the basal slip velocities $u^b$ and $v^b$ as $\left(u^s-u^b \right) = \chi_u u^b$ and $\left(v^s-v^b \right)= \chi_v v^b$, where $\chi_u > 0$ and $\chi_v > 0$ are some quantities. Note that $\chi_u \to 0$ and $\chi_v \to 0$ (typically, say 0.1) correspond to the situation closer to the basal slip, because, for this, $u^s \to u^b$ and $v^s \to v^b$, while $\chi_u \to 1$ and $\chi_v \to 1$ represent strong shearing through the depth, because, for this, $u^s \to 2 u^b$ and $v^s \to 2 v^b$. Other admissible values of $\chi_u$ and $\chi_v$ can model different scenarios. With this, (\ref{Eqn_13}) can be simplified to yield:
 \begin{eqnarray}
 	\begin{array}{lll}
 		\displaystyle{\overline{Q^+} = \frac{1}{2}\l_s\,\overline{\rho_{ri}}g^z{h}\,\overline{\alpha_{ri}}\,\overline{\mu_{ri}} 
 			\left (\chi_u\frac{\overline{u}}{|{\overline{\bf u}|}}u^b 			       	  +\chi_v\frac{\overline{v}}{|{\overline{\bf u}|}}v^b   \right).}
 		\label{Eqn_13aa}
 	\end{array}    
 \end{eqnarray}
  Because of the presence of the mixture properties $\overline{\rho_{ri}}$, $\overline{\alpha_{ri}}$ and $\overline{\mu_{ri}}$, the formally derived heat source in (\ref{Eqn_13aa}) is an extension to the previously considered frictional energy dissipation in Sosio et al. (2012), and its application to the two-phase rock-ice avalanche model and simulation by Pudasaini and Krautblatter (2014). But, the new heat source involves a factor $\chi/2$, where $1/2$ appears from the formal depth-averaging of the shear stress, and $\chi$ originates due to the in-depth shearing of the velocity. In this respect, even for a single-phase flow, the heat source in the previous models (Leloup et al., 1999; Sosio et al., 2012) are twice as much as what we formally derived here showing the substantial difference between the present model and the previous models which appear to overestimate the heat production.
 However, such a formal and general derivation and use of the frictional heat generation in the rock-ice avalanche, and its application to the temperature evolution (see later) of the truly multi-phase rock-ice-fluid debris avalanche is novel.

\subsubsection{Frictional heat dissipation in ice melting}

The frictional heat is dissipated as the ice melts. We assume that such a heat dissipation is proportional to the rate of ice melt (per unit density) due to friction (Pudasaini and Krautblatter, 2014). 
Then, from  (\ref{Eqn_13a}) 
\begin{eqnarray}
\begin{array}{lll}
\displaystyle{M_i = \frac{\l_s}{L_i\rho_i} \left[
	\tau_{xz}\frac{\partial u}{\partial z} + \tau_{yz}\frac{\partial v}{\partial z} 
	\right]},
\label{Eqn_14_0}
\end{array}    
\end{eqnarray}
which is a function of the latent heat $L_i$ (Andersland and Ladanyi, 2003) and involves the ice density $\rho_i$. So, the heat dissipation in melting is given by: 
\begin{eqnarray}
\begin{array}{lll}
	\displaystyle{Q^- =  \frac{\l_s\,C_i}{L_i}\frac{\rho_{ri}}{\rho_i}g^z(s-z)\mu_{ri} \alpha_{ri} \left (\frac{u}{|{\bf u}|}\frac{\partial u}{\partial z} + \frac{v}{|{\bf u}|}\frac{\partial v}{\partial z}\right)}.
\label{Eqn_14a}
\end{array}    
\end{eqnarray}
where, 
 the coefficient $C_i$ emerges due to the dimensional consistency, and has the dimension of $L_i \rho_i$ (with plausibly minimal values probably closer to unity). For ease of notation, we combine $\l_s$ and $C_i$ together, and simply write as $C_{\l_s} = \l_s C_i$. We call $C_{\l_s}$ the frictional ice melt coefficient
  due to shearing.
Depth-averaging (\ref{Eqn_14a}), and following the procedure as in (\ref{Eqn_13}) and (\ref{Eqn_13aa}), leads to
\begin{eqnarray}
\begin{array}{lll}
\displaystyle{\overline{Q^-} = 
\frac{1}{2}\frac{\overline{C_{\l_s}}}{{L_i}}\frac{\overline{\rho_{ri}}}{\overline{{\rho_i}}}g^z h \, \overline{\alpha_{ri}}\, \overline{\mu_{ri}} \left (\chi_u\frac{\overline{u}}{|\overline{{\bf u}}|}u^b + \chi_v\frac{\overline{v}}{|\overline{{\bf u}}|}v^b\right)	
	 }.
\label{Eqn_14}
\end{array}    
\end{eqnarray}
The heat dissipation in ice melting given by (\ref{Eqn_14}) is an extension to the heat dissipation utilized in Sosio et al. (2012) and its use in the mechanical modeling and dynamic simulation of the two-phase rock-ice avalanche by Pudasaini and Krautblatter (2014). However, the formal, consistent and the general derivation (\ref{Eqn_14}) of the heat dissipation in relation to the temperature evolution (see later) in truly multi-phase rock-ice-fluid avalanche is new. This is so, because, (\ref{Eqn_14}) includes: (i) the heat dissipation in both the flow directions, (ii) the formal mean
indicated by $\displaystyle{ \frac{1}{2}\left (\chi_u\frac{\overline{u}}{|\overline{{\bf u}}|}u^b + \chi_v\frac{\overline{v}}{|\overline{{\bf u}}|}v^b\right)}$
 of the slope normal load producing heat, and (iii) the mechanical consistency coefficient $C_i$, whereas  (\ref{Eqn_14}) characteristically involves the latent heat $L_i$, and the density of ice $\rho_i$, and $C_{\l_s}$ probably also contains the physical properties of the mixture. Moreover, the heat dissipation in ice melting (\ref{Eqn_14}) is rigorously derived, and is based on the mechanics of the rock-ice avalanche.

\subsubsection{Ice melt rates}

{\bf I. Ice melt rate due to frictional shearing}
\\[1mm]
 From the derivations above, we obtain the depth-averaged ice melt rate, or the fluid mass production rate, due to frictional shearing (indicated by the superscript $^s$ in $\mathcal M_p^s$) as:
\begin{eqnarray}
\begin{array}{lll}
\displaystyle{\mathcal M_p^s \!=\!\left[{\rho_i}\frac{1}{2}\frac{\l_s}{L_i}\frac{{\rho_{ri}}}{{{\rho_i}}}g^z{\mu_{ri}}\, {\alpha_{ri}} h\left (\!\chi_u\frac{u}{|{\bf u}|}u^b \!+\! \chi_v\frac{v}{|{\bf u}|}v^b\right)\!\right]
\!=\!  M_{pf}^s\left[{\alpha_{ri}}h\left (\!\chi_u\frac{u}{|{\bf u}|}u^b + \chi_v\frac{v}{|{\bf u}|}v^b\right)\right], M_{pf}^s\! =\! \frac{1}{2}\frac{{\l_s}}{L_i}{{\rho_{ri}}}g^z {\mu_{ri}}
},
\label{Eqn_17}
\end{array}    
\end{eqnarray}
where, $M_{pf}^s$ is the fluid mass production rate factor associated with the frictional shearing. These rate and factor are written in the structure that already appear in Pudasaini and Krautblatter (2014), but here, they are formally derived for the multi-phase mixture, and are in general form. 
\\[3mm]
{\bf II. Ice melt rate due to changing temperature}
\\[1mm]
The ice melt rate can also be related to the state of temperature. 
The amount of available energy for melting (denoted by the subscript $_m$ in $Q_m$) of frozen water is given by
\begin{eqnarray}
\begin{array}{lll}
Q_m = \rho_{i} c_i \left (T_i -T_{ph}\right),
\label{Eqn_melt_1}
\end{array}    
\end{eqnarray}
where, $T_{ph}$ is the temperature of the phase transition (Kelleners et al., 2016; Heinze, 2021), typically 0$^\circ$C (freezing/melting
point), and (\ref{Eqn_melt_1}) is applied when $T_i > 0$. In general, one may also replace $T_i$ in  (\ref{Eqn_melt_1}) by $T_{si}$, indicating the temperature of the material surrounding the ice.
Then, the amount of frozen water per unit volume that can be melted with this energy turns out to be (Kelleners et al., 2016; Heinze, 2021):
\begin{eqnarray}
\begin{array}{lll}
 -\displaystyle{\frac{Q_m}{L_i}  = -\frac{1}{L_i} \rho_{i} c_i \left (T_i -T_{ph}\right)}.
\label{Eqn_melt_2}
\end{array}    
\end{eqnarray}
Depth-averaging (\ref{Eqn_melt_2}) yields
\begin{eqnarray}
\begin{array}{lll}
 \displaystyle{-\overline{\rho_{i}}\frac{\overline{c_i}}{{L_i}} \left (\overline{T_i} -\overline{T_{ph}}\right)h}.
\label{Eqn_melt_3}
\end{array}    
\end{eqnarray}
 So, the time rate of change of the frozen water becomes:
\begin{eqnarray}
\begin{array}{lll}
\displaystyle{-\frac{\partial}{\partial t}\left[\overline{\rho_{i}}\frac{\overline{c_i}}{{L_i}} \left (\overline{T_i} -\overline{T_{ph}}\right)h\right]}.
\label{Eqn_melt_4}
\end{array}    
\end{eqnarray}
With this, the depth-averaged ice melt rate, or the fluid mass production rate, due to change in temperature (indicated by the superscript $^t$ in $\mathcal M_p^t$) is given by:
\begin{eqnarray}
\begin{array}{lll}
\displaystyle{\mathcal M_p^t =  \frac{\partial}{\partial t}\left[\frac{1}{{L_i}}{\rho_{i}}{c_i}\left ({T_i}h - {T_{ph}}h\right)\right]},
\label{Eqn_melt_5}
\end{array}    
\end{eqnarray}
where, as before, for simplicity, the over bars are removed. Heinze (2021) provides a numerical procedure on how to deal with $T_{ph}$ for the time difference form of (\ref{Eqn_melt_5}).  
\\[3mm]
Although, at a first glance, (\ref{Eqn_melt_5}) is very appealing, from a dynamical point of view, it bears its intricacy. There are two main reasons for that. First, it involves the temperature of the ice fraction $T_i$, rather than the bulk temperature $T$ as in the temperature equation (see Section 2.3). We may tackle this problem by assuming that the ice temperature and the bulk temperature is connected by a relation of the type $T_i = \lambda_i T$, where $\lambda_i$ needs to be parameterized, or given a proper value, probably closer to unity if the rock-ice mixture and the ice has similar temperature. 
 This can be perceived as a reasonable undertaking because, as the ice melts, the surrounding temperature declines resulting in a decreased bulk temperature.
Otherwise, its value may deviate away from unity. With this, (\ref{Eqn_melt_5}) can be simplified to yield
\begin{eqnarray}
\begin{array}{lll}
\displaystyle{\mathcal M_p^t =  \frac{\partial}{\partial t}\left[\frac{1}{{L_i}}{\rho_{i}}{c_i}\left (\lambda_i h {T} - h{T_{ph}}\right)\right]}.
\label{Eqn_melt_6}
\end{array}    
\end{eqnarray}
 Second, unlike (\ref{Eqn_17}), which does not involve any time or spatial rates of the field variables, (\ref{Eqn_melt_5}), or (\ref{Eqn_melt_6}) involves a time rate. So, whereas (\ref{Eqn_17}) is put (with proper sign, + for the fluid production rate, and - for the ice loss rate) on the right hand side of the balance equations for fluid and ice phases, due to its structure, (\ref{Eqn_melt_6}) needs to be united (with proper sign) with the time rate of changes of the fluid and ice masses on the left hand side of the respective mass balance equations (Pudasaini and Krautblatter, 2014). This way, (\ref{Eqn_melt_6}) can be well embedded into the standard phase mass balance equations considering the mass production and loss for the fluid and the ice phase in the rock-ice avalanche. The same applies to the momentum productions for the phases associated with the ice melting.
 \\[3mm]
 {\bf III. The effective ice melt rate}
 \\[1mm]
We now have two differently derived models to calculate the ice melt rate or the fluid production rate in the rock-ice avalanche: (i) as given by (\ref{Eqn_17}) due to shear heating and, (ii) given by (\ref{Eqn_melt_6}) due to the changing temperature, respectively. However, structurally and thermo-mechanically, they are essentially distinct, and they influence the mass and momentum balances differently. This is important to perceive. In principle, the mechanism of (\ref{Eqn_17}) and its impact in the balance equations are known (Pudasaini and Krautblatter, 2014). However, as (\ref{Eqn_melt_6}) is novel, its embedding into the mass and momentum balance equations needs to be analyzed carefully. Moreover, selection of (\ref{Eqn_17}) or (\ref{Eqn_melt_6}), or probably even better their combination, is a technical question for practitioners. This also helps to scrutinize the practical appropriateness of these models and their combination. 
\\[3mm] 
For fast moving landslides, the frictional shear heating may dominate the ice melting process over the ice melting due to the changes in temperature (from within or external sources). However, for slow moving landslides, the frictional shear heating may be small, but the ice melting due to the changing temperature of the landslide body (resulting from within or external sources, e.g., the heat fluxes across the basal surface and the free-surface) may be crucial. So, it is preferred to combine both of the malting processes.  Therefore, for general purpose, we suggest to take a linear combination between the ice melt rate due to the shear heating and the changing temperature. Let, the effective (total) ice melt rate be denoted by $\mathcal M_p$, and $\mathcal P_m \in [0, 1]$ provides a linear combination between $\mathcal M_p^s$ and $\mathcal M_p^t$. Then, from (\ref{Eqn_17}) and (\ref{Eqn_melt_6}), the effective and unified ice melt rate is given by
\begin{eqnarray}
\begin{array}{lll}
\displaystyle{\mathcal M_p =  
{\mathcal P_m}\left[\frac{1}{2}\frac{\l_s}{L_i}{{\rho_{ri}}}g^z{\mu_{ri}}\, {\alpha_{ri}} h\left (\!\chi_u\frac{u}{|{\bf u}|}u^b \!+\! \chi_v\frac{v}{|{\bf u}|}v^b\right)\!\right]
+\left ( 1- \mathcal P_m\right )\frac{\partial}{\partial t}\left[\frac{1}{{L_i}}{\rho_{i}}{c_i}\left (\lambda_i h {T} - h{T_{ph}}\right)\right]},
\label{Eqn_melt_7}
\end{array}    
\end{eqnarray}
where ${\mathcal P_m}$ closer to 1 represents the dominantly frictional heat induced ice melting, whereas ${\mathcal P_m}$ closer to 0 represents the ice melting rate principally dominated by the changes in the temperature. 
However, without loss of generality, one may also choose ${\mathcal P_m} = 0.5$, the mean value between $\mathcal M_p^s$ and $\mathcal M_p^t$, providing a simple possibility.
\\[3mm]
This completes the depth-averaging and developing the necessary closures associated with the thermo-mechanical processes associated with the energy conservation of the rock-ice avalanche.

\subsubsection{Internal mass and momentum exchanges}

As derived above, $-\mathcal M_p$ and $\mathcal M_p$ are the effective mass loss and production rates for the ice (solid) and fluid phases in the mixture. Multiplying these with the respective  ice (solid) and fluid velocities result in the internal momentum exchange between the ice and fluid.
\\[3mm]
However, unlike the basal erosion rate $E_t$, $-\mathcal M_p$ and $\mathcal M_p$ are associated with the ice and fluid, internal to the rock-ice avalanche. This means, there are two mass and momentum productions: (i) Internal to the mixture, the internal mass and momentum exchanges between the ice and fluid in the rock-ice avalanche as modeled by the  production and loss rates associated with $\mathcal M_p$. Nonetheless, this does not (substantially) change the mass (volume) of the rock-ice avalanche.  And, (ii) the mass and momentum production rates due to the entrainment of the basal material through the basal erosion rate $E_t$ (Pudasaini and Fischer, 2020; Pudasaini, 2025). This amplifies (changes) the mass of the rock-ice avalanche in proportion to the newly entrained basal mass into the avalanching material. So, these two mass and momentum productions are fundamentally different processes in the rock-ice avalanche. While the second process can occur in any erosive mass transport, the first process is a very special characteristic property of the rock-ice avalanche for which Pudasaini and Krautblatter (2014) proposed the first two-phase simulation model, that is extended here for the multi-phase rock-ice avalanche.  
However, as evident, depending on their inherent processes, either (i) or (ii) or both may play important role in the dynamics of the rock-ice avalanche.

\subsection{The temperature equation for rock-ice avalanche}

 With the entities, and the closures developed in (\ref{Eqn_6}), (\ref{Eqn_9}), (\ref{Eqn_10}), (\ref{Eqn_12_total}), (\ref{Eqn_13aa}), (\ref{Eqn_14}) and (\ref{Eqn_melt_6}), the temperature equation (\ref{Eqn_2}) takes the form:
 \begin{align}
\label{Eqn_16}
&\displaystyle{\frac{\partial}{\partial t} \left (\rho c\, h {T}\right) 
+ \frac{\partial}{\partial x}\left (\rho c \,h {uT}\right) 
+ \frac{\partial}{\partial y}\left (\rho c \,h {vT}\right) }\\\notag
&\displaystyle{= {\kappa} h\left\{\frac{\partial^2 {T}}{\partial x^2} 
                                  +\frac{\partial^2 {T}}{\partial y^2}\right\} 
+ h\left\{\frac{\partial {\kappa}}{\partial x}\frac{\partial {T}}{\partial x}
+ \frac{\partial {\kappa}}{\partial y}\frac{\partial {T}}{\partial y}\right\}
}\\\notag
&-\displaystyle{\left [{\mathcal P_s}\Lambda^s (1+\lambda_v) A^s + \left (1 - {\mathcal P_s}\right)\frac{\kappa^s}{h_s} \right]\left ( T - T^s\right)
  +\left [{\mathcal P_b}\Lambda^b A^b (1+\lambda_v)  +\left (1-{\mathcal P_b} \right) \frac{\kappa^b}{h_b}\right]\left ( T^b - T\right) }\\\notag
& \displaystyle{+\frac{1}{2}{\rho_{ri}}g^z{\alpha_{ri}}\,{\mu_{ri}}{h}\left[\l_s\left ( 1-{\mathcal P_m}\frac{C_i}{L_i \rho_i}\right)  \right]\left (\chi_u\frac{u}{|{\bf u}|}u^b + \chi_v\frac{v}{|{\bf u}|}v^b \right)}\\\notag
 & \displaystyle{-\left ( 1 - {\mathcal P_m}\right)\l_t\frac{\partial}{\partial t}\left[{\rho_{i}}{c_i}\left (\lambda_i h {T} - h{T_{ph}}\right)\right] }
 +  \rho^b c^b\, T^bE_t,
\end{align}
where, for notational convenience, the over bars have been omitted, and $\mathcal S_i =\displaystyle{\left[\l_s\left ( 1-{\mathcal P_m}\frac{C_i}{L_i \rho_i}\right)  \right]}$ is the net heat production factor due to frictional shear,
 and, $\l_t \in (0, 1)$ is a number indicating that the fraction of heat loss due to ice melt associated with the changing temperature is proportional to the ice melt rate due to changing temperature. One may simplify the situation by assuming $\l_t \to 1$, if not, it remains a parameter as stated. 
Moreover, even for $\chi_u \approx 1$ and  $\chi_v \approx 1$ we should retain the structure $\displaystyle{\left(\chi_v\frac{u}{|{\bf u}|}u^b + \chi_v\frac{v}{|{\bf u}|}v^b \right)}$ in this form without further simplification, because this preserves the information that shear resistances act against the landslide motions in the $xz$ and $yz$ planes. 
Furthermore, in (\ref{Eqn_16}), the temperature (heat) flux across the boundary, and the entrainment terms are kept as they are to retain the information of the boundary fluxes. Otherwise, these terms would disappear, or, do not properly represent the complex processes at the boundary, loosing the crucial  physics of rock-ice avalanches. 
 We call (\ref{Eqn_16}) the general temperature equation for rock-ice avalanche, the first of this kind in the simulation of relevant mass transport. 
\\[3mm]
The model (\ref{Eqn_16}) explains the advection-diffusion of the temperature $T$ of a column of rock-ice mass per unit (basal) area including the heat exchange across the avalanche body, basal heat conduction, the production and loss of heat due to the frictional shear heating and ice melting, and the enhancement of the temperature associated with the entrainment of the basal material. 
\\[3mm]
 It is important to note that as the deformation intensifies and the rock-ice mass rapidly propagates along the slope (Pudasaini and Krautblatter, 2014; Mergili et al., 2018; Shugar et al., 2021), its thermodynamic properties (namely $\rho, c, \kappa$) can evolve quickly. In this situation, we must very carefully consider the temperature equation (\ref{Eqn_16}) with respect to the appearance of these bulk parameters in their proper places.  

\subsection{Parameter estimates}

Most of the physical parameters  involved in the general temperature equation for rock-ice avalanche (\ref{Eqn_16}) are known or can be obtained from the literature (Robertson, 1988; Crisp and Baloga, 1990; Andersland and Ladanyi, 2003; Costa and Macedonio, 2005; Schneider et al., 2011a,2011b; Sosio et al., 2012; Pudasaini and Krautblatter, 2014; Young et al., 2022; Heinze, 2024). We list here some of the plausible parameter values:
densities: $\rho_r = 2900$, $\rho_i = 907, \rho_f = 1100$ [kgm$^{-3}$];
thermal conductivities:  $\kappa_r = 3.36$, $\kappa_i =  2.25$, $\kappa_f = 0.55$ [Wm$^{-1}$K$^{-1}$];
 specific heat capacities: $c_r =1000 , c_i =2000 , c_f = 4000$ [Jkg$^{-1}$ $^\circ C^{-1}$];
friction coefficients: $\mu_r = 0.8, \mu_i = 0.4 \,[-]$;
 and the latent heat for ice: $L_i = 333.7$ [KJm$^{-3}$].
 Based on Crisp and Baloga (1990) and Costa and Macedonio (2005):
 the heat exchange coefficient at the free-surface: $\Lambda^s = 70$ [Wm$^{-2}$K$^{-1}$], and at the bed $\Lambda^b$ can be assumed reasonably lower than $\Lambda^s$;
  and the fraction of areas exposed to the heat exchange at the free-surface and the sliding bed: $A^s, A^b  \in (0.001, 0.1) \,[-]$.
 Yet, the numerical values for $\Lambda^s$ and $A^s, A^b$ are derived from lava flows, and thus must be properly adjusted (may differ, or reduce, substantially) for the rock-ice avalanche motion. Moreover, due to the inclusion of the parameter $\lambda_v$ incorporating the velocity of the sliding mass, in reality, the parameter values mentioned above for $\Lambda^s$ and $\Lambda^b$ may be substantially lower.
  Furthermore, the atmospheric and ground temperatures $T^s, T^b$ should be properly prescribed, e.g., in the range [-10, 20]; ice melt coefficient: $C_l \approx 1.0$ (or, otherwise, see Section 2.5); and the
erosion rate $E_t$ (determined mechanically, Pudasaini and Fischer, 2020; Pudasaini, 2025), or selected empirically (Mergili et al., 2018). 
 Andersland and Ladanyi (2003) provides a collective list of thermal parameters for different geo-materials. Thermal conductivities: 
for water at 0$^\circ$C and 10$^\circ$C, $\kappa_w = 0.56$ and $0.58$;
 for ice at 0$^\circ$C, $\kappa_i = 2.21$; 
for sand gravel mixture $\kappa_{sg}= 1.3 - 1.7$;
and for quartz, $\kappa_q = 8.4$ [Wm$^{-1}$K$^{-1}$].
Similarly, the specific heat capacities are as follows: 
for water at 0$^\circ$C, $c_w = 4217$; 
for ice $c_i = 2090$; 
for the sand gravel mixture, $c_{sg} = 890$; 
and for quartz, $c_{q} = 733$ [Jkg$^{-1}$ $^\circ C^{-1}$], respectively.

\subsection{Mechanism and essence of the temperature equation} 

The temperature equation (\ref{Eqn_16}) for rock-ice avalanches is structurally interesting and possesses many important thermo-mechanical and dynamical properties of rock-ice avalanches. 
\\[3mm]
(i) It is mathematically consistent and physically-explained.
\\[3mm]
(ii) For constant temperature, the equation vanishes identically as the heat production and heat loss balance each other. From this identity, we may obtain the upper bound of $C_i$: $C_i \approx L_i\rho_i$. So, $C_i \in \left (1, L_i \rho_i\right)$ provides a plausible range for $C_i$. However, $C_i$ may also lie outside of this range as it may depend on the material and thermal properties of the rock-ice mass.
\\[3mm]
(iii) 
 The term $\displaystyle{\left[ {h}\left\{\frac{\partial {\kappa}}{\partial x}\frac{\partial {T}}{\partial x}
	+ \frac{\partial {\kappa}}{\partial y}\frac{\partial {T}}{\partial y}\right\}\right]}$
	on the right hand side of (\ref{Eqn_16})
	constitutes a very special structure and is probably one of the most important innovations here. It emerged due to the variation of the mixture (thermal) conductivity $\kappa$. As the constituent volume fractions evolve due to the fragmentation of the rock and ice, and melting of the ice, leading to the changes in the fractions of the rock, ice and the fluid in the mixture, $\kappa$ evolves accordingly. The intensity of fragmentation and melting determines the rates of changes of $\kappa$, i.e., $\partial \kappa/\partial x$ and $\partial \kappa/\partial y$. So, this special structure tells that it contains two coupled dynamics in it: the rate of changes of $\kappa$ (i.e.,  $\partial \kappa/\partial x$ and $\partial \kappa/\partial y$), and the rate of changes of temperature $T$ (i.e.,  $\partial T/\partial x$ and $\partial T/\partial y$). In contrast to the term associated with  classical diffusion 
	$ \displaystyle{\left[ {\kappa h} \left\{\frac{\partial^2 {T}}{\partial x^2} 
	+\frac{\partial^2 {T}}{\partial y^2}\right\}	\right]}$
	and advection 
	$\displaystyle{\left[\frac{\partial}{\partial x}\left (\rho c\, h {uT}\right) 
		+ \frac{\partial}{\partial y}\left (\rho c\, h {vT}\right)\right]}$
	processes, 
$\displaystyle{\left[{h} \left\{\frac{\partial {\kappa}}{\partial x}\frac{\partial {T}}{\partial x}
	+ \frac{\partial {\kappa}}{\partial y}\frac{\partial {T}}{\partial y}\right\}
	\right]}$
 is the only term that involves variations of both of $\kappa$ and $T$. So, this term characterizes the special thermo-mechanical processes and their dynamical consequences in rock-ice avalanche propagation in a coupled way as it describes the changing state of the moisture content of the rock-ice mass connected to the latent heat.
 \\[3mm]
 (iv) It includes the heat production due to shearing in both the flow directions.	
\\[3mm]
(v) The heat exchanges through the boundary of the rock-ice avalanche with its surrounding play crucial role in the temperature evolution. 
The model (\ref{Eqn_16}) includes the velocity dependent heat transfer uniquely combining both the convective and conductive heat transfer processes at the free-surface and the bed.
 If the heat entrainment (or detrainment) across the avalanche boundary is substantial, then, the temperature changes are rapid. 
\\[3mm]
(vi) Heat production is proportional to the normal load of the avalanche. So, thicker (heavier) avalanches produce more heat than the thinner (lighter) avalanches. 
\\[3mm]
(vii) Depending on the shearing in the longitudinal or the lateral direction, either one or both of the longitudinal or lateral heat productions can play important to dominant role in the temperature evolution. 
 The factor $\displaystyle{\left[\chi_u({u}/{|{\bf u}|})u^b + \chi_v({v}/{|{\bf u}|})v^b \right]}$ in (\ref{Eqn_16}) explains this behavior. Examples include the locally expanding and/or compacting flows in the inception, in the transition to the run-out, or the deposition process.
\\[3mm]
(viii) Fast moving (or deforming) avalanches produce higher amount of heat than the slow moving (deforming) avalanches. This is explained by the shear rates $\partial u/\partial z$ and $\partial v/\partial z$ in (\ref{Eqn_13a}), equivalently,  
$\displaystyle{\left[\chi_u({u}/{|{\bf u}|})u^b \right]}$ and 
$\displaystyle{\left[\chi_v({v}/{|{\bf u}|})v^b \right]}$ in (\ref{Eqn_16}).
\\[3mm]
(ix) The ice melting rate controls the change of temperature of the rock-ice avalanche. Conversely, the changing temperature controls the ice melt rate. Fast ice melting means substantial change in the temperature of the avalanching body. Fluid mass production or ice mass loss also depends on the melting efficiency as described by the unified ice melt rate (\ref{Eqn_melt_7}) which is a remarkable development here.
 In principle, re-freezing of the fluid (water) to ice can take place (Heinze, 2024). However, during the rapid motion of the rock-ice mass (Pudasaini and Krautblatter, 2014; Shugar et al., 2021), this is less likely to happen.
\\[3mm]
(x) Entrainment of the basal material into the propagating body changes the state of temperature of the rock-ice avalanche. The rock-ice avalanche temperature is enhanced if the entrained mass has higher temperature than the avalanche, and the rock-ice avalanche temperature is reduced if the entrained mass has lower temperature  than the avalanche itself. 
\\[3mm]
All these intrinsic properties indicate that rock-ice avalanches are more complex than the other types of avalanches. In principle, none of these (mostly new) aspects can be compromised for rock-ice avalanches.  

\subsection{The model structure, unification and coupling}

The temperature equation (\ref{Eqn_16}) and the effective ice melt rate (\ref{Eqn_melt_7}) are now ready to be combined with the existing multi-phase mass flow model, together with the two-phase rock-ice avalanche model and the multi-phase erosion model. 
 The final stage of the model development involves: (i) considering the multi-phase mass flow model (Pudasaini and Mergili, 2019), considering the rock, ice and fluid as three constituents in the avalanching mixture, (ii) utilizing the principles of ice melting, internal mass and momentum exchanges between the solid (rock and ice) and fluid, basal lubrication and internal fluidization described by the two-phase mechanical rock-ice avalanche model (Pudasaini and Krautblatter, 2014), (iii) incorporating the unified multi-phase entrainment model (Pudasaini, 2025), and (iv) embedding the novel temperature evolution equation and the proposed unified ice melt rate for rock-ice avalanche into the multi-phase equations. 
\\[3mm]
There are nine conservative state variables in the multi-phase mass flow model (Pudasaini and Mergili, 2019). These are:
the phase flow depths: $(h_r = h\alpha_r, h_i=h\alpha_i, h_f=h\alpha_f)$, where $\alpha_r+\alpha_i+\alpha_f = 1$ is the hold-up identity and $h_r+h_i+h_f = h$ (the total flow depth); and the phase fluxes in the down slope and cross slope directions: $(h_ru_r, h_iu_i, h_fu_f)$, $(h_rv_r, h_iv_i, h_fv_f)$, respectively. By definition, the bulk velocities are $u = \alpha_r u_r + \alpha_i u_i + \alpha_f u_f$, and $v = \alpha_r v_r + \alpha_i v_i + \alpha_f v_f$. The same applies to the bulk density, specific heat capacity and thermal conductivity: 
$\rho =  \alpha_r\rho_r +  \alpha_i\rho_i +  \alpha_f\rho_f$, 
$c =  \alpha_r c_r + \alpha_i c_i + \alpha_f c_f$ and 
$\kappa =  \alpha_r\kappa_r +  \alpha_i\kappa_i + \alpha_f\kappa_f$. The temperature equation (\ref{Eqn_16}) and the ice melting rate (\ref{Eqn_melt_7}) contain these quantities. This, together with the other parameters involved in  (\ref{Eqn_melt_7}) and (\ref{Eqn_16}), indicates a strong coupling between the existing mass and the momentum balance equations for the phases and the new temperature equation for the bulk motion of the rock-ice avalanche and the unified ice melt rate. 
 The model (\ref{Eqn_16}) introduces an additional conservative variable $\rho c \,hT$ (for the transport of the bulk temperature) to the set of existing conservative variables in Pudasaini and Mergili (2019). 
So, all together, there are ten conservative (state) variables in the newly proposed multi-phase rock-ice avalanche model.
\\[3mm]
The existing and the widely used mass flow simulation codes, e.g., Pudasaini and Krautblatter (2014), and  r.avaflow (Pudasaini and Mergili, 2019; Mergili and Pudasaini, 2025) can be directly expanded to include the new ice melt rate (\ref{Eqn_melt_7}) and the temperature model (\ref{Eqn_16}) into the computational framework, making it a comprehensive tool for the simulation of rock-ice avalanche. This is a structural advantage.  
 This completes the construction of the novel, multi-phase, thermo-mechanical model for rock-ice avalanches. 
 \\[3mm]
 However, we note that, in principle, with the properly chosen physical parameters, the model developed here may also be applied to the scenario of rock melting.
 
 \section{Temperature evolution: simple model and analytical solutions}
 
 Here, we develop the first, thermo-mechanically described, simple advection-diffusion-decay-source model for the evolution of the temperature of the rock-ice avalanche with the ice melting and entrainment. Then, we construct exact-analytical solutions for the temperature evolution of the propagating rock-ice mass. We reveal important thermo-mechanical and dynamical features of the temperature evolution. This offers fundamentally novel, analytical understanding of the complex process of rock-ice avalanche propagation, flashing the deep insights into the underlying dynamics.

\subsection{A simple thermo-mechanical advection-diffusion-decay-source model for temperature}

\subsubsection{The temperature equation}

By assuming a slab-type mass of constant bulk properties propagating with a terminal velocity, the temperature equation for rock-ice avalanche (\ref{Eqn_16}) can be simplified as:
\begin{eqnarray}
\begin{array}{lll}
\displaystyle{\frac{\partial T}{\partial t} + C\frac{\partial T}{\partial x} -{D} \frac{\partial^2 T}{\partial x^2} = -P_d T + P_s},
\label{Equation_AdvecDiffDecSource}
\end{array}    
\end{eqnarray}
\\[-2mm]
where, $C = {\widetilde C_x}/{\widetilde C_t},\,\,\, D = \widetilde D/{\widetilde C_t}, \,\,\,P_d = {\widetilde P_d}/{\widetilde C_t}, \,\,\,P_s = {\widetilde P_s}/{\widetilde C_t}$
 correspond to the advection, diffusion, decay and source entities for the evolution of the rock-ice mass temperature 
with the definitions:
\begin{eqnarray}
\begin{array}{lll}
\displaystyle{{\widetilde C_t} = \left[ 1+ \left(1 - P_m \right)\l_t\lambda_i\frac{\rho_ic_i}{\rho c}\right]},\,
\displaystyle{{\widetilde C_x} = \left[ u - \frac{1}{\rho c}\frac{\partial \kappa}{\partial x}\right]},\,
\displaystyle{{\widetilde D} = \left[ \frac{\kappa}{\rho c}\right]},\\[4mm]
\displaystyle{{\widetilde P_d} = 
 \left (\left[ P_s\Lambda^s\left ( 1+\lambda_\nu\right)A^s + \left(1-P_s\right)\frac{\kappa^s}{h_s}\right]
+\left[ P_b\Lambda^b\left ( 1+\lambda_\nu\right)A^b + \left(1-P_b\right)\frac{\kappa^b}{h_b}\right]\right)\frac{1}{\rho c h}},\\[4mm]
\displaystyle{{\widetilde P_{sb}} = 
 \left (\left[ P_{s}\Lambda^s\left ( 1+\lambda_\nu\right)A^s + \left(1-P_s\right)\frac{\kappa^s}{h_s}\right]T^s
+\left[ P_{b}\Lambda^b\left ( 1+\lambda_\nu\right)A^b + \left(1-P_b\right)\frac{\kappa^b}{h_b}\right]T^b\right)\frac{1}{\rho c h}},\\[4mm]
\displaystyle{{\widetilde P_{m\chi}} = \frac{1}{2}\rho_{ri}\left[\l_s\left(1 - P_m \right)\frac{c_i}{\l_i\rho_i}\right]\chi_i\frac{\lambda^b u}{\rho c}},\,
\displaystyle{{\widetilde T_E} = \frac{\rho^b c^b}{\rho c h}T^b E_t},\,
{\widetilde P_s} = {\widetilde P_{sb}} + {\widetilde P_{m\chi}} +{\widetilde T_E}.
\label{Equation_AdvecDiffDecSource_Param}
\end{array}    
\end{eqnarray}
Here, the decay (rate) is the first-order amplification of the temperature. 
However, note that, we can keep the advection entity $C$ inside the differential, so we do not necessarily need to assume about the speed of the avalanche. If some functional relations for $u$ in terms of $T$ or $x$ can be prescribed, analytical solutions may still be constructed for that seemingly more complex situation. The same applies for the diffusion entity $D$.
For the sole aim of developing a simple temperature evolution equation for the rock-ice avalanche, we made some justifiable assumptions, such that the resulting model equation (\ref{Equation_AdvecDiffDecSource}) took the  standard form. This later facilitates for the construction of exact analytical solutions in closed form from which we gain some enlightenments on the basic mechanical features of the propagating rock-ice mass with changing temperature. 

\subsubsection{Important features of the temperature equation} 

There are some important mechanical and dynamical features of the temperature evolution model (\ref{Equation_AdvecDiffDecSource}) manifesting the first-ever revelations in connection to the evolution of the temperature of the rock-ice mass.
(i) ${\widetilde C_t} >1$ is a finite quantity, probably closer to 1. We can combine ${\widetilde C_t}$ with $t$ and define $t/{\widetilde C_t} =  {\widetilde t}$. This results in a temperature evolution model equivalent to (\ref{Equation_AdvecDiffDecSource}). After constructing the solution, one can recall the natural time $t$, better for acquiring the temperature in the physical time scale. This implies that, with internal ice melting, the time evolution process of temperature of the rock-ice mass is enhanced.  
(ii) Since $\partial \kappa/\partial x < 0$ is the most probable scenario, the melting of ice enhances the advection of the rock-ice avalanche. 
(iii) ${\widetilde C_t}$, or the internal ice melting, delays the process of advection, diffusion, decay and the effects of the heat exchanges through the avalanche boundaries, shearing and entrainment. 
(iv) Which parameters in (\ref{Equation_AdvecDiffDecSource_Param}) play dominant, substantial or negligible role in the temperature evolution (\ref{Equation_AdvecDiffDecSource}) depends on the rock-ice avalanche dynamics and the magnitude of the physical parameters associated with the state of the considered mass and boundary conditions. These are fundamentally novel understanding of the complex process of rock-ice propagation, flashing the deep insights of the underlying dynamics.

\subsection{Analytical solutions to thermo-mechanical rock-ice avalanche temperature equation}

Within the given circumstances, exact-analytical solutions are the best  solutions for the underlying problem (Pudasaini and Krautblatter, 2022). As such, these solutions are fascinating.  
The model equation (\ref{Equation_AdvecDiffDecSource}) for the evolution of the temperature of the propagating rock-ice mass can be solved exactly for different physically relevant initial and boundary conditions. 
We present two types of extended solutions: 
 (i) fundamental solution in infinite domain, and (ii) series solution in finite domain. The basic procedures for constructing both the solutions are similar (van Genuchten and Alves, 1982; Balluffi et al., 2005;  Kelleners et al., 2016; Cushman-Roisin, 2021). First, look at the simple heat diffusion process and consider the corresponding basic solution for rock-ice mass. Second, apply advection in the solution structure obtained from diffusion by translating its domain by $Ct$ along the down-stream of propagation. Third, amend the diffusion-advection solution by embedding into it the exponential time decay of the instantaneous temperature, $\exp(-P_d t)$. Fourth, inclusion of the temperature source is achieved by blending the linear temporal change in the temperature, $P_st$ with the diffusion-advection-decay solution. This completes the construction process of the full analytical solution for the advection-diffusion-decay-source model for the evolution of the rock-ice avalanche temperature.     

\subsubsection {Fundamental solution in infinite domain}

\begin{figure}[t!]
\begin{center}
  \includegraphics[width=18cm]{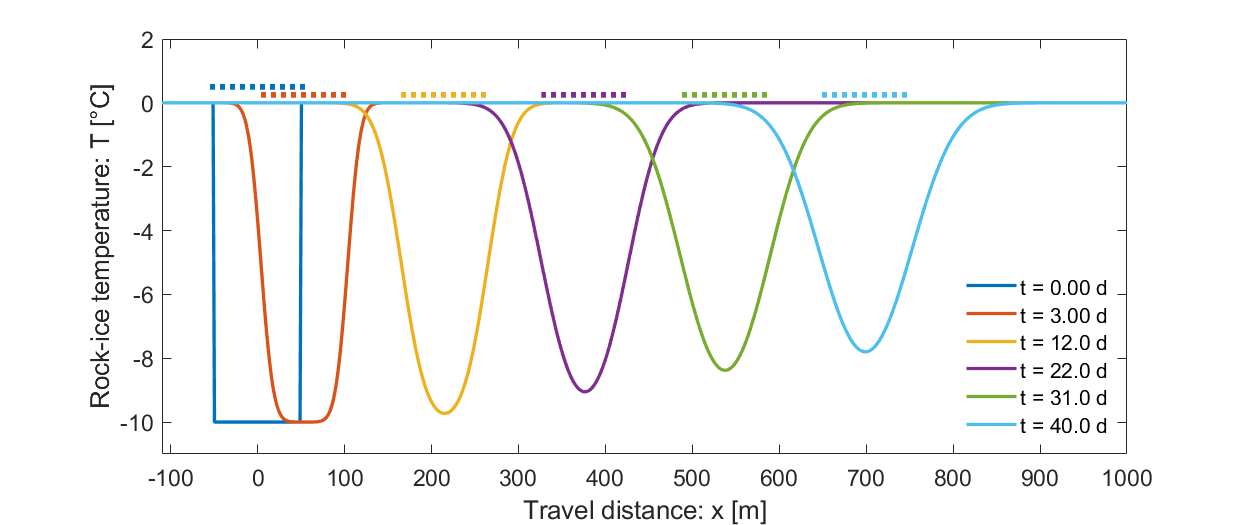}\\
  \includegraphics[width=18cm]{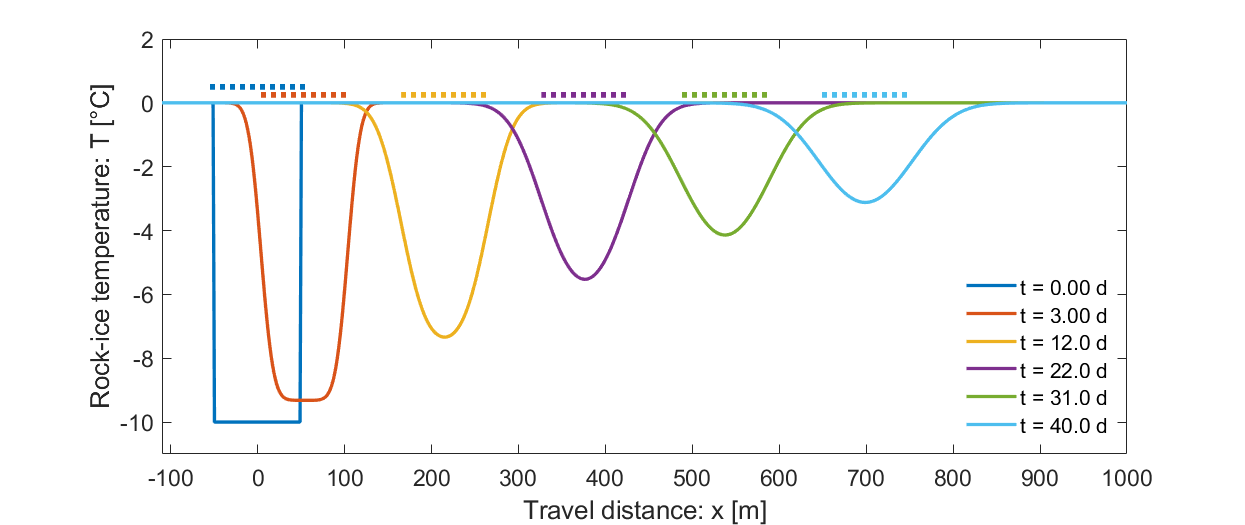}
  \end{center}
  \vspace{-5mm}
  \caption[]{Temperature evolution in rock-ice avalanche given by the solution (\ref{Equation_AdvecDiffDecSource_Soln}), involving: (A) Advection-diffusion, (B) advection-diffusion-decay processes, respectively. The top dotted-line segments indicate the actual positions of the 100 m long rock-ice mass propagating at the speed of $43.2$ md$^{-1}$.}
  \label{Equation_AdvecDiffDecSource_Fig1}
  \begin{picture}(0,0)
\put(450,462){\bf A}     
\put(450,245){\bf B}    
\end{picture}
\end{figure}
\begin{figure}[]
\begin{center}
  \includegraphics[width=13.5cm]{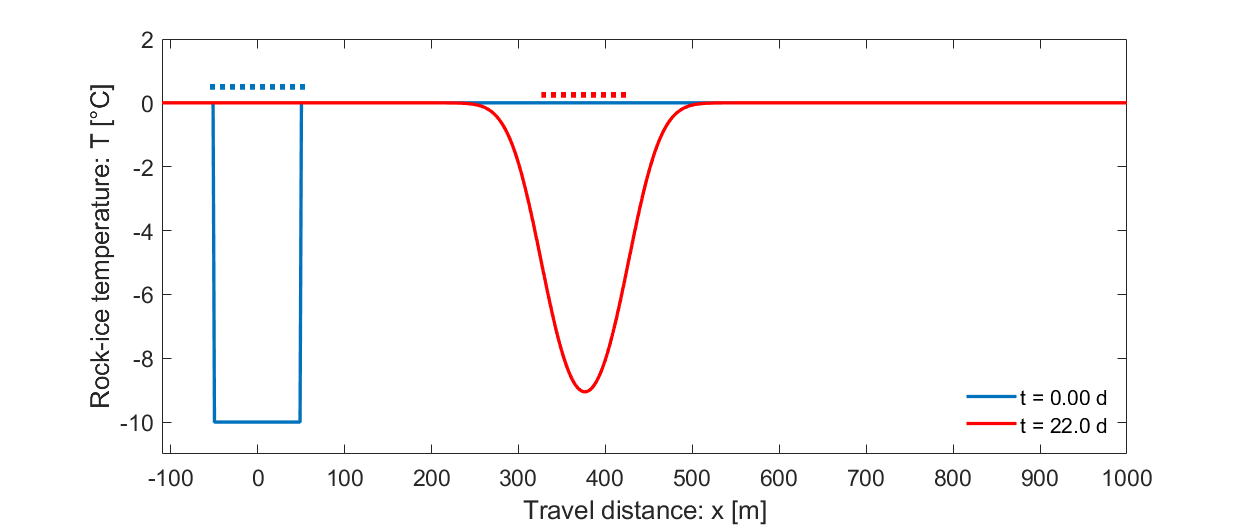}
  \includegraphics[width=13.5cm]{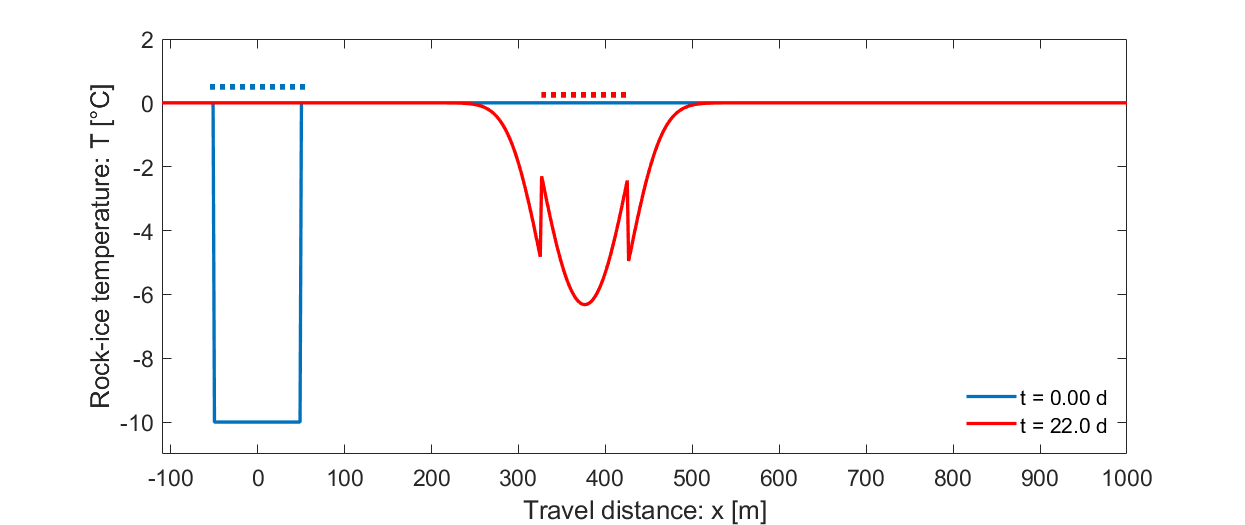}
  \includegraphics[width=13.5cm]{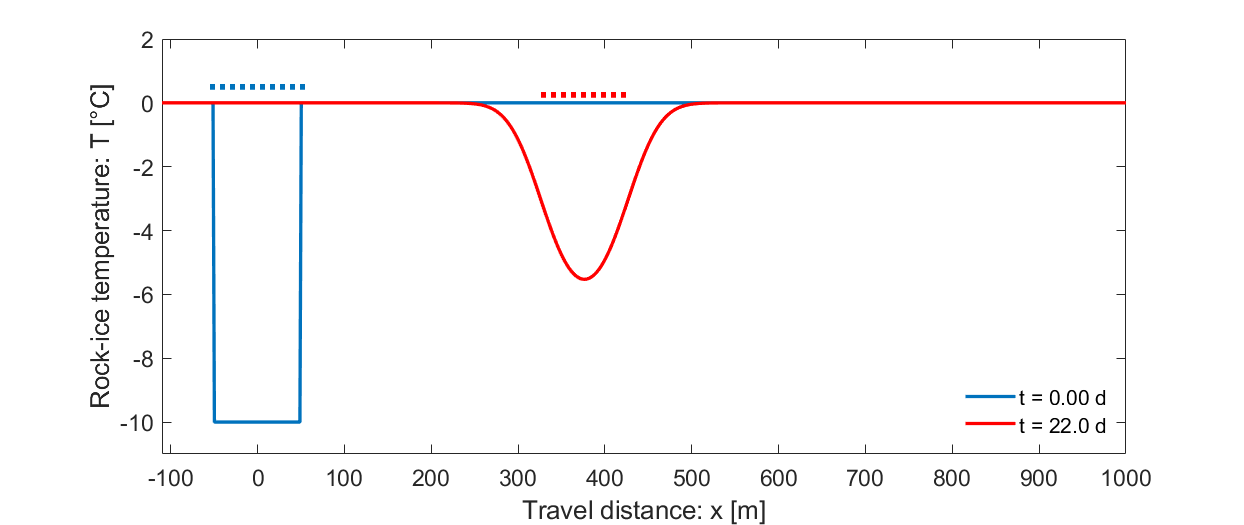}
  \includegraphics[width=13.5cm]{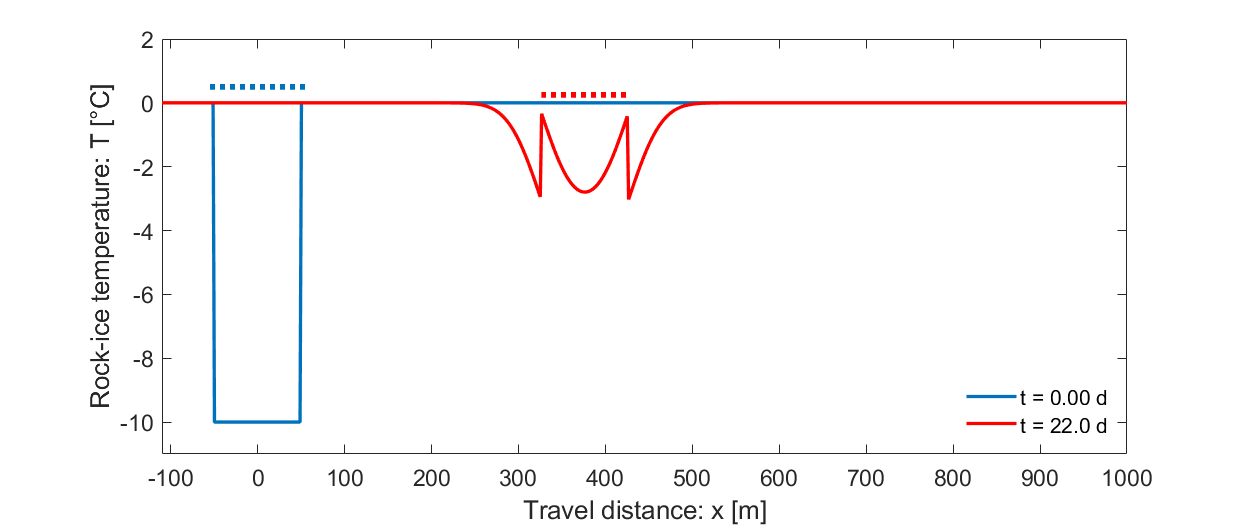}
  \end{center}
  \vspace{-5mm}
  \caption[]{Temperature evolution in rock-ice avalanche given by the solution (\ref{Equation_AdvecDiffDecSource_Soln}): (A) Advection-diffusion, (B) advection-diffusion-source, (C) advection-diffusion-decay, (D) advection-diffusion-decay-source, respectively. The top dotted-line segments indicate the actual positions of the propagating rock-ice mass.}
  \label{Equation_AdvecDiffDecSource_Fig1S}
  \begin{picture}(0,0)
\put(401,689){\bf A}     
\put(401,525){\bf B}    
\put(401,362){\bf C}  
\put(401,199){\bf D}
\end{picture}
\end{figure}
 The analytical solution with the advection-diffusion-decay-source ($C, D, P_d, P_s$) for the evolution of the temperature (\ref{Equation_AdvecDiffDecSource}) in the avalanching rock-ice mass of length $L$ and the thickness $h$ involving the fundamental error function $erf$ takes the form:
 {\small
\begin{eqnarray}
\begin{array}{lll}
\displaystyle{T (x,t)= \frac{T_0}{2}\!\left[erf\!\left ( \frac{x+\frac{L}{2}- C\,t}{\sqrt{4\,D\,t}}\right) -
                                  erf\!\left ( \frac{x-\frac{L}{2}- C\,t}{\sqrt{4\,D\,t}}\right)\!\right]\!\exp{(-P_d\,t)}
            +\left[{\mathcal H}\left(x+\frac{L}{2}- C\,t\right)-{\mathcal H}\left(x-\frac{L}{2}- C\,t\right)\right] \!P_s t},
\label{Equation_AdvecDiffDecSource_Soln}
\end{array}    
\end{eqnarray}
}
\hspace{-3.mm}
where ${\mathcal H}$ is the Heaviside function. Note that, by definition, $\left[{\mathcal H}\left(x+\frac{L}{2}- C\,t\right)-{\mathcal H}\left(x-\frac{L}{2}- C\,t\right)\right] P_s \equiv P_s$ on the domain of influence as its support rests on the avalanche domain of interest. This legitimately filters out the irrelevant region as it comes to the source of heat production. Moreover, the derivative of the Heaviside function ${\mathcal H(\chi)}$ is the Dirac delta functional, $\delta(\chi)$, and from the perspective of the domain of influence of the heat source, one may perceive $\delta(a)-\delta(b) \equiv 0$ for any $a, b$. 
Otherwise, the Heaviside function can be lifted.
It can be routinely proven laboriously that (\ref{Equation_AdvecDiffDecSource_Soln}) identically satisfies (\ref{Equation_AdvecDiffDecSource}).
So, the fundamental structure (\ref{Equation_AdvecDiffDecSource_Soln}) offers the first entire analytical solution for the advection-diffusion-decay-source problem (\ref{Equation_AdvecDiffDecSource}).  
\\[3mm]
Results of the model solution (\ref{Equation_AdvecDiffDecSource_Soln}) are presented in Fig. \ref{Equation_AdvecDiffDecSource_Fig1} for different situations with increasing complexity for a propagating mass of length $L = 100$ m with the initial temperature $T_0 = -10^\circ$C. This includes (A) advection-diffusion and (B) advection-diffusion-decay,
respectively. These two scenarios represent fundamentally different dynamics. The location of the center of mass $x = 0$ represents the reference point as $t = 0$ indicates the incipient motion for which the rock-ice mass has the temperature $T_0$.
Based on the parameters described at Section 2.4, and without loss of generality, the collective advection, diffusion and decay parameters are plausibly estimated as: $C = 2.0084\times10^{-4}, D = 2.3894\times10^{-4}, P_d = 2.6319\times10^{-7}$. 
However, depending on the real flow situation, these parameters can take wide range of values controlling the dynamics as the diffusion process can be substantially slower than the one used here to speed-up this process.
\\[3mm]
The fundamental solution assumes the constant temperature ($0^\circ$C) sufficiently far upstream and down-stream of the rock-ice mass. The mass is propagating in a relatively warm area of the mountain valley possibly in the lower region of the track or the run-out in the foot-hill.  
However, these scenarios can be adjusted in-line-with the field situations that may require structural amendments to (\ref{Equation_AdvecDiffDecSource_Soln}), a technical-mathematical aspect. 
\\[3mm]
The evolution of the temperature profile of the propagating rock-ice mass with the advection-diffusion process is shown in Fig. \ref{Equation_AdvecDiffDecSource_Fig1}A for more than a month time. As the mass propagates downslope, the temperature increases slowly, faster around the end points of the rock-ice body, but relatively slowly towards its interior. There are three different processes involved here. (i) The advection of the rock-ice mass as represented by its spatial location as the time proceeds. (ii) The heat flux through the boundaries. And, (iii) the diffusion of heat across the rock-ice mass.
\\[3mm]
Now, the curiosity is how would the decay process influences the temperature evolution. To analyze this, Fig.~\ref{Equation_AdvecDiffDecSource_Fig1}B depicts the temperature evolution that considers advection, diffusion and decay. For the applied decay rate of $P_d = 2.6319\times10^{-7}$, the temperature profiles are extraordinarily different than those without considering the decay, as for now the temperature increases dramatically quickly as the rock-ice slab temperature tends to stabilize to the ambient temperature of $0^\circ$C in 40 days. However, in the same time duration, without decay, the temperature change was relatively small. This implies that the decay of temperature can be a dominating process in the temperature evolution of the rock-ice avalanche as revealed here with our novel rock-ice avalanche temperature evolution equation (\ref{Equation_AdvecDiffDecSource}) and its analytical solution  (\ref{Equation_AdvecDiffDecSource_Soln}).
\\[3mm]
However, the source must be handled very carefully as this involves three fundamentally different aspects of heat productions: at the rock-ice boundary, internal to it, and via the entrainment of the basal material.  The question arises, what happens when the diffusively travelling temperature wave without or with decay suddenly encounters a heat source on the way of the rock-ice mass. 
Fig.~\ref{Equation_AdvecDiffDecSource_Fig1S} presents some typical solutions at $t = 22.0$~d with the influence of the heat source term with moderate value of $P_s = 1.4574\times10^{-6}$. It is evident that in the domain of influence of the source, both with advection-diffusion and advection-diffusion-decay, the heat source substantially to dominantly elevates the instantaneous temperature of the propagating rock-ice mass in a relatively short time (panel B and panel D). Clearly, outside of the heat source, the temperature profile is defined by the other processes of advection-diffusion (panel A) and advection-diffusion-decay (panel C) only. This causes jumps in the previously smooth temperature at the margins of the rock-ice mass. Such complex, but spectacular behavior, of the temperature evolution of the rock-ice mass is disclosed here for the first time with our novel temperature equation (\ref{Equation_AdvecDiffDecSource}) and its analytical solution (\ref{Equation_AdvecDiffDecSource_Soln}).

\subsubsection {Series solution in finite domain}

By prescribing the temperature at the end boundaries (for simplicity $0^\circ$C) of the propagating rock-ice mass of length $L$, we can construct the infinite series solution for the finite domain of support, for the advection-diffusion-decay-source of the temperature, as follows: 
{\footnotesize
\begin{eqnarray}
\begin{array}{lll}
{{T (x,t)= \frac{4 T_0}{\pi}\!\sum_{j = 0}^{\infty} {\frac{1}{2 j +1}\sin\!\left [(2 j +1) \left ( x-Ct\right)\!\frac{\pi}{L}\right ]}\exp \!\left [ -(2 j +1)^2 D t\frac{\pi^2}{L^2}\right ]\!\exp{(-P_d\,t)}
+\left[{\mathcal H}\!\left(x+\frac{L}{2}- C\,t\right)-\!{\mathcal H}\!\left(x-\frac{L}{2}- C\,t\right)\right]\!P_s t}}.
\label{Equation_AdvecDiffDec_Soln2}
\end{array}    
\end{eqnarray}
}
\hspace{-4.mm}
\begin{figure}[t!]
\begin{center}
  \includegraphics[width=18cm]{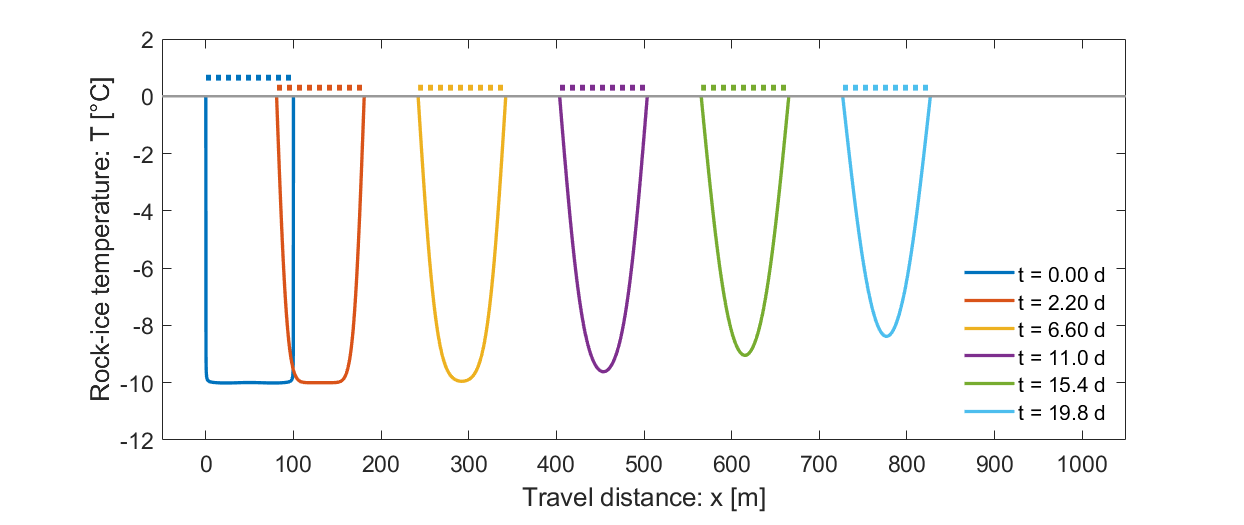}
  \includegraphics[width=18cm]{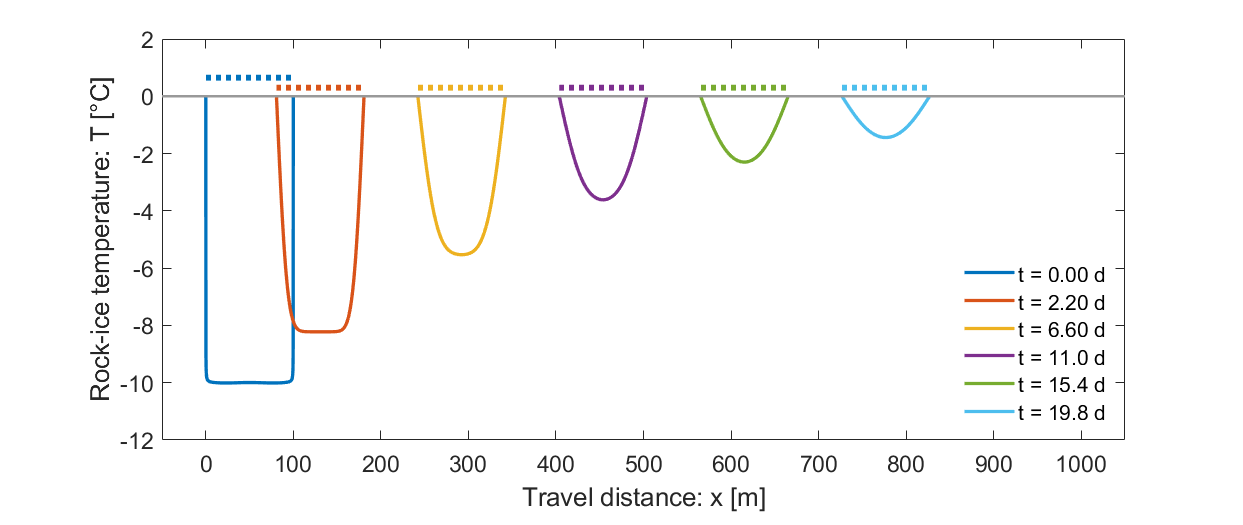}
  \end{center}
  \vspace{-5mm}
  \caption[]{Temperature evolution in rock-ice avalanche of length 100 m given by the solution (\ref{Equation_AdvecDiffDec_Soln2}), involving: (A) Advection-diffusion, (B) advection-diffusion-decay processes, respectively. The horizontal gray line represents the ambient temperature. Top dotted-line segments indicate the actual positions of propagating rock-ice mass.}
  \label{Equation_AdvecDiffDecSource_Fig2}
  \begin{picture}(0,0)
\put(451,463){\bf A}     
\put(451,246){\bf B}    
\end{picture}
\end{figure}
\begin{figure}[]
\begin{center}
  \includegraphics[width=13.5cm]{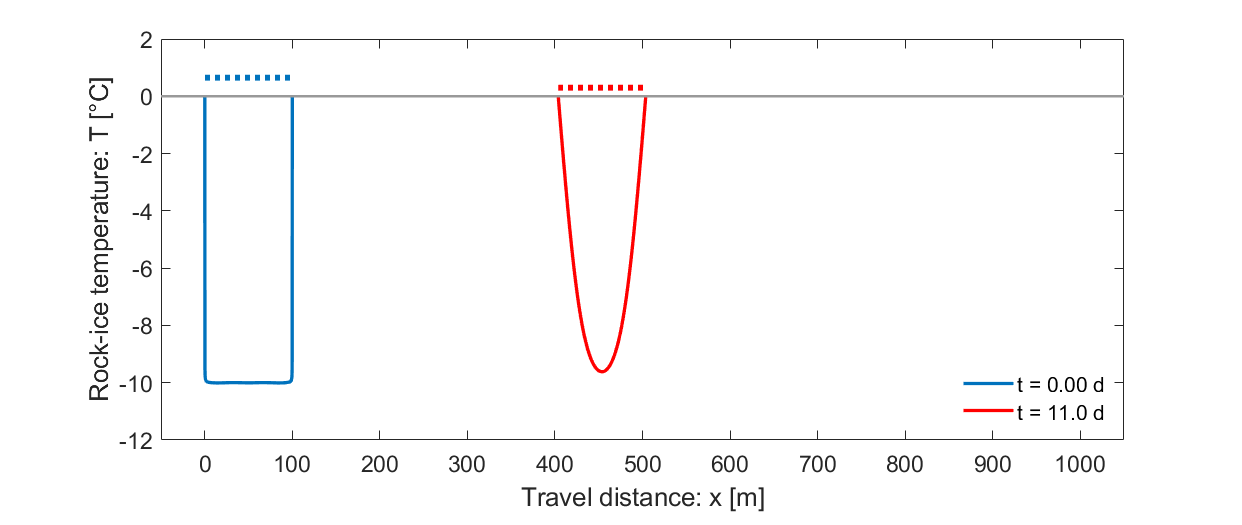}
  \includegraphics[width=13.5cm]{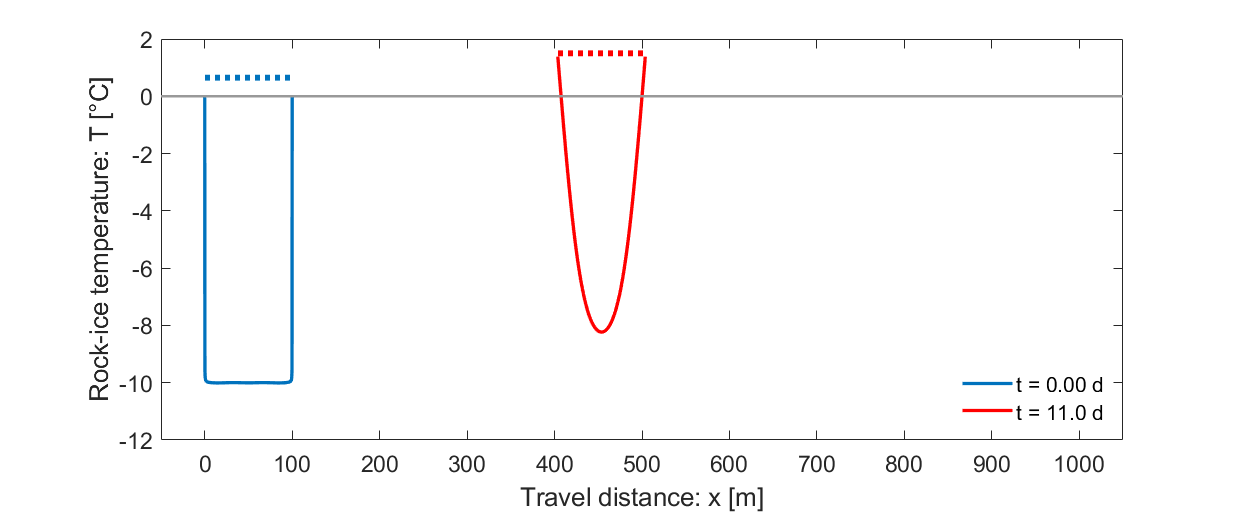}
  \includegraphics[width=13.5cm]{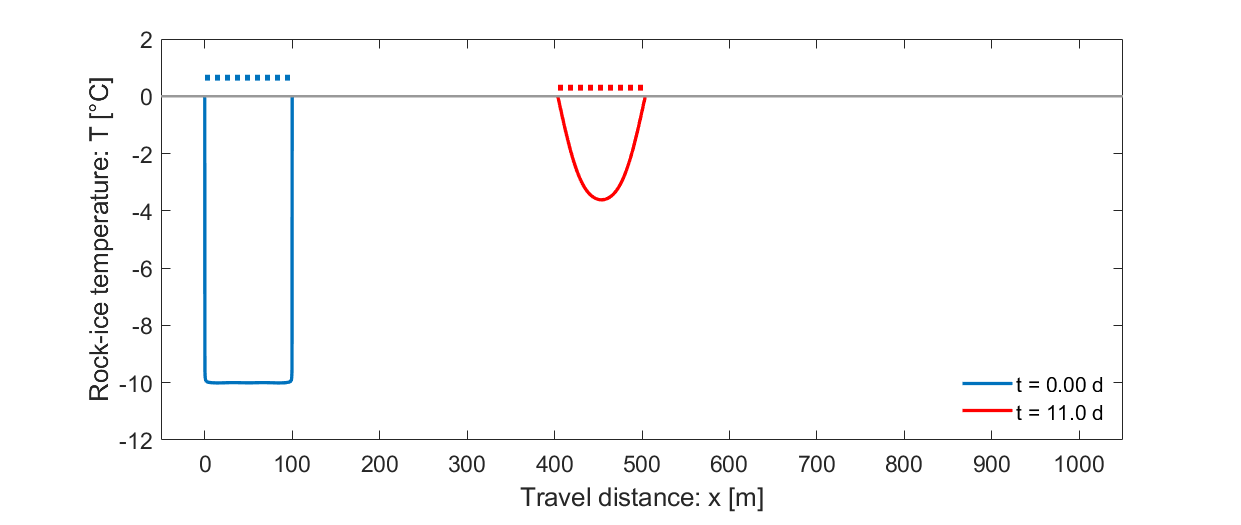}
  \includegraphics[width=13.5cm]{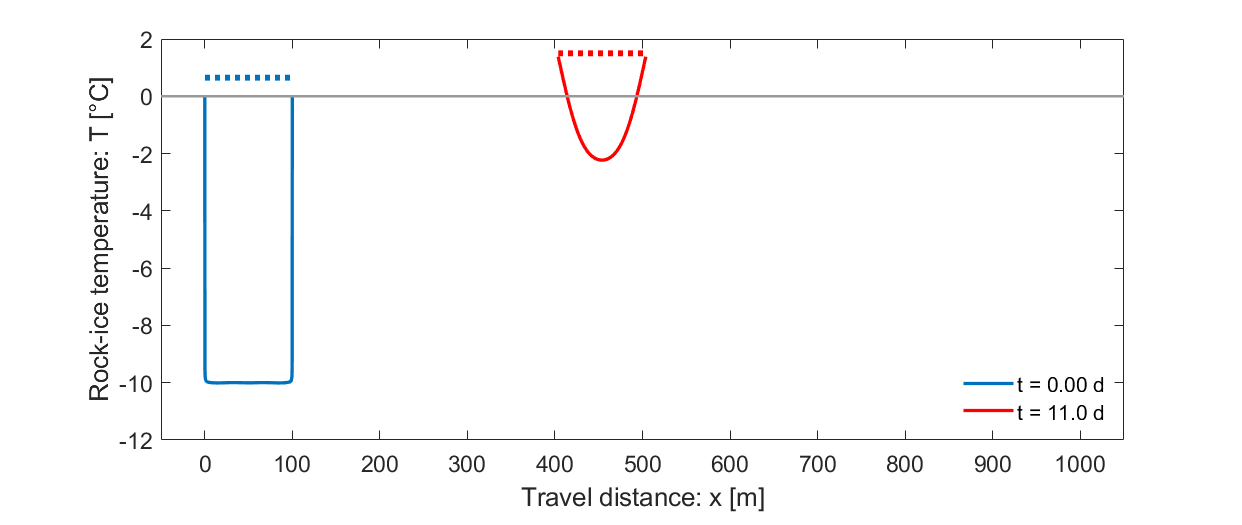}
  \end{center}
  \vspace{-5mm}
  \caption[]{Temperature evolution in rock-ice avalanche given by the solution (\ref{Equation_AdvecDiffDec_Soln2}): (A) Advection-diffusion, (B) advection-diffusion-source, (C) advection-diffusion-decay, (D) advection-diffusion-decay-source, respectively. The horizontal gray line represents the ambient temperature. The top dotted-line segments indicate the actual positions of the propagating rock-ice mass.}
  \label{Equation_AdvecDiffDecSource_Fig2S}
  \begin{picture}(0,0)
\put(401,702){\bf A}     
\put(401,539){\bf B}    
\put(401,375){\bf C}  
\put(401,212){\bf D}
\end{picture}
\end{figure}
The series structure (\ref{Equation_AdvecDiffDec_Soln2}) offers the first entire analytical solution for the advection-diffusion-decay-source problem~(\ref{Equation_AdvecDiffDecSource}).  
As in the previous solution, it can be routinely proven laboriously that (\ref{Equation_AdvecDiffDec_Soln2}) identically satisfies~(\ref{Equation_AdvecDiffDecSource}).
Figure~\ref{Equation_AdvecDiffDecSource_Fig2} displays the dynamics of the rock-ice avalanche temperature as given by the solution (\ref{Equation_AdvecDiffDec_Soln2}) for $j = 500$. 
It is assumed that the cold rock-ice mass advects with $C = 36.72$ md$^{-1}$ and $P_d = 1.0303\times10^{-7}$
in a relatively warm region of the mountain foot-hill (or the gentle slopped run-out) of $0^\circ$C as prescribed at the end boundaries.
\\[3mm]
The effects of advection and diffusion are shown in Fig.~\ref{Equation_AdvecDiffDecSource_Fig2}A while Fig.~\ref{Equation_AdvecDiffDecSource_Fig2}B also includes the decay process. As in Fig.~\ref{Equation_AdvecDiffDecSource_Fig1} the effects of advection, diffusion and decay are clearly visible, with substantial to strong influence of the decay process. However, compared with Fig.~\ref{Equation_AdvecDiffDecSource_Fig1}, the heat fluxes across the boundaries (the end points) are fundamentally different in the model solution  (\ref{Equation_AdvecDiffDec_Soln2}) used in Fig.~\ref{Equation_AdvecDiffDecSource_Fig2}. Whereas in Fig.~\ref{Equation_AdvecDiffDecSource_Fig1} the solution boundaries extend to infinity attaining boundary temperature of $0^\circ$C, but the end points of the rock-ice mass remain substantially colder than the distant far ambient temperature of $0^\circ$C, in Fig.~\ref{Equation_AdvecDiffDecSource_Fig2}, however, the end points of the rock-ice mass always remain at $0^\circ$C, so much higher than in Fig.~\ref{Equation_AdvecDiffDecSource_Fig1}.
 These fundamental differences between the solutions (\ref{Equation_AdvecDiffDecSource_Soln}) and (\ref{Equation_AdvecDiffDec_Soln2}) used in Fig.~\ref{Equation_AdvecDiffDecSource_Fig1} and Fig.~\ref{Equation_AdvecDiffDecSource_Fig2} demonstrate the nature of the physical-mathematical processes and the employed boundary conditions therein. 
\\[3mm]
Now, we complete the analysis on how the solution behaviour changes as the temperature wave described by the advection-diffusion (Fig.~\ref{Equation_AdvecDiffDecSource_Fig2S}A) and the advection-diffusion-decay (Fig.~\ref{Equation_AdvecDiffDecSource_Fig2S}C) process meet with the local heat source somewhere in the downstream of the sliding path. Unlike the results presented in Fig.~\ref{Equation_AdvecDiffDecSource_Fig2} with the solution (\ref{Equation_AdvecDiffDecSource_Soln}), now the solutions with (\ref{Equation_AdvecDiffDec_Soln2}) remain smooth as before even after encountering the heat source, the complete solution is merely elevated by the amount of the heat produced by the source, $P_s t$. The results are presented in Fig.~\ref{Equation_AdvecDiffDecSource_Fig2S}B for advection-diffusion-source and in Fig.~\ref{Equation_AdvecDiffDecSource_Fig2S}D for advection-diffusion-decay-source, respectively, for the time point $t = 6.6$ d. The main reason of maintaining of the solution structure, though with different amplitude, is because of the  common domain of all the four processes of heat production and transportation, the advection, diffusion, decay and source.  
\\[3mm]
Figure \ref{Equation_AdvecDiffDecSource_Fig1S} and Fig.~\ref{Equation_AdvecDiffDecSource_Fig2S} demonstrate the fundamentally different physical processes represented by the solutions (\ref{Equation_AdvecDiffDecSource_Soln}) and (\ref{Equation_AdvecDiffDec_Soln2}) to the temperature evolution equation (\ref{Equation_AdvecDiffDecSource}) to the rock-ice avalanche motion. 

\subsubsection{Selecting the solution}

The relations presented in (\ref{Equation_AdvecDiffDecSource_Soln}) and (\ref{Equation_AdvecDiffDec_Soln2}) are the first simple exact analytical solutions in closed form for the temperature evolution of the rock-ice avalanche. These solutions consider the broad and complex physical, thermo-mechanical and dynamical processes of the propagating mass. 
\\[3mm]
The dynamics presented in Fig.~\ref{Equation_AdvecDiffDecSource_Fig1} and  Fig.~\ref{Equation_AdvecDiffDecSource_Fig1S}, and Fig.~\ref{Equation_AdvecDiffDecSource_Fig2} and Fig.~\ref{Equation_AdvecDiffDecSource_Fig2S} associated with the analytical solutions (\ref{Equation_AdvecDiffDecSource_Soln}) and (\ref{Equation_AdvecDiffDec_Soln2}) represent more like the long time temperature evolution in the run-out, and thus, the behaviour close to the mass halting. However, these analytical solutions are valid for any situation, from the inception of motion, through the track, to the run-out. 
\\[3mm]
It is important to note that, depending on the physical situations and the dynamics of propagating rock-ice mass, either the fundamental solution or the series solution may better represent the scenario. So, the practitioners should first understand the real situation, and then, properly apply one of these solutions. However, we note that the solutions  (\ref{Equation_AdvecDiffDecSource_Soln}) and (\ref{Equation_AdvecDiffDec_Soln2}) are based on the physically and geometrically reduced model equation (\ref{Equation_AdvecDiffDecSource}). For the fully coupled simulations of the rock-ice avalanche, we must use the full temperature evolution model (\ref{Eqn_16}), without any assumptions, together with the mass and momentum balance equations for the multi-phase mass flows (Pudasaini and Mergili, 2019), and numerically solve them with the advanced efficient simulation tools, such as r.avaflow (Mergili and Pudasaini, 2025). 

 \section{Potential applications to field events}

Considering the catastrophic 2021 Chamoli rock-ice avalanche event (Shugar et al., 2021; Van Wyk de Vries et al., 2022), we offer some basic, academic simulations of this event with our new multi-phase, thermo-mechanical rock-ice avalanche model. Below, four different scenarios are simulated on how the rock-ice avalanche dynamics is controlled by: the ice-melt-efficiency; the fraction of ice in the initial rock-ice mass; ice quality (friction); and rock quality (friction). The purpose is to illustrate the essential functionality of the proposed thermo-mechanical rock-ice avalanche model from an academic point of view, however, not to directly validate or back calculate the event. The focus is on how these aspects govern the process of melting, flow transformation, mass propagation, spreading and the mobility in terms of the inundation area and the flow reach. We argue that these simulations provide the essential computational tool for the earth surface scientists, civil engineers, or practitioners in simulating the purpose-based real rock-ice avalanche events.

\subsection{Simulations with r.avaflow}

The model equations are solved with r.avaflow v4, a widely used, GIS-based open-source mass flow simulation framework (Mergili and Pudasaini, 2025). It employs the three-phase mass flow model of Pudasaini and Mergili (2019) consisting of mass and momentum balance equations, with a solid, a fine-solid, and a fluid phase, each associated with their rheological properties. This expanded computational tool includes the thermo-mechanics of rock-ice avalanche presented in Section 2. The interfacial momentum transfers; e.g., buoyancy, virtual mass, and drag; explain the interactions between different phases. The high-resolution, efficient, TVD-NOC numerical scheme (Tai et al., 2002; Pudasaini and Hutter, 2007) solves the multi-phase, thermo-mechanical model equations.

\subsection{The control of ice-melt-efficiency}

One of the most important aspects of rock-ice avalanches is how the heat produced by sharing is transmitted into the rock-ice avalanche body causing the ice to melt. In the net ice melt rate (\ref{Eqn_melt_7}), this is modelled by the melt-rate efficiency parameter $l_s$, whose value is on the order of 0.1. We choose three different values of $l_s$: 0.0, 0.1, 0.2, respectively, no melt, relatively low melt and moderate melt. Here, the focus is on the ice melt process in the early stage of mass collapse and movement (until $t = 105$ s), flow transformation from the solid-type slide; effectively single-phase rock-ice mass; to fluid-type truly multi-phase motion of rock-ice-fluid debris avalanche; to two-phase rock-fluid debris mass flow, or the hyperconcentrated debris flood; and the ice-melt-induced flow mobility. Such explicit simulations are presented here for the first time with the multi-phase, thermo-hydro-mechanical model for rock-ice avalanche. Unless otherwise specified, the basic parameters are: initial rock and ice volume fractions: $\alpha_r = 0.5, \alpha_i = 0.5$; their frictions: $\delta_r = 20^\circ$, $\delta_i = 15^\circ$ (basal); $\phi_r = 40^\circ$, $\phi_i = 40^\circ$ (internal); $\chi \approx 0.1$ (minimal shearing in the flow depth direction, compatible with the depth averaged models); the fluid-friction $ff = 0.25$ (relatively high for the purpose of focusing the flow in the area closer to the mass release); and the densities $\rho_r = 2700, \rho_i = 1000, \rho_f = 1000$. These parameters are needed for the simulation of multi-phase mass flows (Pudasaini and Mergili, 2019). Other parameters are specified at Section 2, in Mergili and Pudasaini (2025) and in Shugar et al. (2021). In simulations, the rock, ice and the fluid phases are displayed with red, green and blue color maps, respectively, in their entirety. Otherwise, the intermediate composite color maps are determined by the local fractions of the relevant phases.
\\[3mm]
\begin{figure}[t!]
\begin{center}
  \includegraphics[width=6cm]{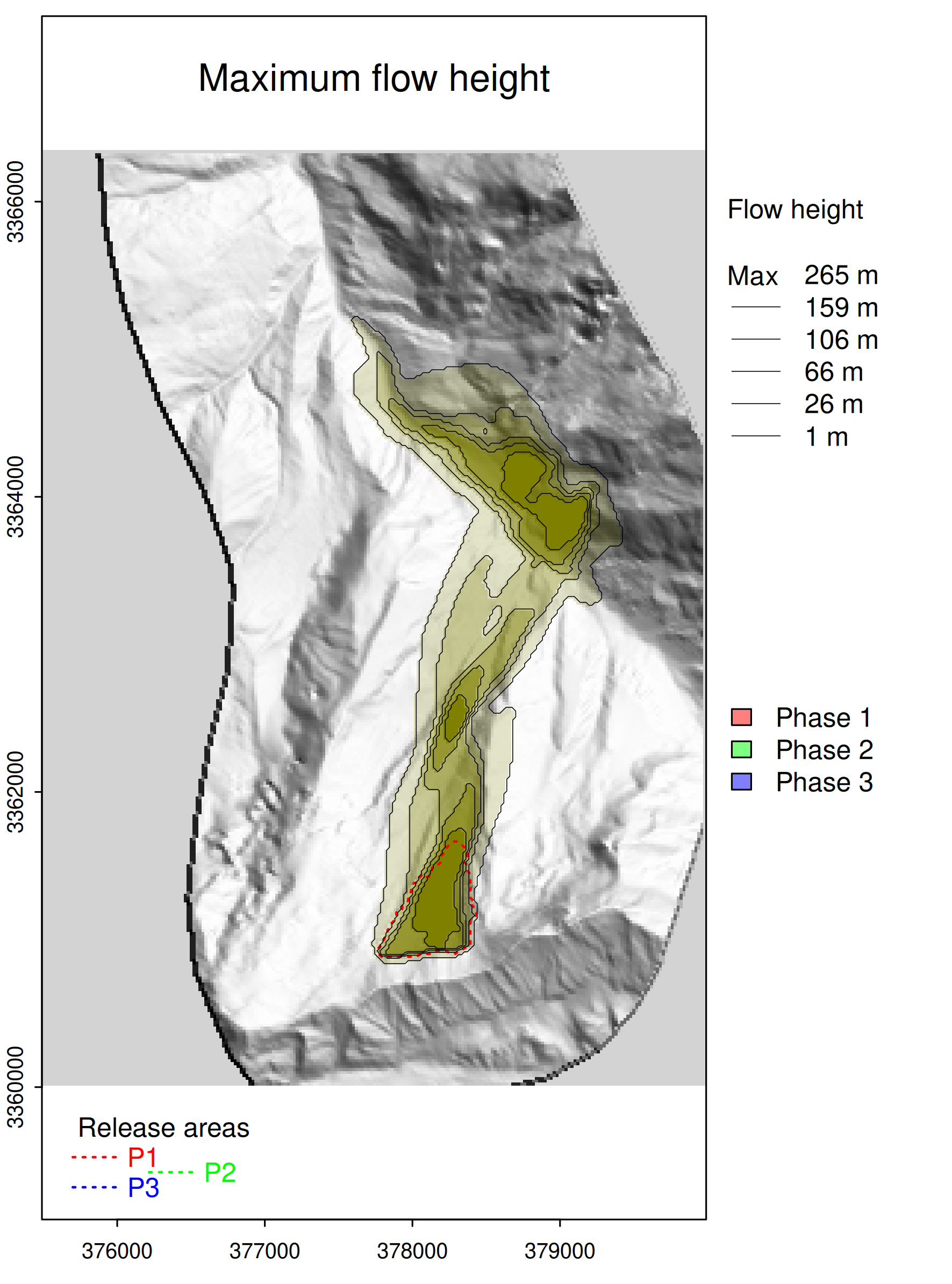}
  \includegraphics[width=6cm]{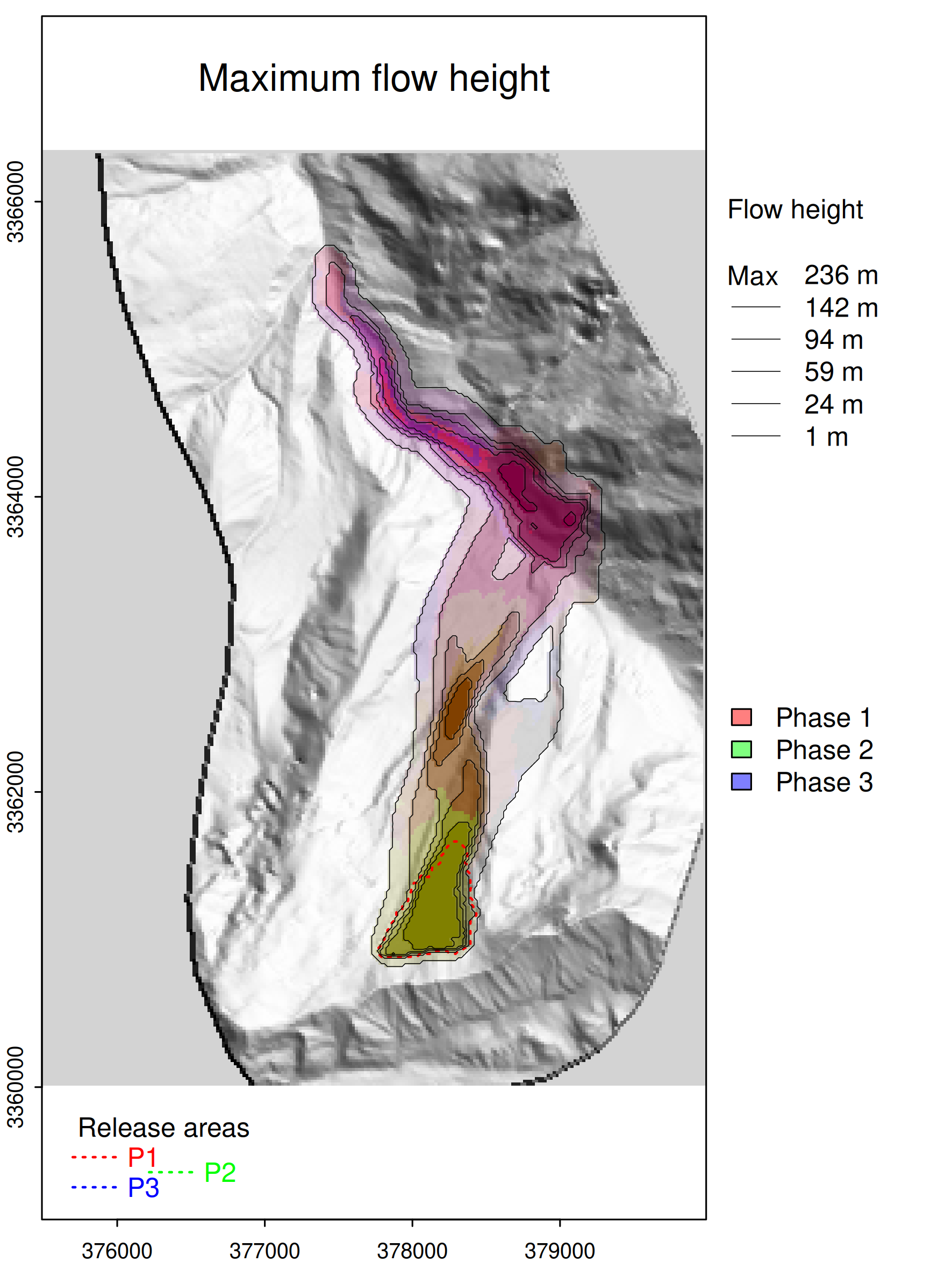}
  \includegraphics[width=6cm]{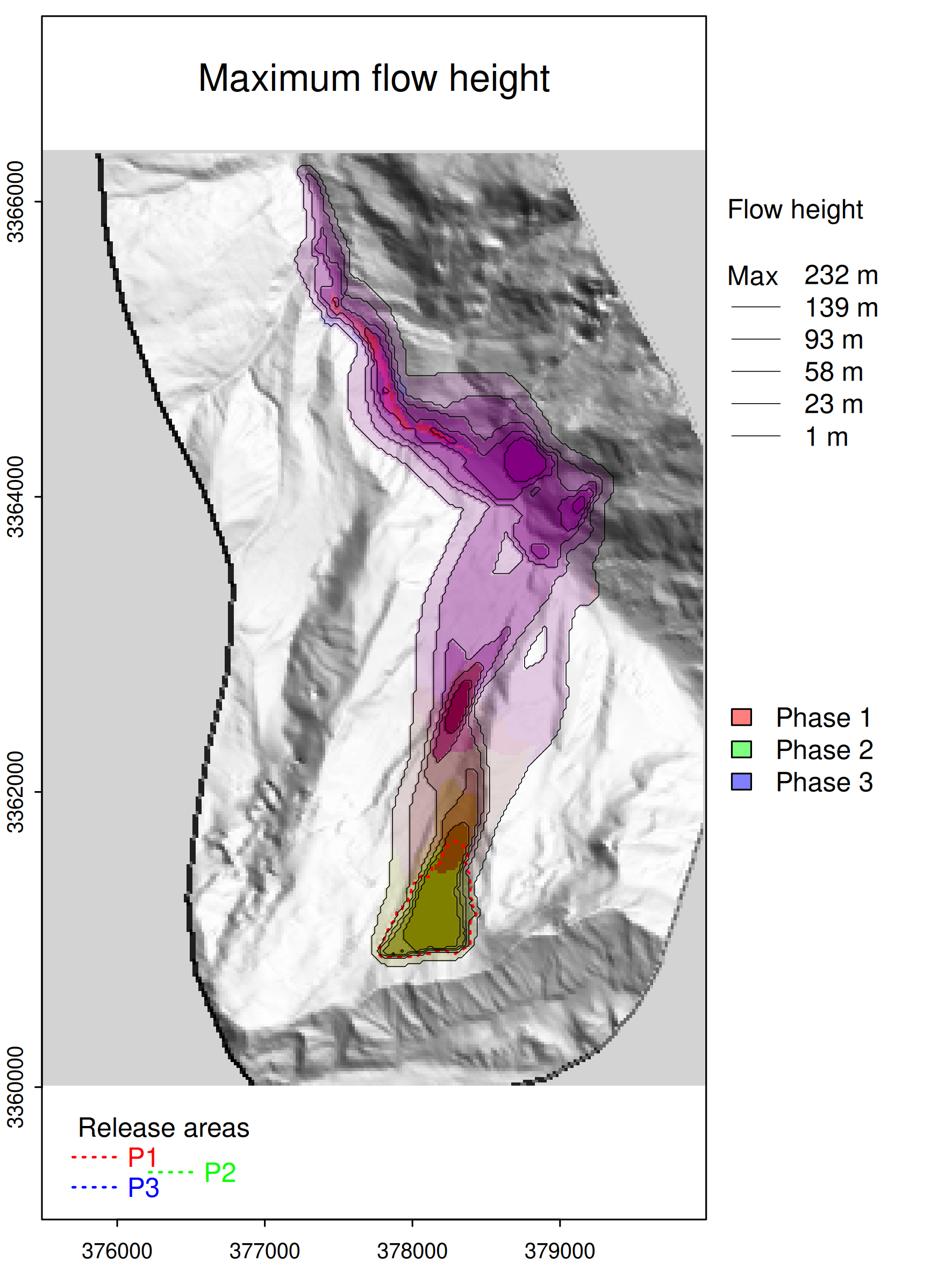}\\[1mm]
  \includegraphics[width=6cm]{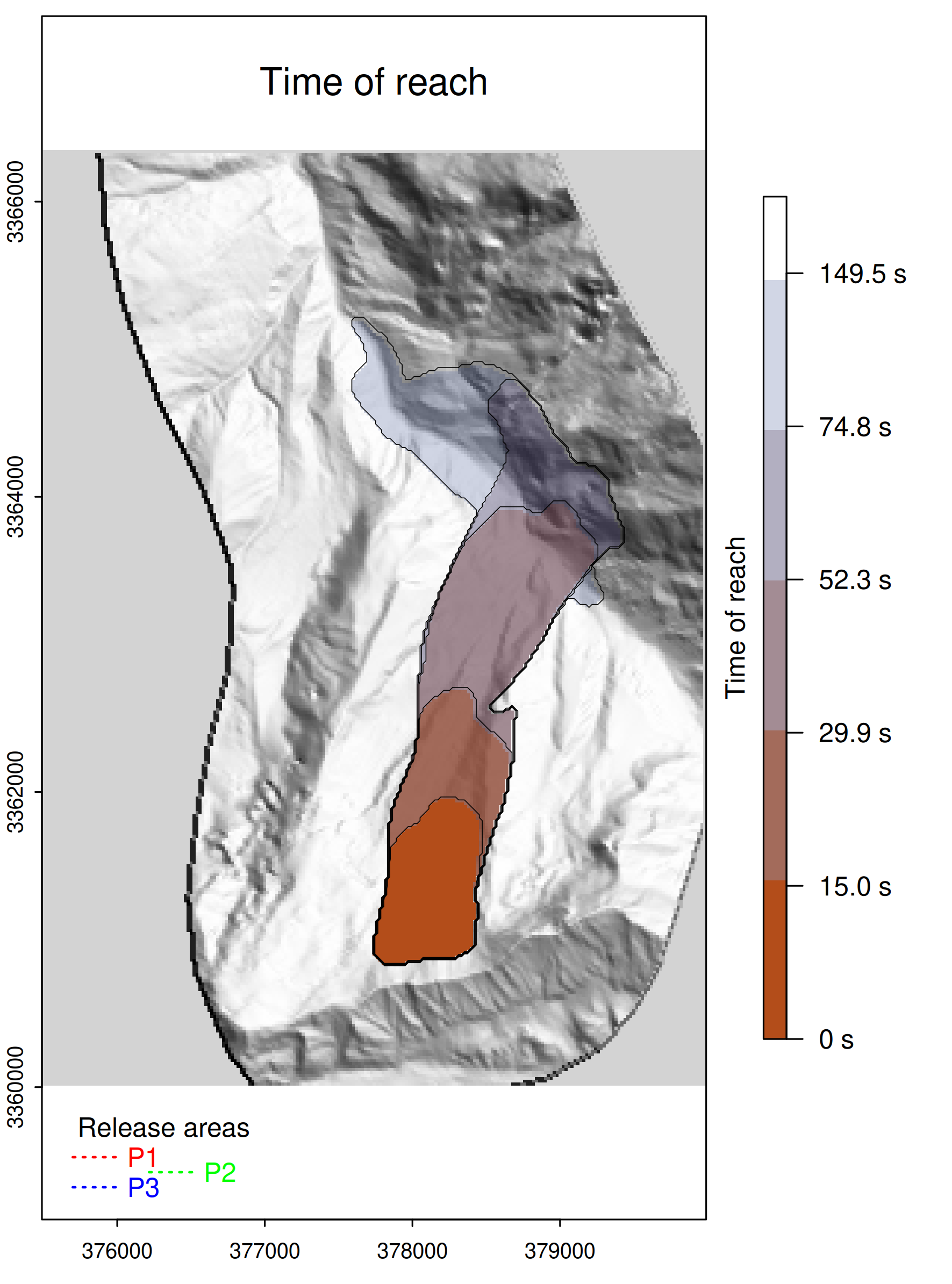}
  \includegraphics[width=6cm]{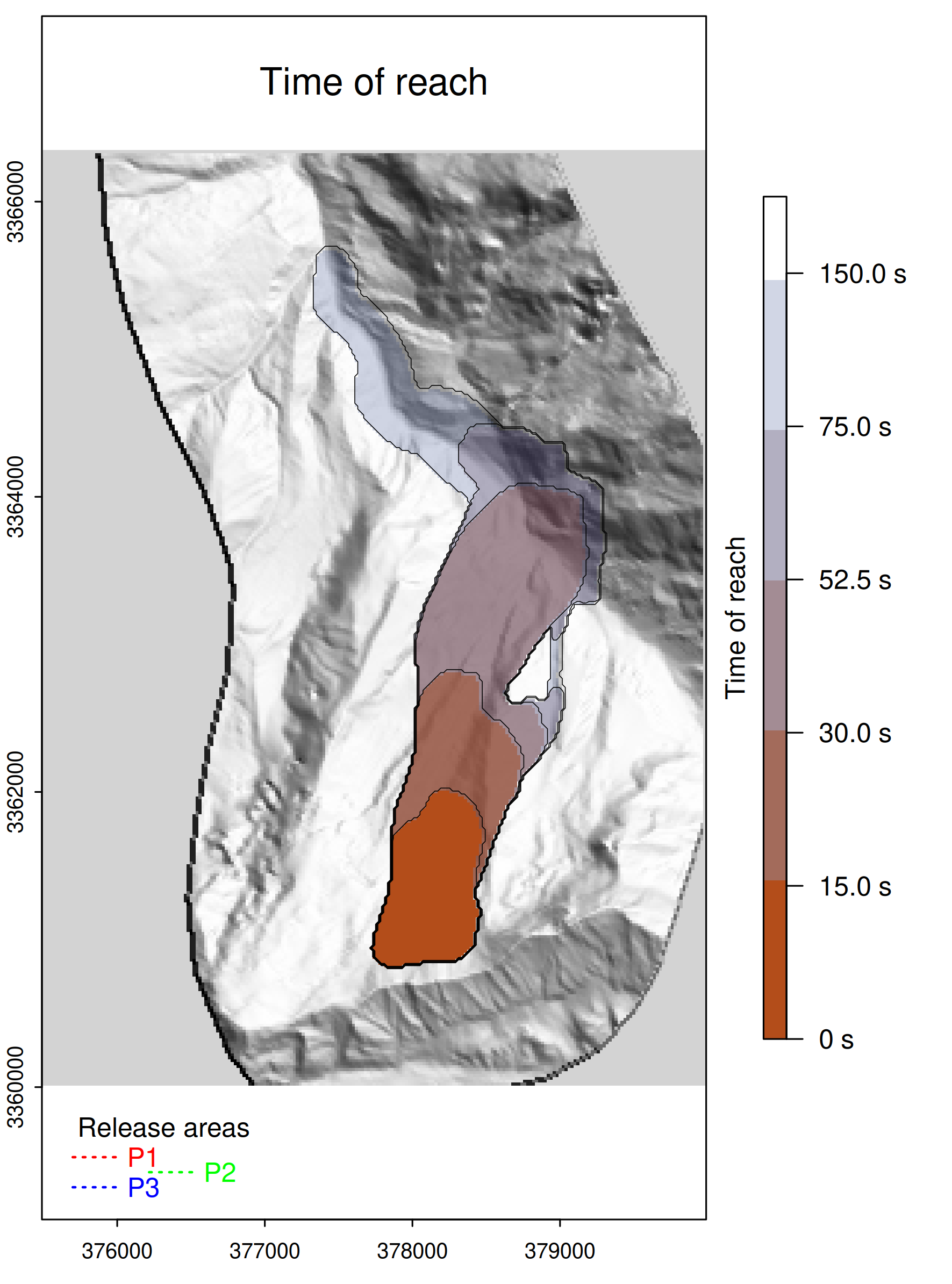}
  \includegraphics[width=6cm]{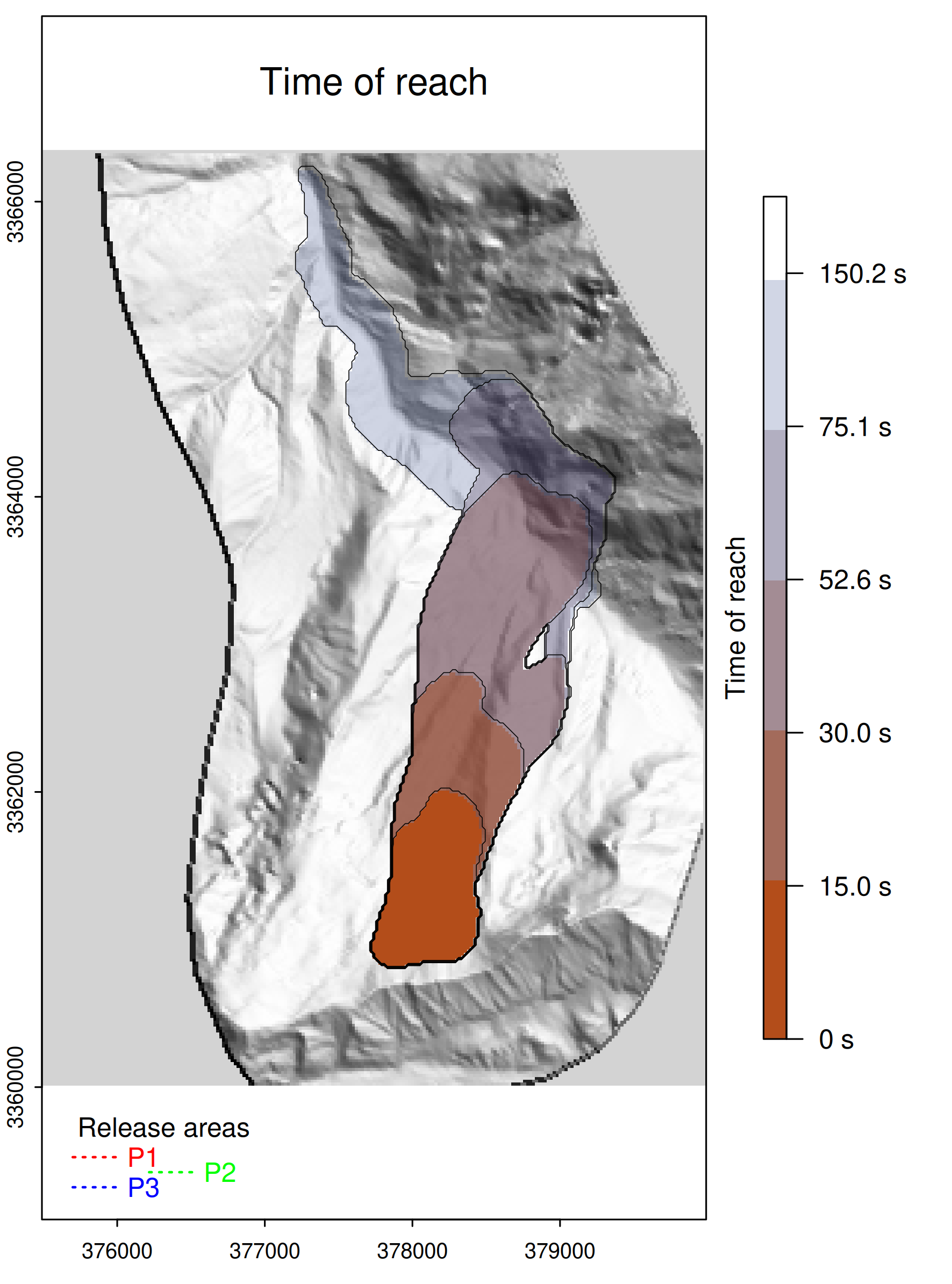}
  \end{center}
  \caption[]{The control of ice-melt-efficiency $l_s$ on the dynamics, spreading and mobility of the rock-ice avalanche. Increased melt-efficiency results in enhanced flow spreading and mobility. $P_1, P_2, P_3$ indicate the actual rock, ice and fluid fractions with red, green and blue color maps. Also displayed are the corresponding reach time.}
  \label{Fig-Ch1}
\begin{picture}(0,0)
\put(22, 545){$l_s = 0.0$}
\put(195,545){$l_s = 0.1$}
\put(370,545){$l_s = 0.2$}
\put(117,545){A}
\put(290,545){B}
\put(465,545){C}
\end{picture}
\end{figure}
The results are presented in Fig. \ref{Fig-Ch1} in the form of the inundation area with the maximum flow height and the reach time, providing the comprehensive picture of the dynamics. Although, recently several rock-ice avalanche experiments have been conducted in the laboratory settings (Yang et al., 2019; Ren et al., 2021; Feng et al., 2023; Fan et al., 2025; Peng et al., 2025), they effectively do not produce melt water, a situation simulated in Fig.~\ref{Fig-Ch1}A with $l_s = 0.0$. However, from the natural perspective, where the rock-ice avalanches produce substantial to large amount of melt water, the result in Fig. \ref{Fig-Ch1}A cannot be physically appreciated as it only takes into account the reduced friction of ice, ignoring the fundamental contribution from the ice melting.  The simulation with the low ice-melt ($l_s = 0.1$) as presented in Fig. \ref{Fig-Ch1}B displays the process of ice melt, converting ice (in green) to melt water (in blue). Due to the low melting efficiency, it takes substantial amount of time until most of the ice is melted, and the process continues as long as the mass reaches the foothill as indicated by the reddish-green (brown) color. Due to the melt-produced water (fluid), the travel distance and the inundation area is now longer and wider than in Fig. \ref{Fig-Ch1}A without melt. Fig. \ref{Fig-Ch1}C shows the simulation with the higher (moderate) ice-melt efficiency, $l_s = 0.2$. As expected, now, the melting process is much quicker, inundation area is much wider and the mass reaches much farther downstream than in Fig. \ref{Fig-Ch1}A and Fig. \ref{Fig-Ch1}B. This, for the first time, demonstrates the ice-melt-induced flow mobility; as the melt-rate increases, the portion of the frictional ice mass decreases as it is converted into fluid, as a whole, the mixture material becomes frictionally weaker, resulting in the wider and longer flow reach. Consequently, the maximum flow height reduces from 265 m to 236 m to 232 m, respectively, in Fig. \ref{Fig-Ch1}A, Fig. \ref{Fig-Ch1}B and Fig. \ref{Fig-Ch1}C. So, even with the low melt-rate, the rate of decrease of the flow height is quick, that rate slows down as the melt rate increases.

\subsection{The control of ice fraction}

\begin{figure}[t!]
\begin{center}
 \includegraphics[width=6cm]{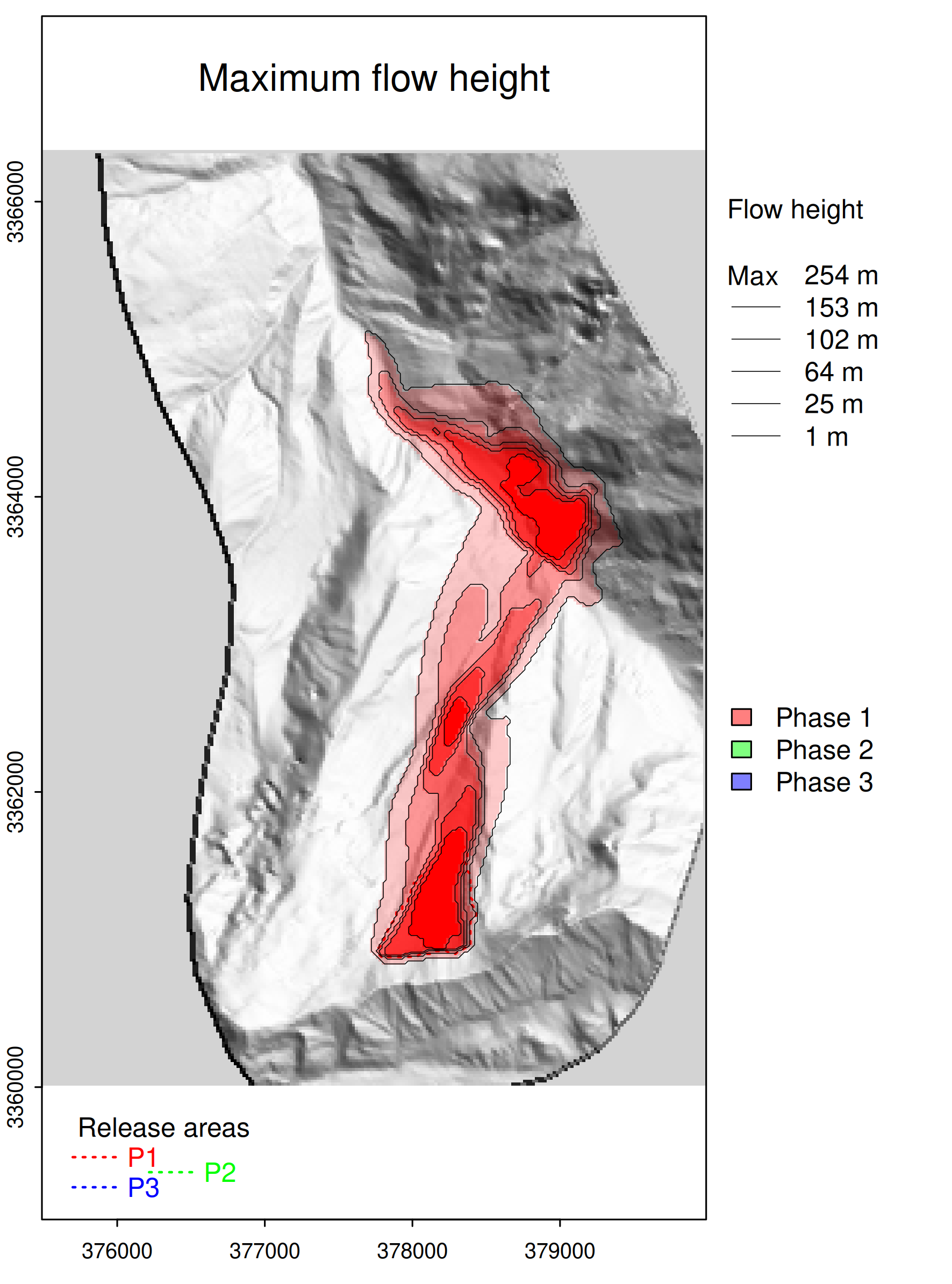}
 \includegraphics[width=6cm]{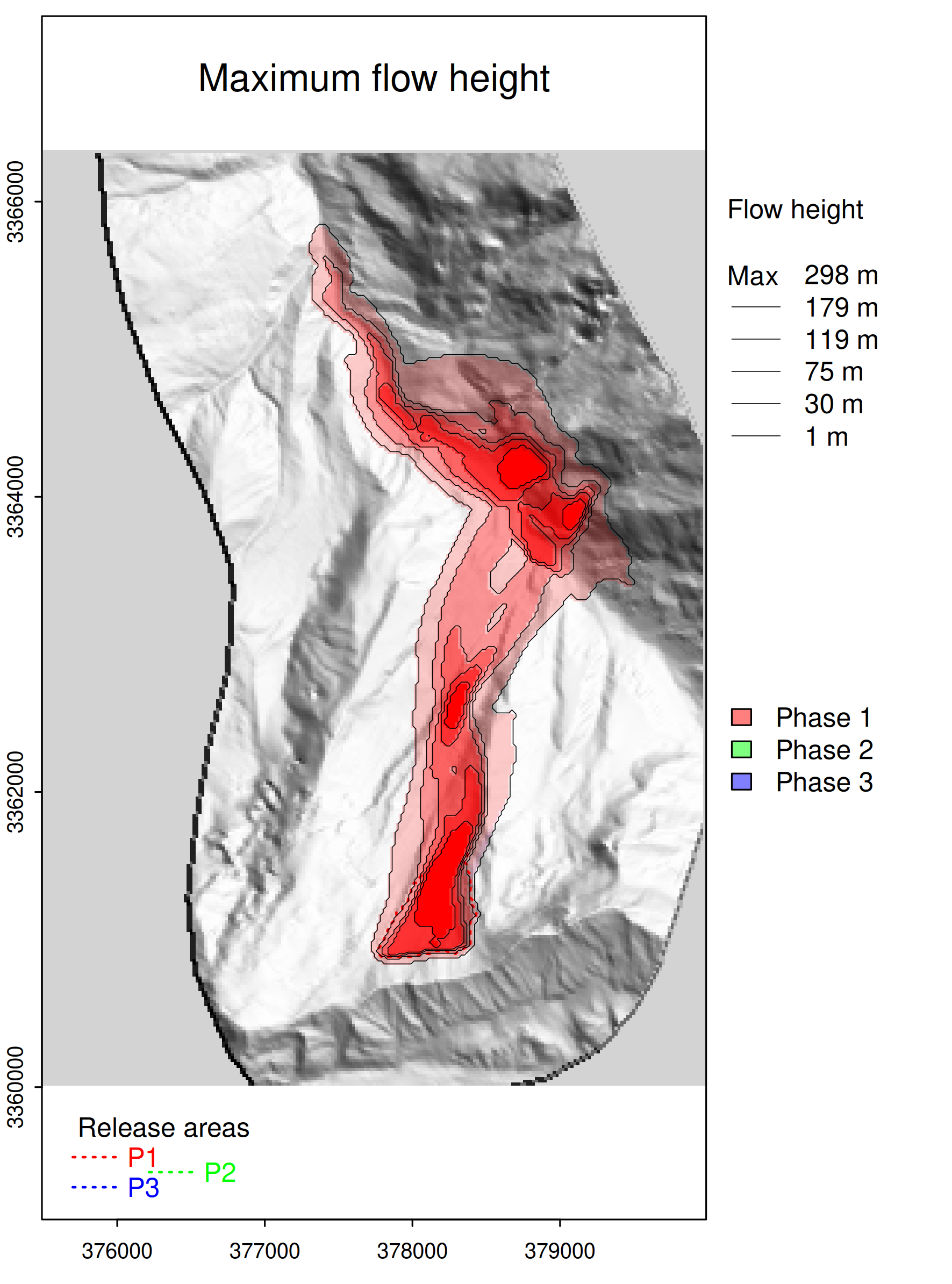}
 \includegraphics[width=6cm]{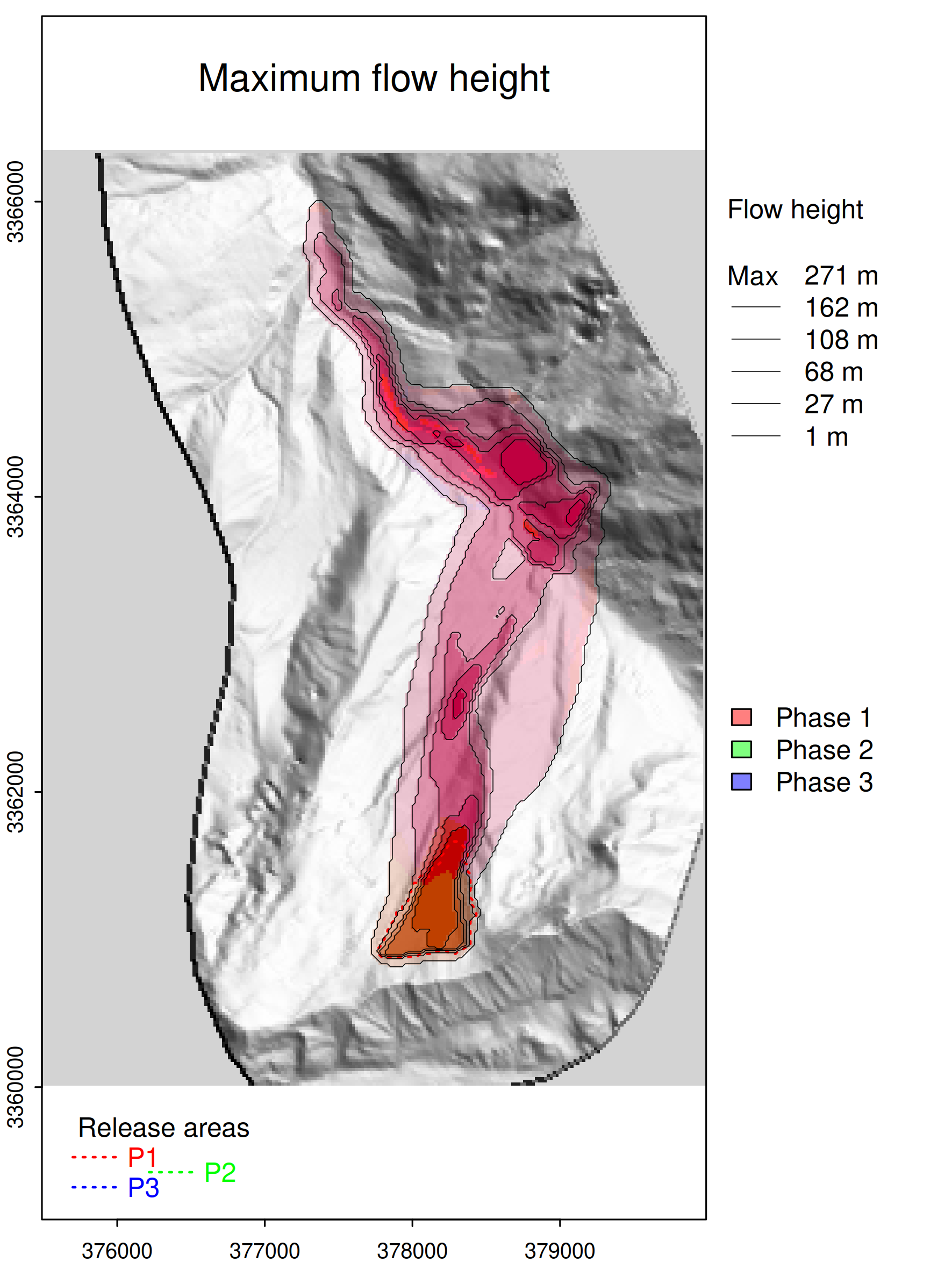}\\[7mm]
 \includegraphics[width=6cm]{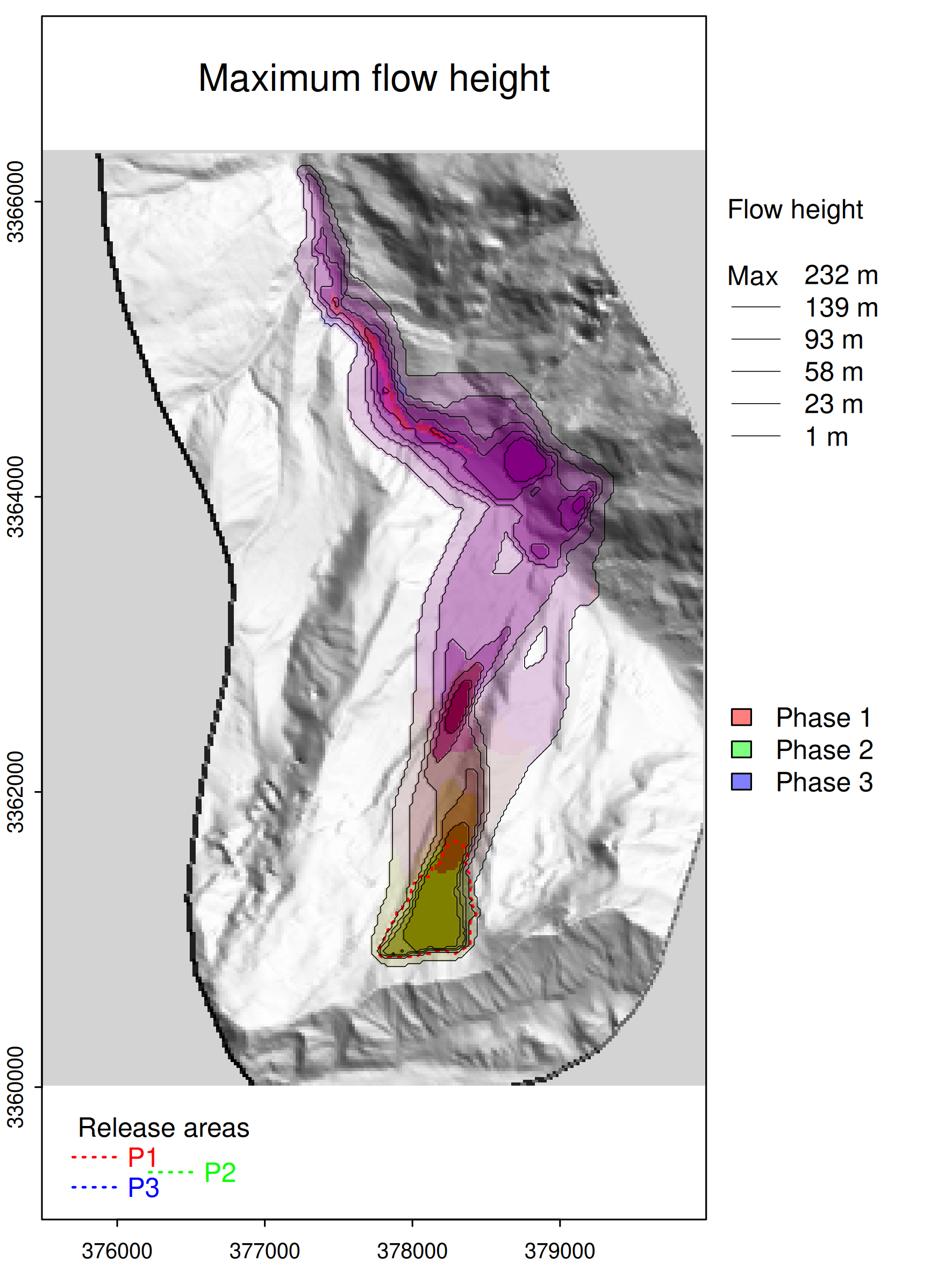}
 \includegraphics[width=6cm]{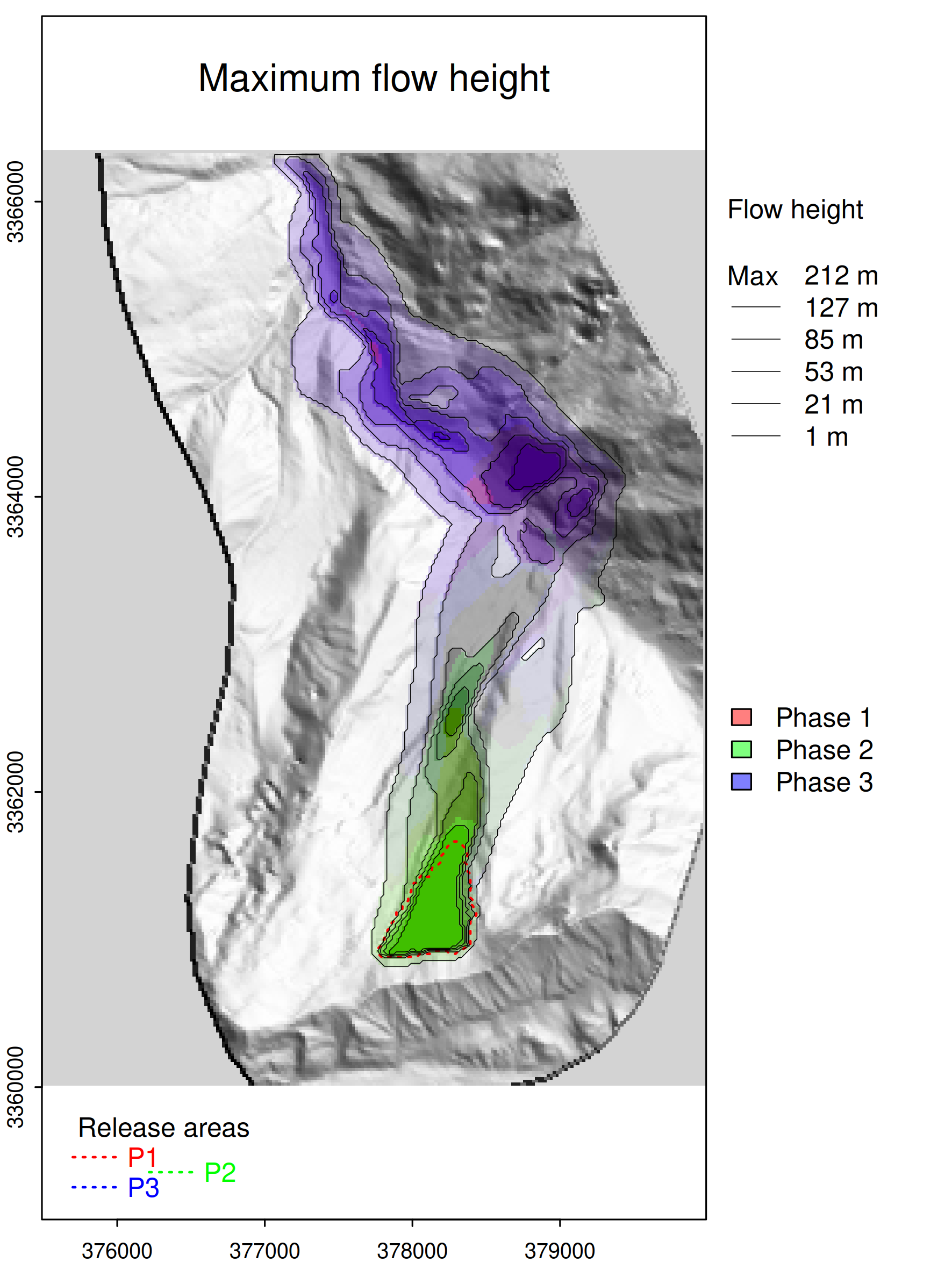}
 \includegraphics[width=6cm]{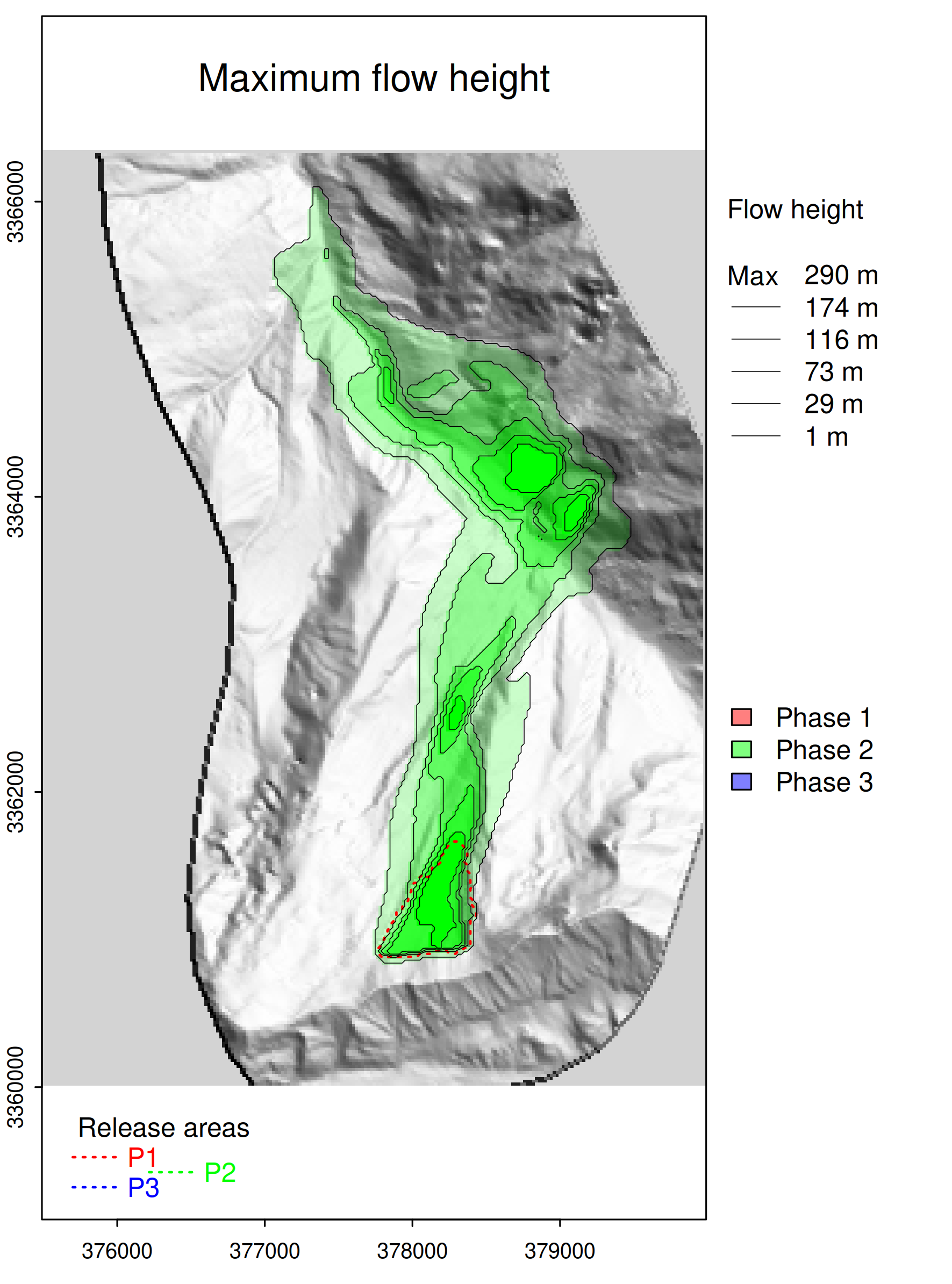}
  \end{center}
  \caption[]{The control of ice volume fraction $\alpha_i$ on the dynamics, spreading and mobility of rock-ice avalanche. Increased ice volume fraction with ice melting results in enhanced flow spreading and mobility. For comparison, also shown are the scenarios of effectively rock-avalanche (A), and effectively ice-avalanche (F) without melting, which are physically pointless. The reach time are displayed in Fig. \ref{Fig-Ch2Reach}.}
  \label{Fig-Ch2}
  \begin{picture}(0,0)
\put(22, 575){$\alpha_i = 0.1; l_s = 0.0$}
\put(195,575){$\alpha_i = 0.1$}
\put(360,575){$\alpha_i = 0.3$}
\put(117,575){A}
\put(290,575){B}
\put(465,575){C}
\put(22, 325){$\alpha_i = 0.5$}
\put(195,325){$\alpha_i = 0.7$}
\put(370,325){$\alpha_i = 0.9; l_s = 0.0$}
\put(117,325){D}
\put(290,325){E}
\put(465,325){F}
\end{picture}
\end{figure}
\begin{figure}[t!]
\begin{center}
 \includegraphics[width=6cm]{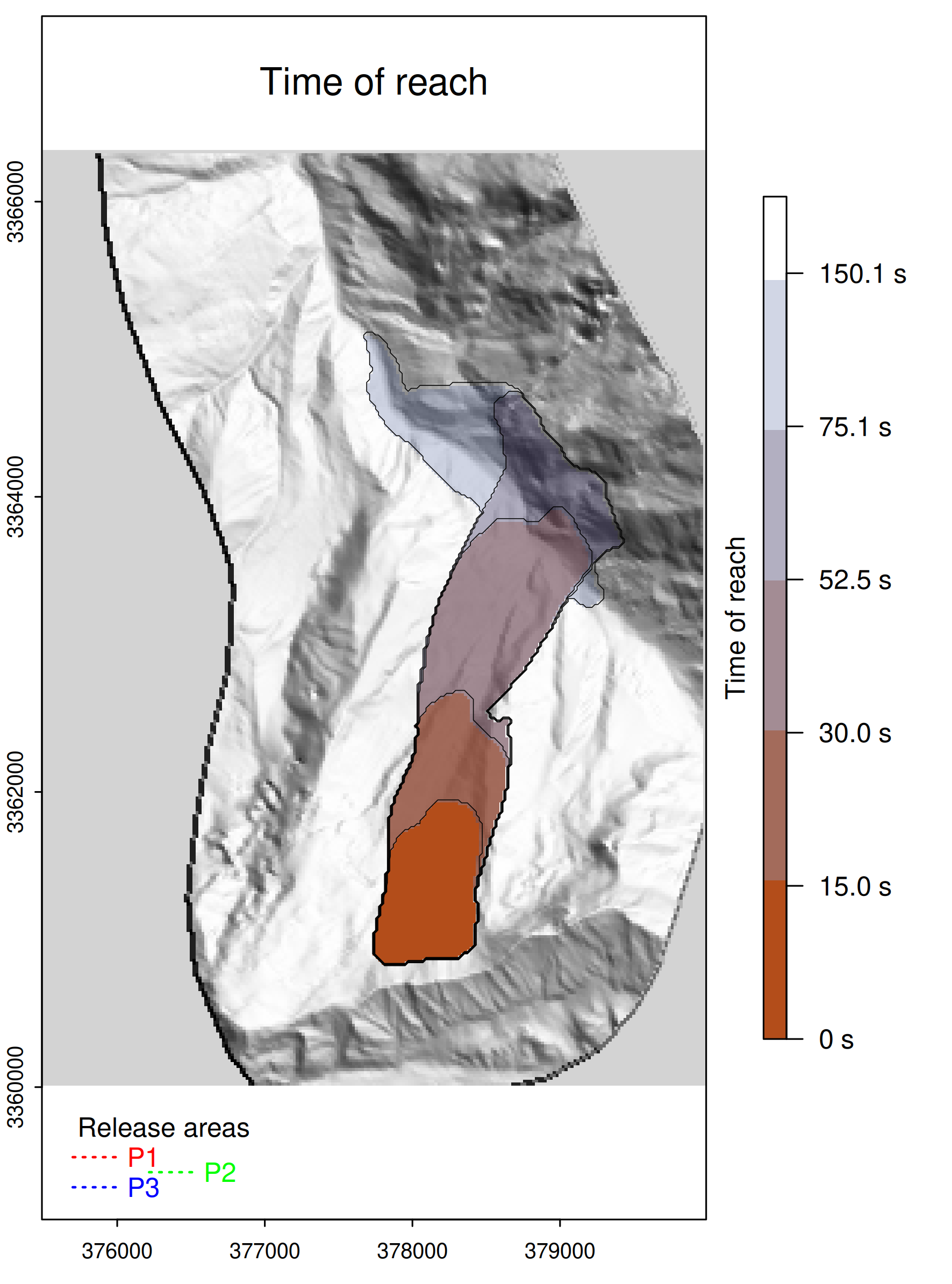}
 \includegraphics[width=6cm]{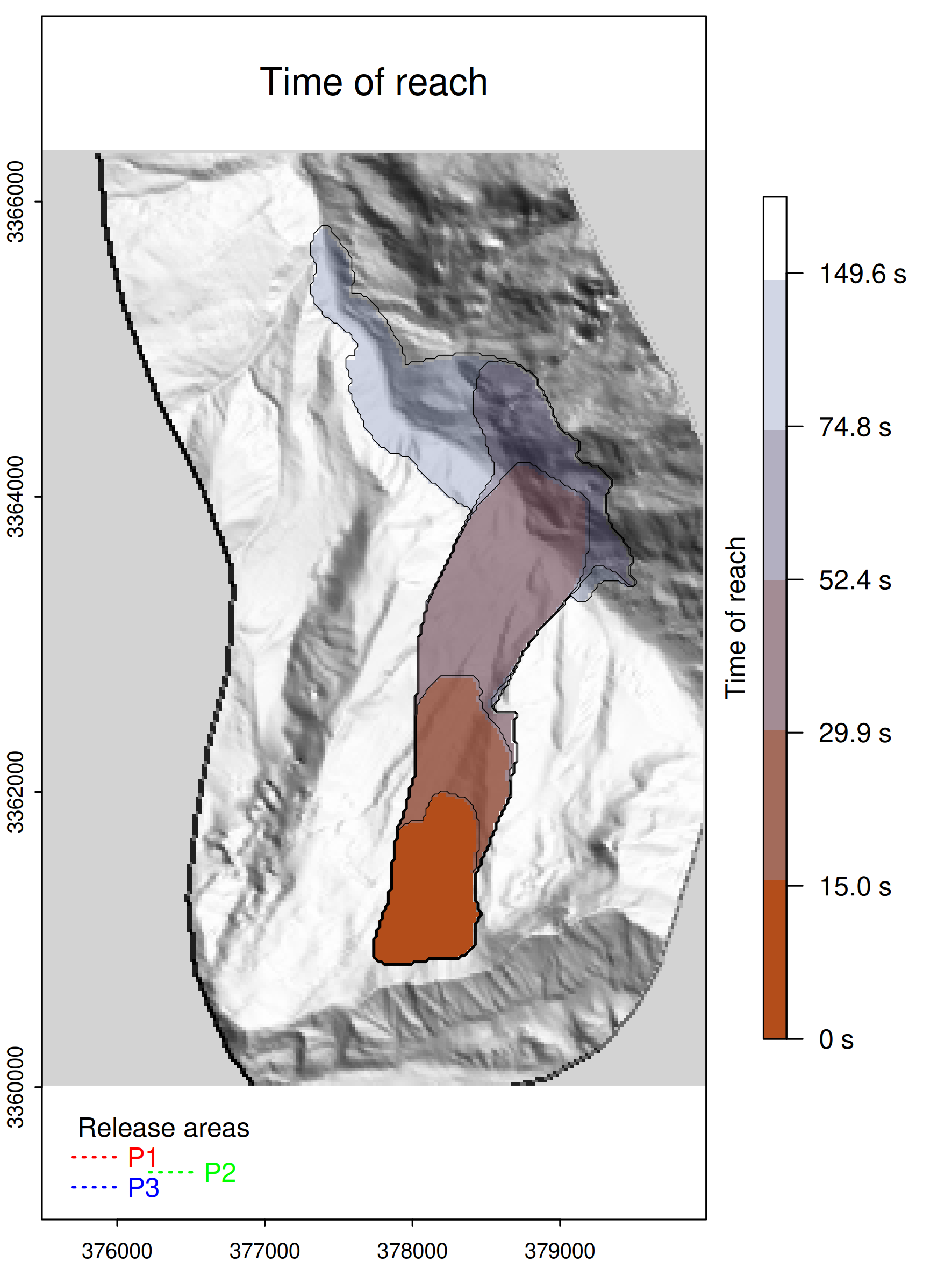}
 \includegraphics[width=6cm]{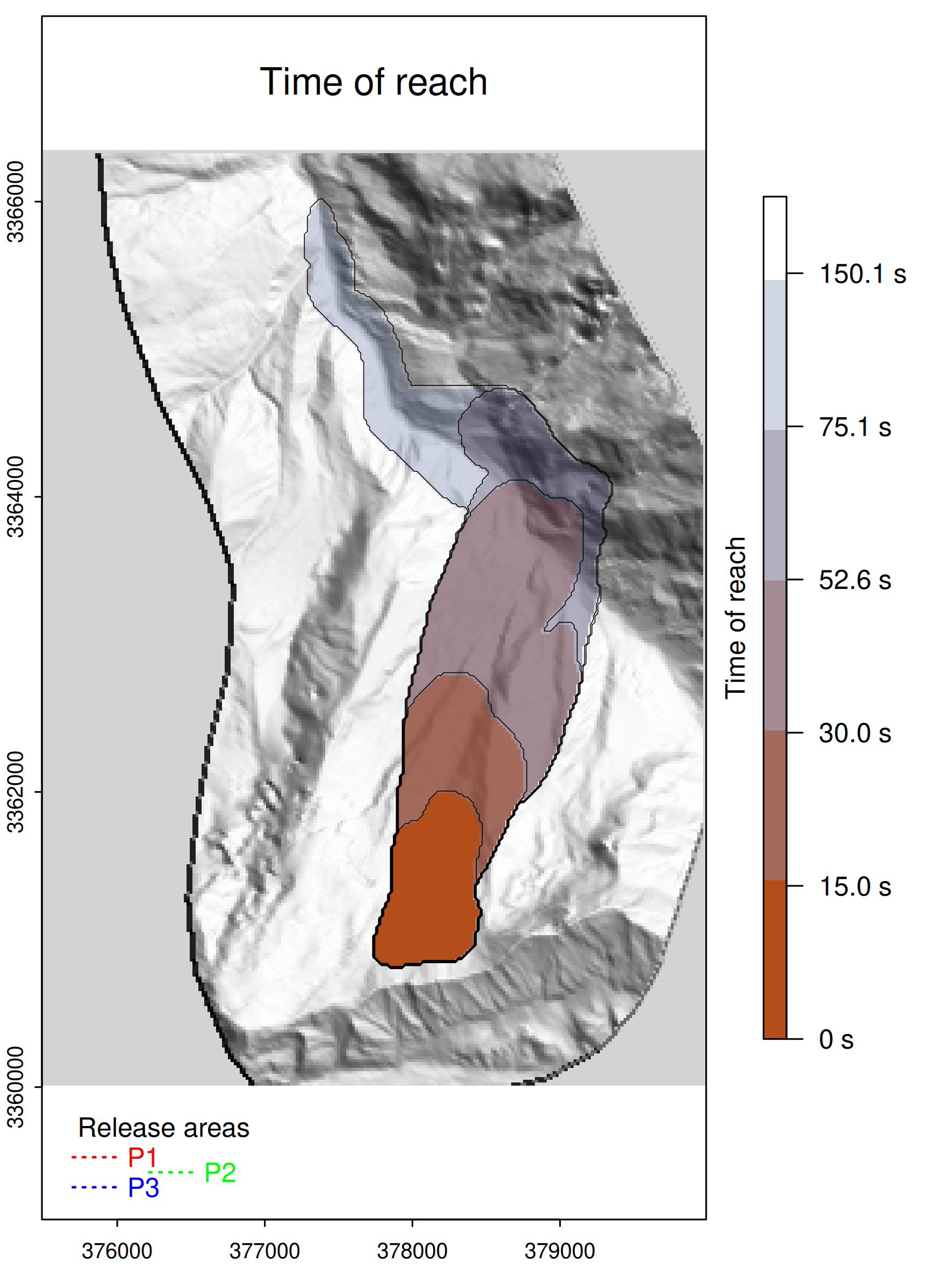}\\[7mm]
 \includegraphics[width=6cm]{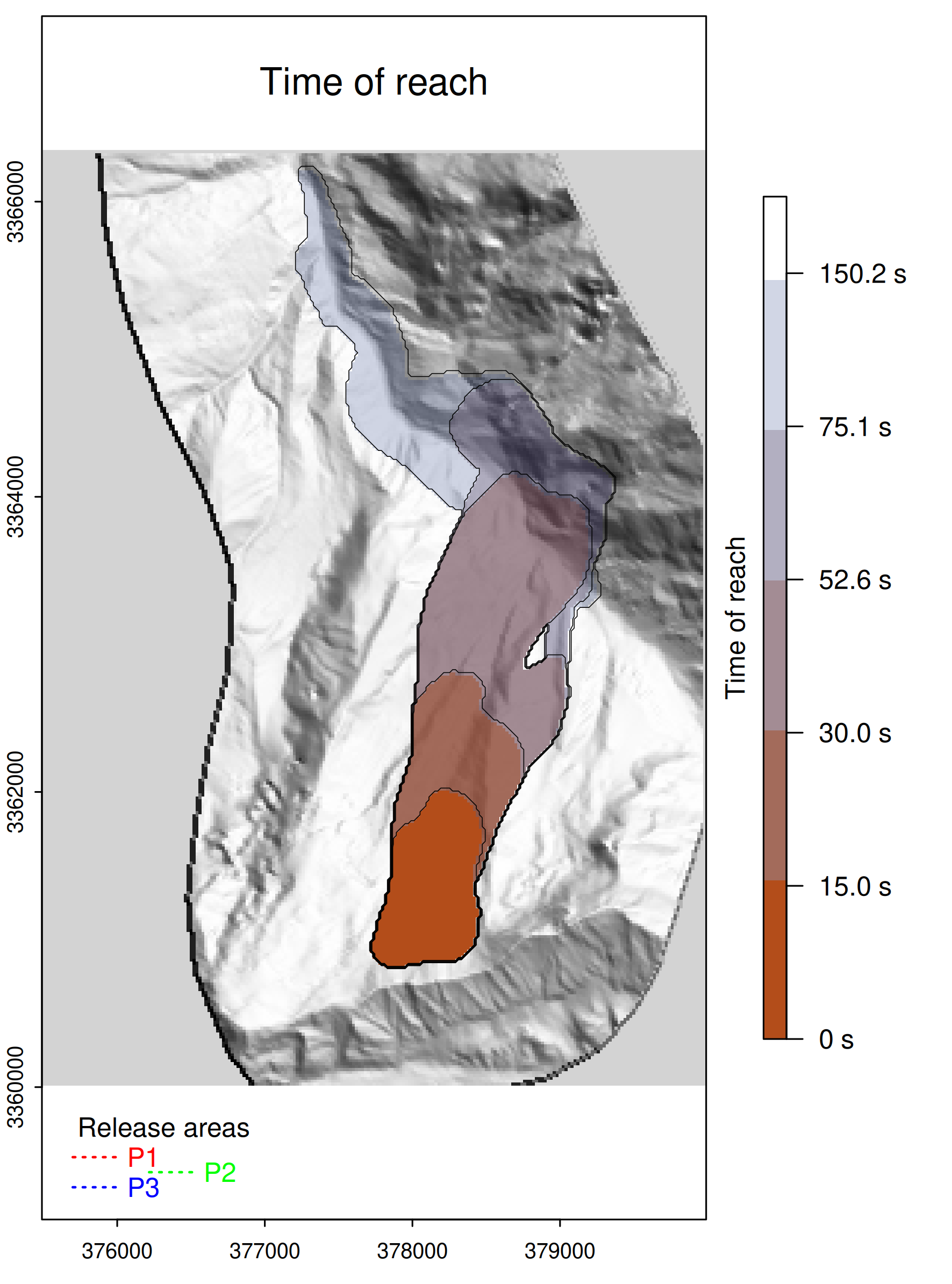}
 \includegraphics[width=6cm]{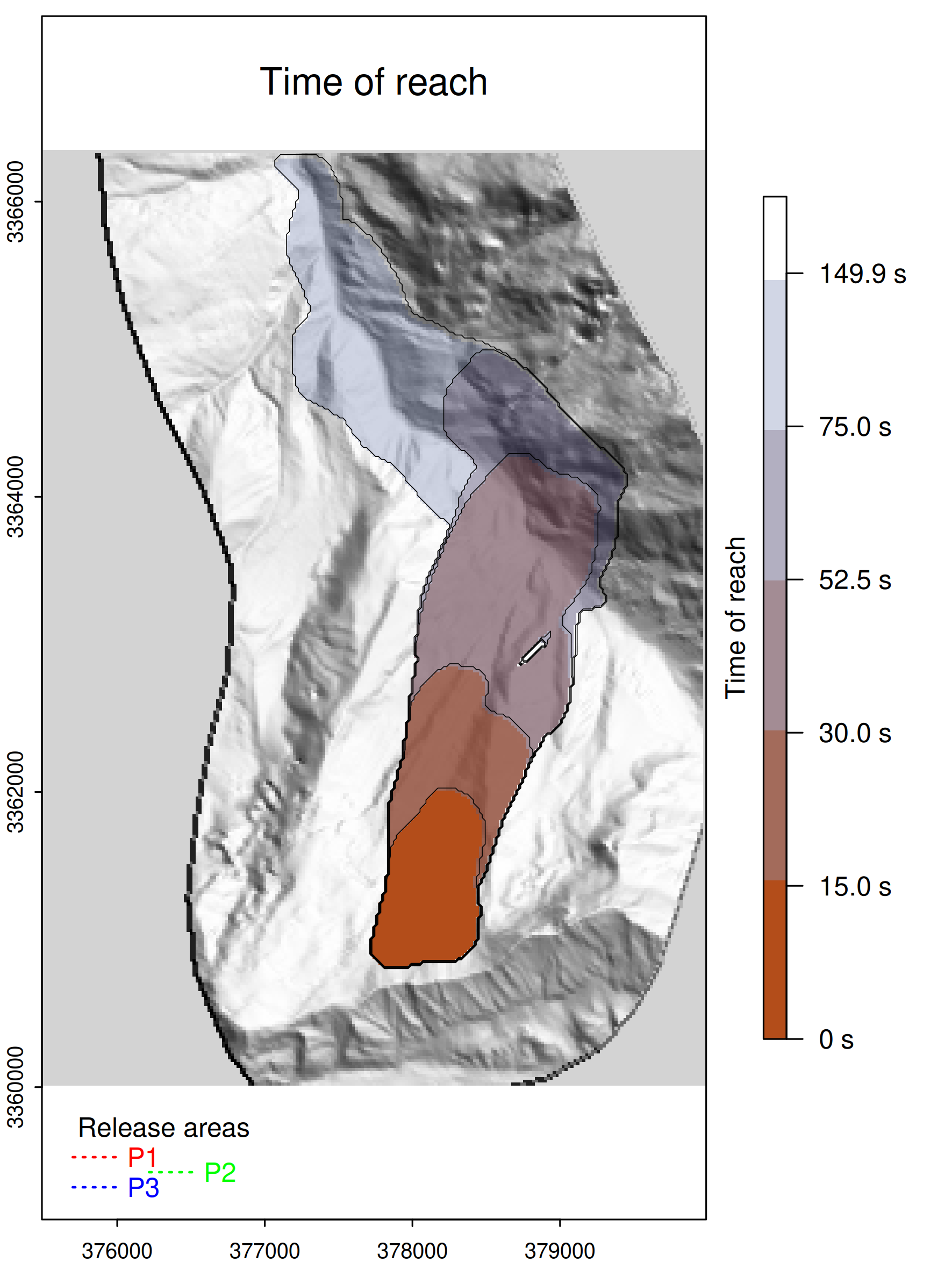}
 \includegraphics[width=6cm]{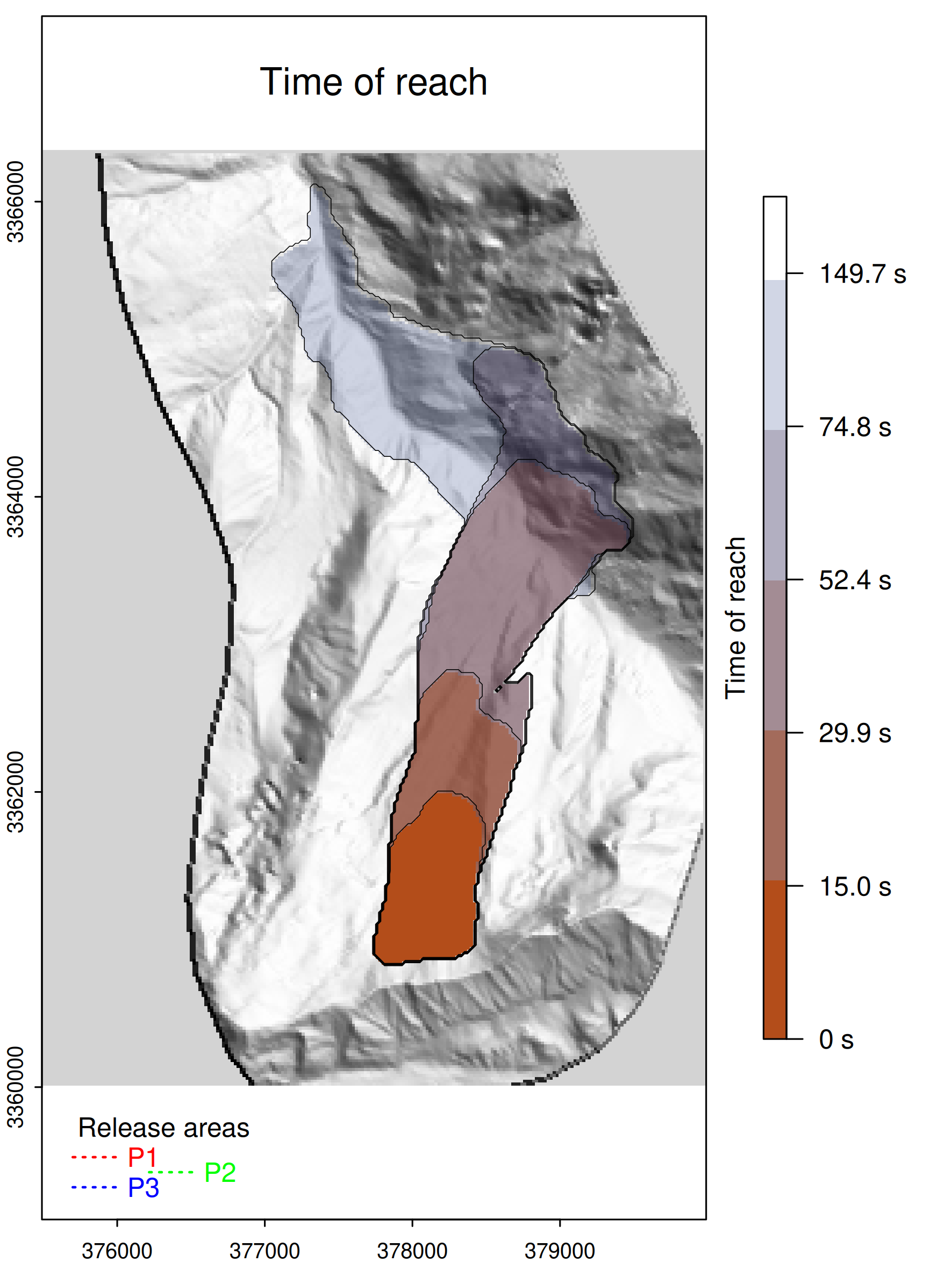}
  \end{center}
  \caption[]{The reach time for simulations in Fig. \ref{Fig-Ch2}. The control of ice volume fraction $\alpha_i$ on the dynamics, spreading and mobility of rock-ice avalanche. Increased ice volume fraction with ice melting results in enhanced flow spreading and mobility. For comparison, also shown are the scenarios of effectively rock-avalanche (A), and effectively ice-avalanche (F) without melting, which are physically irrelevant.}
  \label{Fig-Ch2Reach}
  \begin{picture}(0,0)
\put(22, 575){$\alpha_i = 0.1; l_s = 0.0$}
\put(195,575){$\alpha_i = 0.1$}
\put(370,575){$\alpha_i = 0.3$}
\put(117,575){A}
\put(290,575){B}
\put(465,575){C}
\put(22, 325){$\alpha_i = 0.5$}
\put(195,325){$\alpha_i = 0.7$}
\put(370,325){$\alpha_i = 0.9; l_s = 0.0$}
\put(117,325){D}
\put(290,325){E}
\put(465,325){F}
\end{picture}
\end{figure}
Principally, the most important aspect in the motion of the rock-ice mass is the ice content, that, during the propagation phase, melts and produces fluid, significantly fluidizing the initially frictional granular material. So, our next focus is on how the initial ice content controls the melting process, flow transformation, mass propagation and the mobility in terms of the inundation area and the flow reach.  For this, we consider different scenarios: the ice contents of $\alpha_i$ = 0.1, 0.3, 0.5, 0.7, and 0.9 in the initial rock-ice wall. Results are depicted in Fig. \ref{Fig-Ch2}. For reference, Fig. \ref{Fig-Ch2}A shows the rock-ice avalanche simulation with $\alpha_i = 0.1$, however, without considering the ice melt ($l_s = 0.0$).  This means that, only the effectively single-phase solid (granular) mass slide (akin to rock avalanche) is considered.  So, as it only considers the ice as frictionally reduced solid material, this is physically not meaningful. Fig. \ref{Fig-Ch2}B is as Fig. \ref{Fig-Ch2}A ($\alpha_i = 0.1$), but it considers the ice melt ($l_s = 0.1$). As the melt is activated, the flow coverage area widens, mass slides faster, and brings the ice to the foothill of the slope; yet, with much less melt water production due to low ice content. This is why the colors in Fig. \ref{Fig-Ch2}A and Fig. \ref{Fig-Ch2}B are somewhat similar. However, as the ice melts, it lubricates the frictionally critically weak (heavily rock-dominated) rock-ice avalanching mass. It appears that this also produced locally higher mass height in the foothill than before. In Fig. \ref{Fig-Ch2}C, the initial ice content is increased to $\alpha_i = 0.3$ ($l_s = 0.1$). This causes a significant ice-melt, produces higher amount of water, resulting in the wider flow coverage area and faster flow, reaching farther downslope than with the lower ice content in Fig. \ref{Fig-Ch2}B. Further increase of the ice content to $\alpha_i = 0.5$ ($l_s = 0.2$) in Fig. \ref{Fig-Ch2}D produces much higher amount of fluid than before, yet, it takes relatively longer time for the ice to melt due to the higher ice content available. Fig. \ref{Fig-Ch2}E considers even higher ice content, $\alpha_i = 0.7$ ($l_s = 0.2$), and produces the most fluid of all the scenarios. Consequently, the flow turns into hyperconcentrated rock-ice-debris flow to debris flood with the farthest reach and the widest cross-slope expansion of the debris as the material quickly becomes frictionally weaker due to the ice-melt produced fluid.
\\[3mm]
Compared to the (relatively) lower ice contents in the previous panels, in Fig. \ref{Fig-Ch2}E, large amount of water is produced by ice-melt, leading to the widest flow coverage and the longest flow reach in the farther downstream. As the ice content increases, the maximum flow depth decreases continuously from Fig. \ref{Fig-Ch2}B to Fig. \ref{Fig-Ch2}E. This can be explained hydro-mechanically. Because, as the higher amount of fluid is produced with the increased ice fraction, the sliding mass becomes mechanically weaker, resulting in higher mass spreading, thus causing the reduction of the flow depth. The successive panels with higher ice content in the rock-ice avalanche demonstrates the ice-melt-induced mobility of rock-ice avalanche. Moreover, Fig. \ref{Fig-Ch2}F presents the simulation result for the ice avalanche ($\alpha_i = 0.9$) without the ice melt ($l_s = 0.0$). However, this scenario is physically absurd, only shown here for the completeness for hypothetical reason, showing what would happen if the ice avalanche would slide without melt. The time of reach are shown in Fig. \ref{Fig-Ch2Reach}.

\subsection{The control of ice friction}

\begin{figure}[t!]
\begin{center}
  \includegraphics[width=6cm]{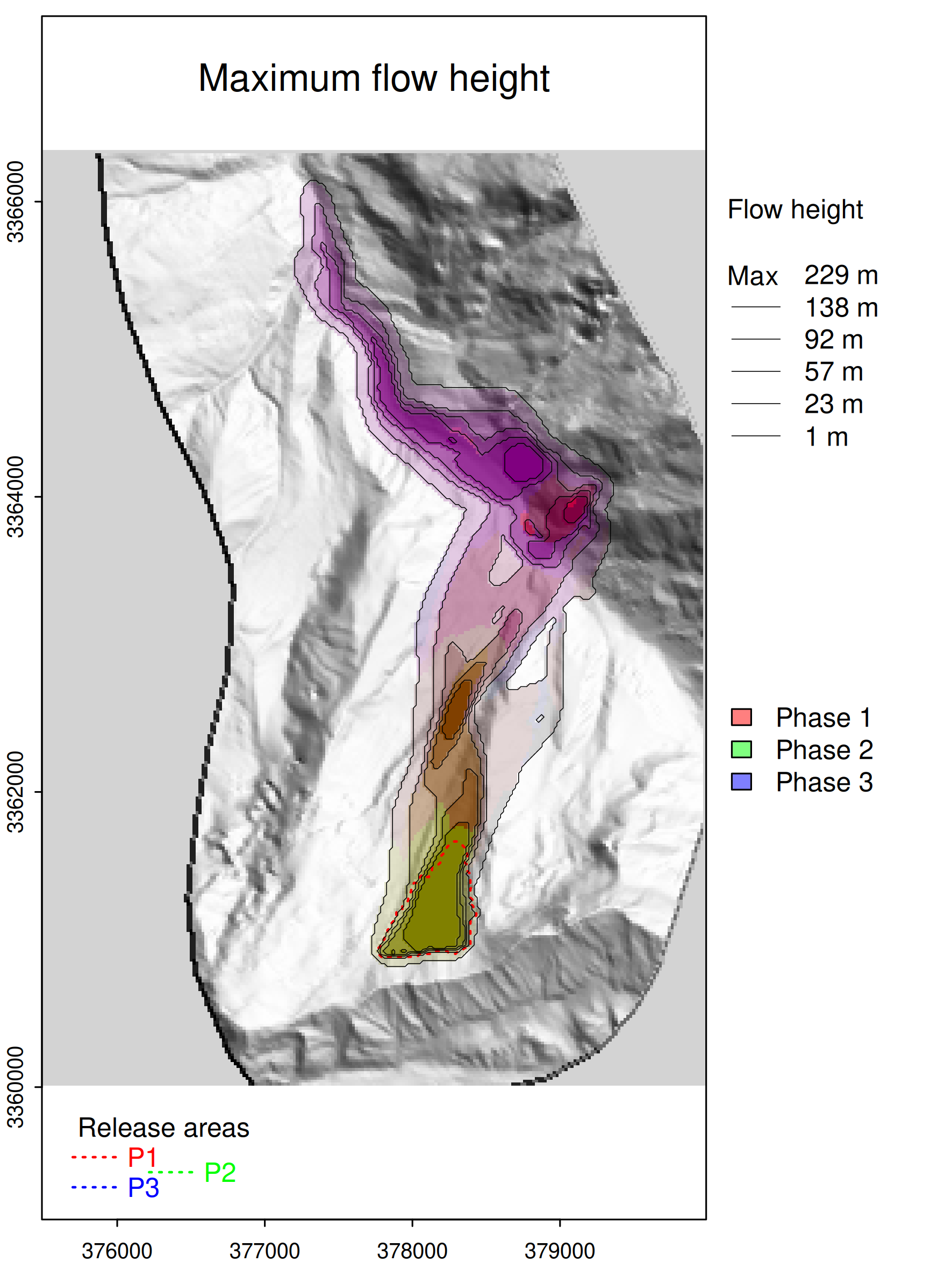}
  \includegraphics[width=6cm]{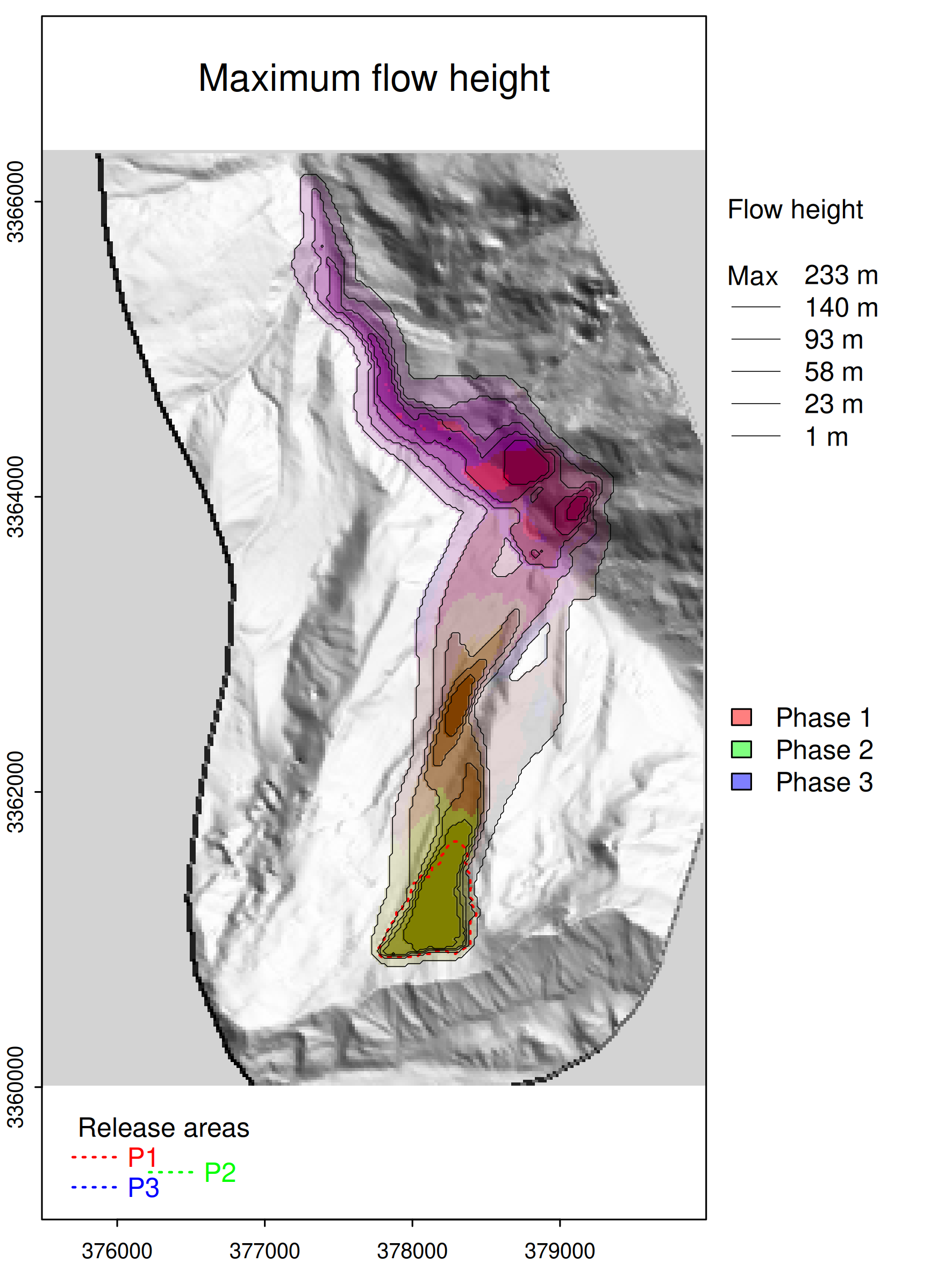}
  \includegraphics[width=6cm]{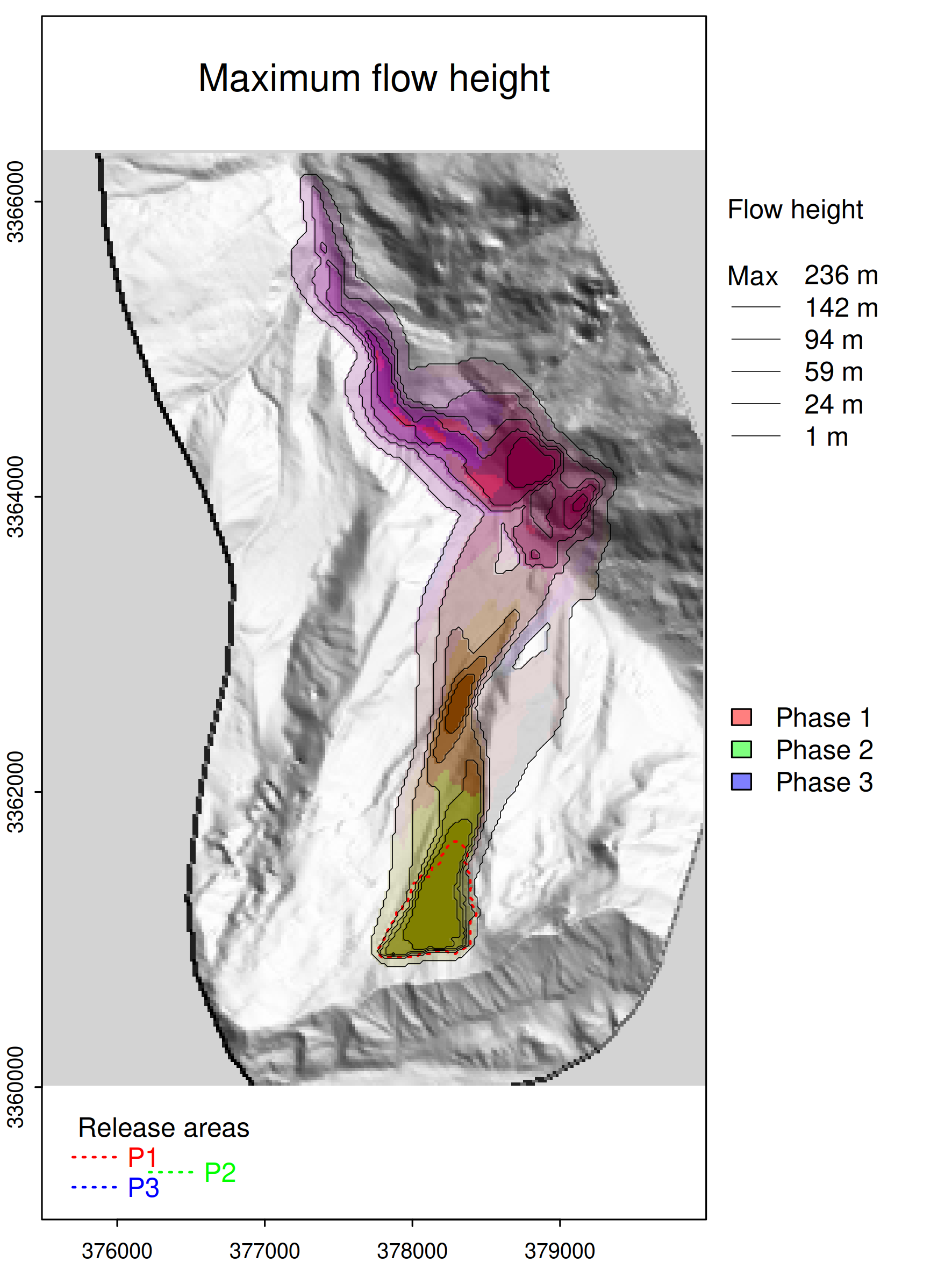}\\[1mm]
  \includegraphics[width=6cm]{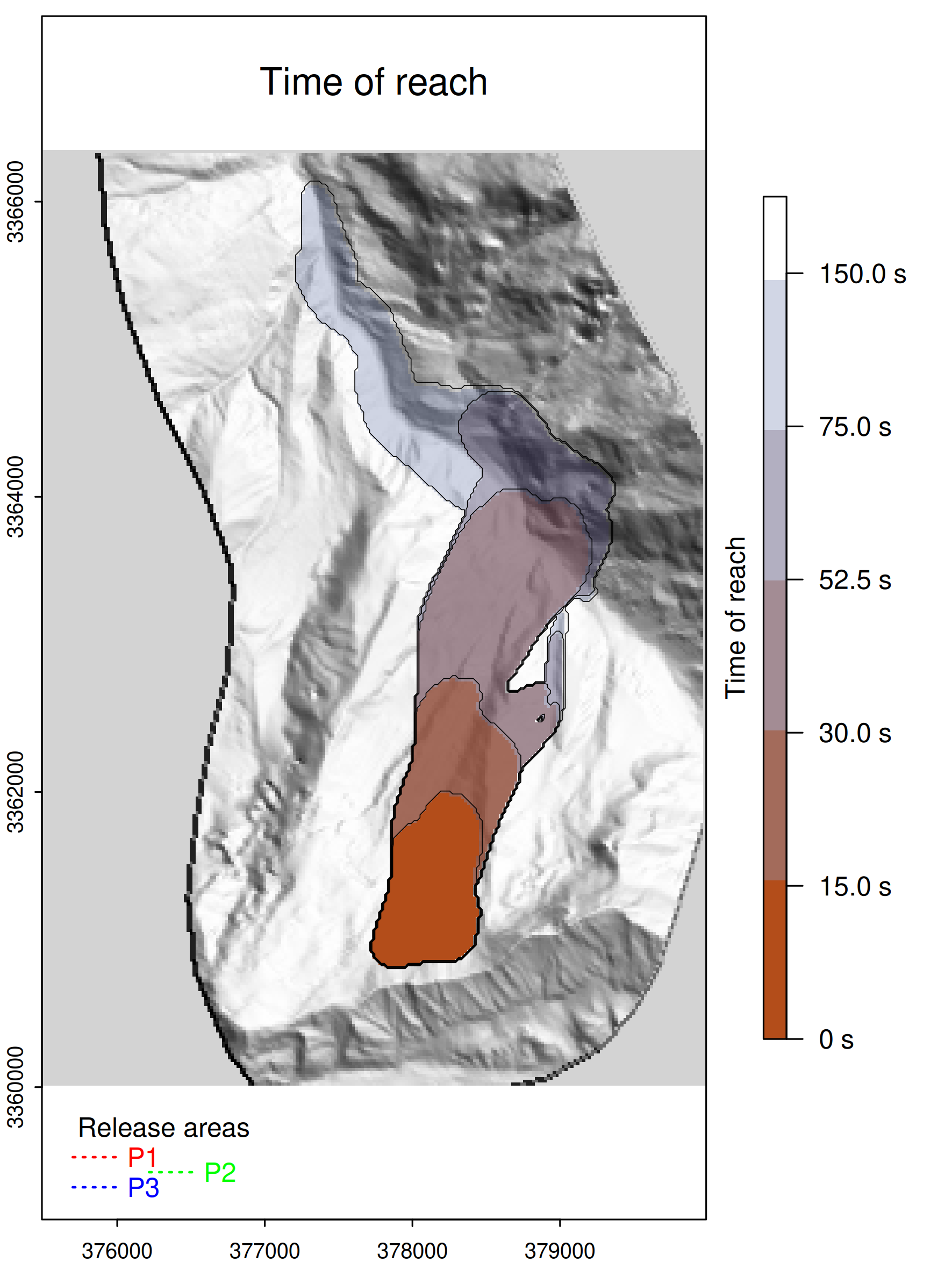}
  \includegraphics[width=6cm]{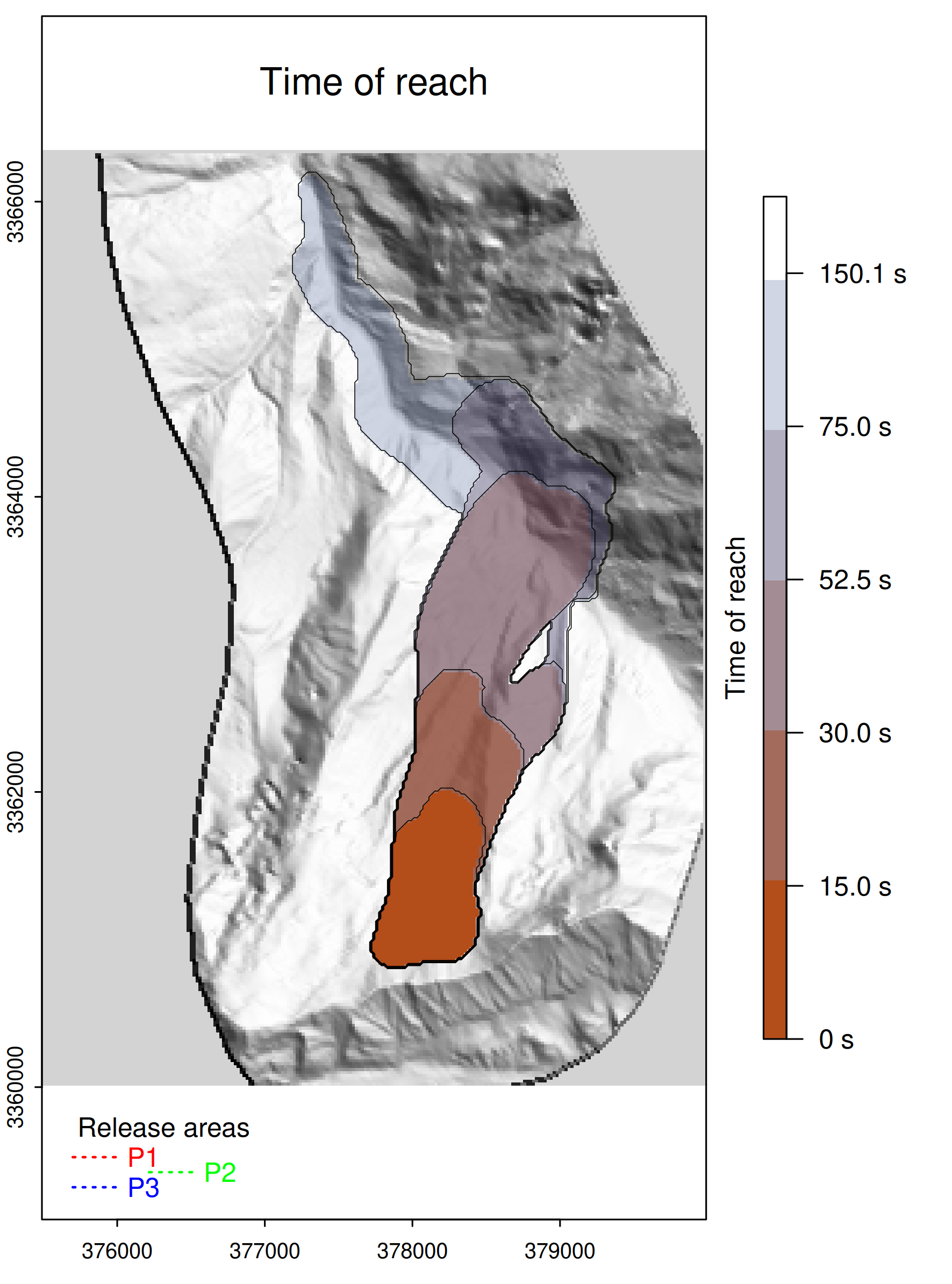}
  \includegraphics[width=6cm]{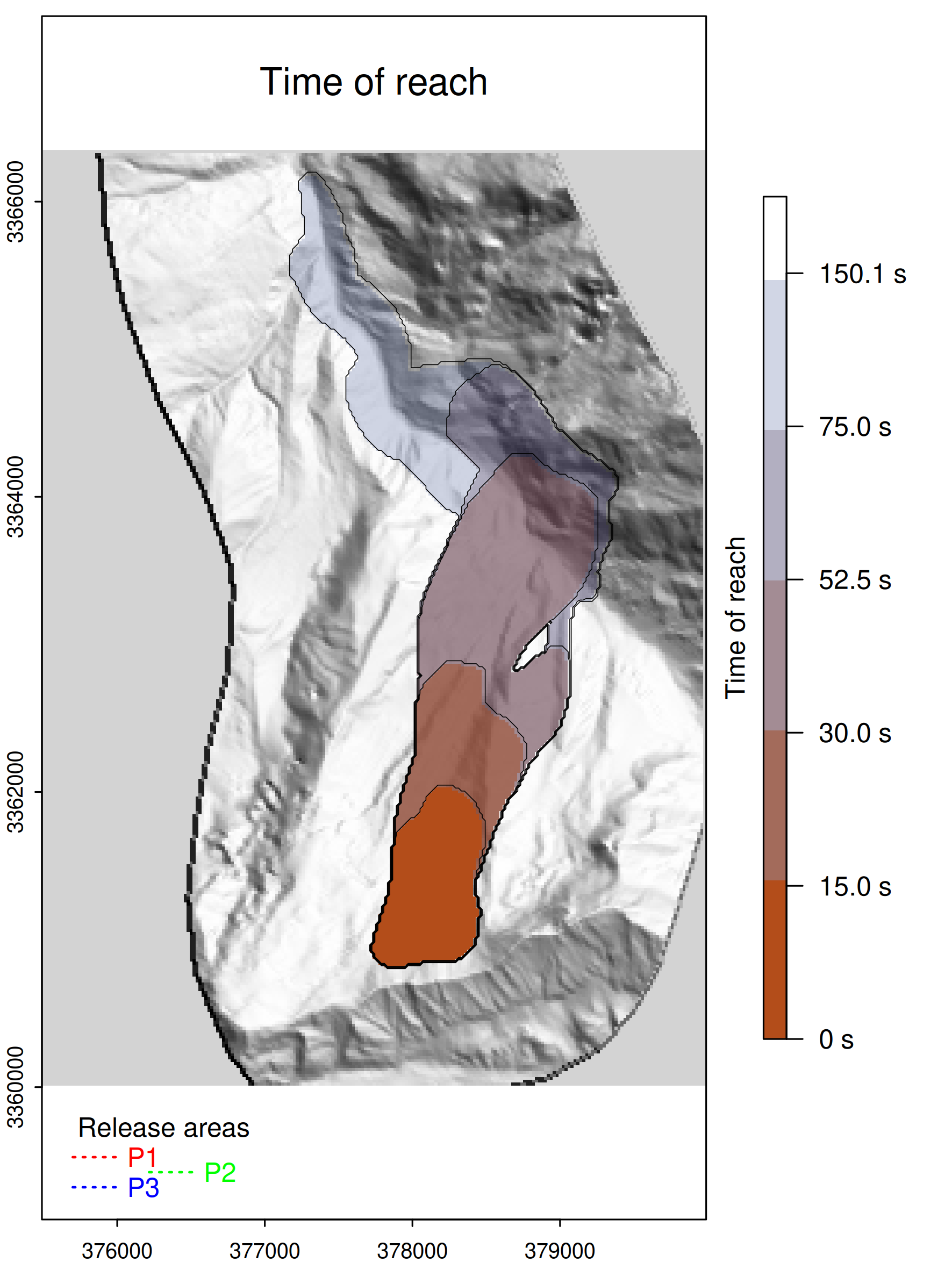}
  \end{center}
  \caption[]{The control of ice friction on the dynamics, spreading and mobility of rock-ice avalanche with ice melting. Decreased ice friction results in changed ice-melt process, however, does not significantly enhance the net flow spreading and mobility. Also displayed are the reach time.}
  \label{Fig-Ch3}
\begin{picture}(0,0)
\put(22, 545){$\delta_i = 20^\circ$}
\put(195,545){$\delta_i = 15^\circ$}
\put(370,545){$\delta_i = 10^\circ$}
\put(117,545){A}
\put(290,545){B}
\put(465,545){C}
\end{picture}
\end{figure}

Ice in rock-ice avalanche or the rock-ice moraine dam may not be a pure ice. It may contain small rock or soil particles and fine solids. Depending on the content of the other solid particles in the ice, or on the nature of the surface where it slides, the effective ice friction may differ substantially. For this reason, next, we present simulation results for three different (basal) frictions for ice. The question we want to address is: with the reduced ice friction, what happens to the thermo-mechanics, dynamics and the mobility of rock-ice avalanche?
\\[3mm]
From the classical perception, the answer would be simple: with decreased friction, the flow mobility would be enhanced. However, this does not happen for the rock-ice avalanche. As discovered here, the ice melting brings an unprecedented thermo-mechanical play, distinctly characterizing the rock-ice avalanche motion . The results are presented in Fig. \ref{Fig-Ch3} with the ice frictions $\delta_i = 20, 15, 10^\circ$, respectively ($\alpha_i = 0.5$, $l_s = 0.1$). As the friction decreases, the ice mass propagates further downslope, this is consistent with the decreased energy dissipation associated with the friction reduction. However, surprisingly, the ice-melting process is different and is delayed with the decreased ice friction. We observe that the reduced ice friction tends to enhance the mobility, but, at the same time, the reduced ice friction reduces ice melting-rate. This is clear from the thermo-mechanics contained in the net ice melt rate (\ref{Eqn_melt_7}), collectively appearing through the effective frictional load that is proportional to $\rho_{ri}\mu_{ri}\alpha_{ri}$. However, as these counter-processes somehow balance each other, there is not that much enhancement in the net mass flow mobility. Even the maximum flow depth increases with the decrease in friction, though only slightly. Therefore, simulation results reveal that there is a complex counter-play between the frictional weakening and the melt-rate weakening in controlling the net flow mobility. These characteristic features could be properly described with the unified, multi-phase thermo-mechanical rock-ice avalanche model.

\subsection{The control of rock friction}

\begin{figure}[t!]
\begin{center}
  \includegraphics[width=6cm]{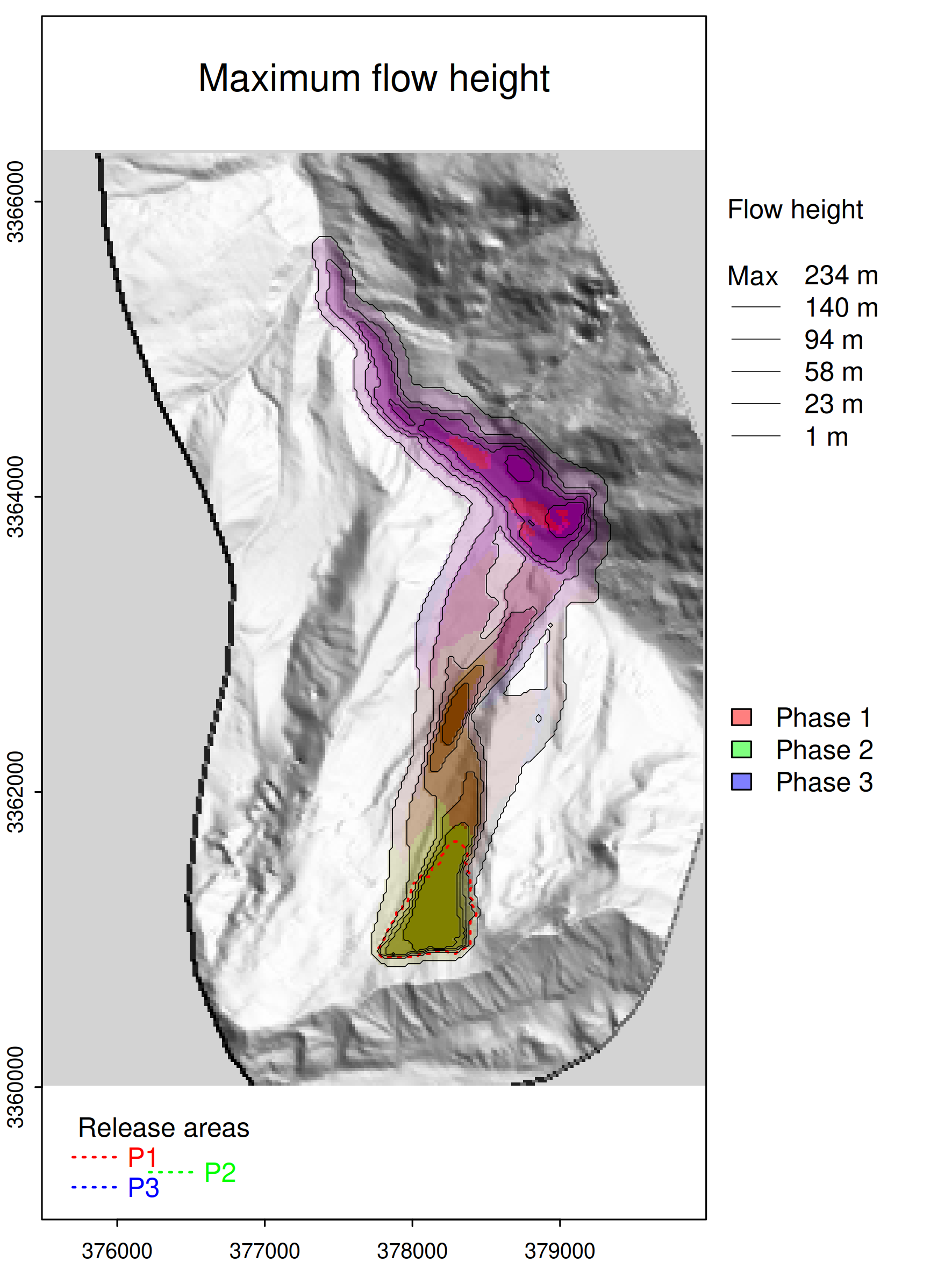}
  \includegraphics[width=6cm]{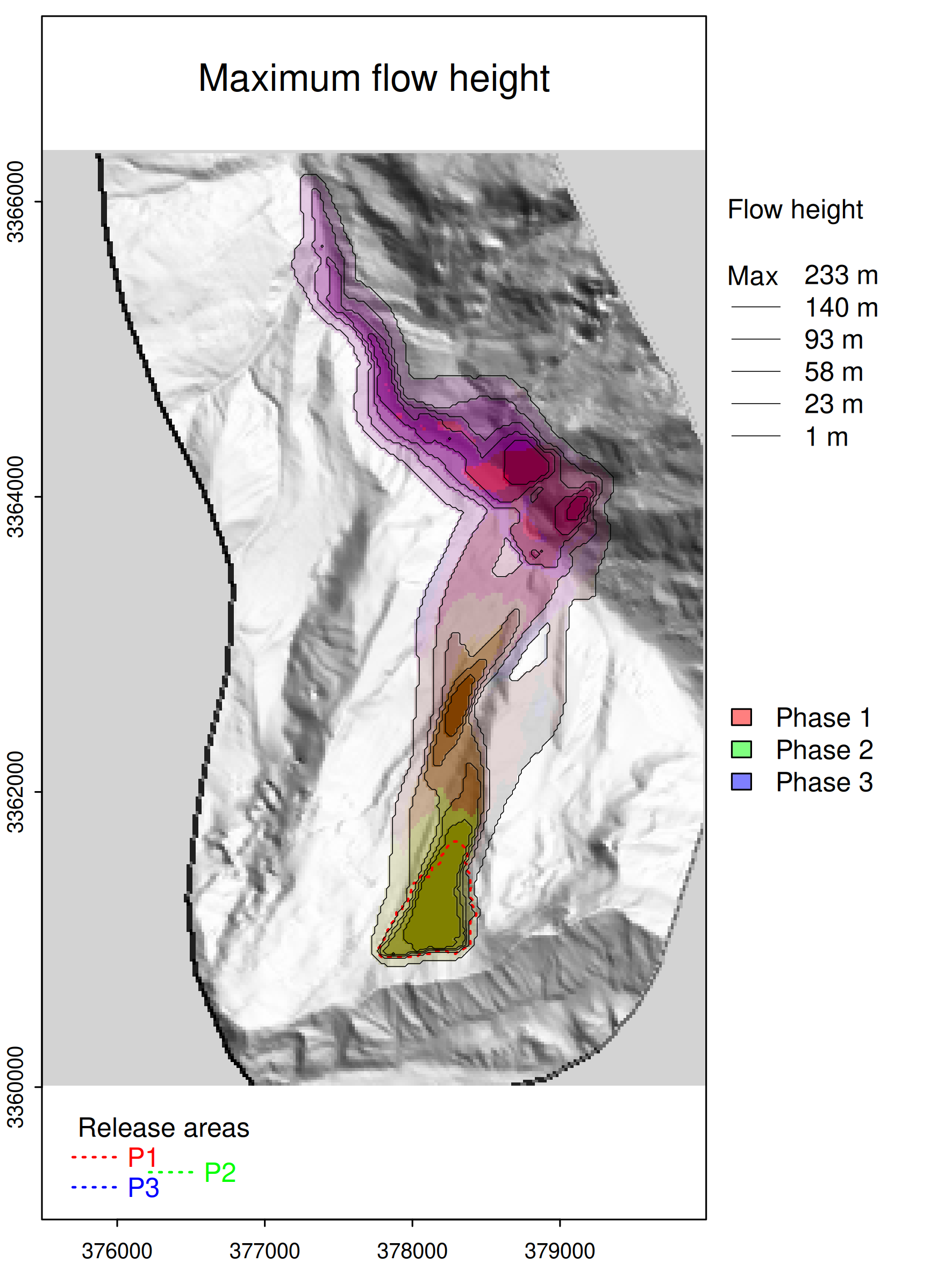}
  \includegraphics[width=6cm]{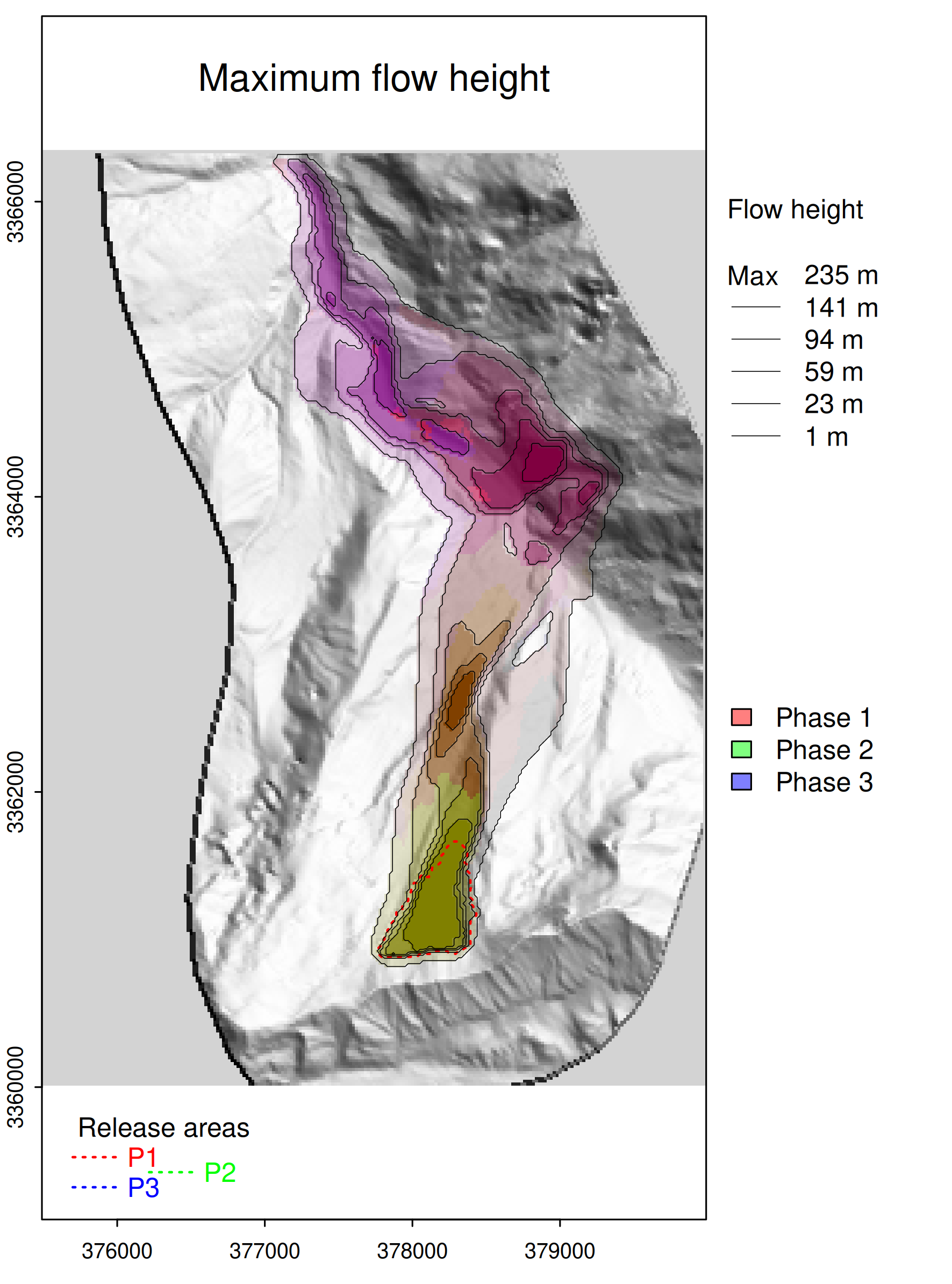}\\[1mm]
  \includegraphics[width=6cm]{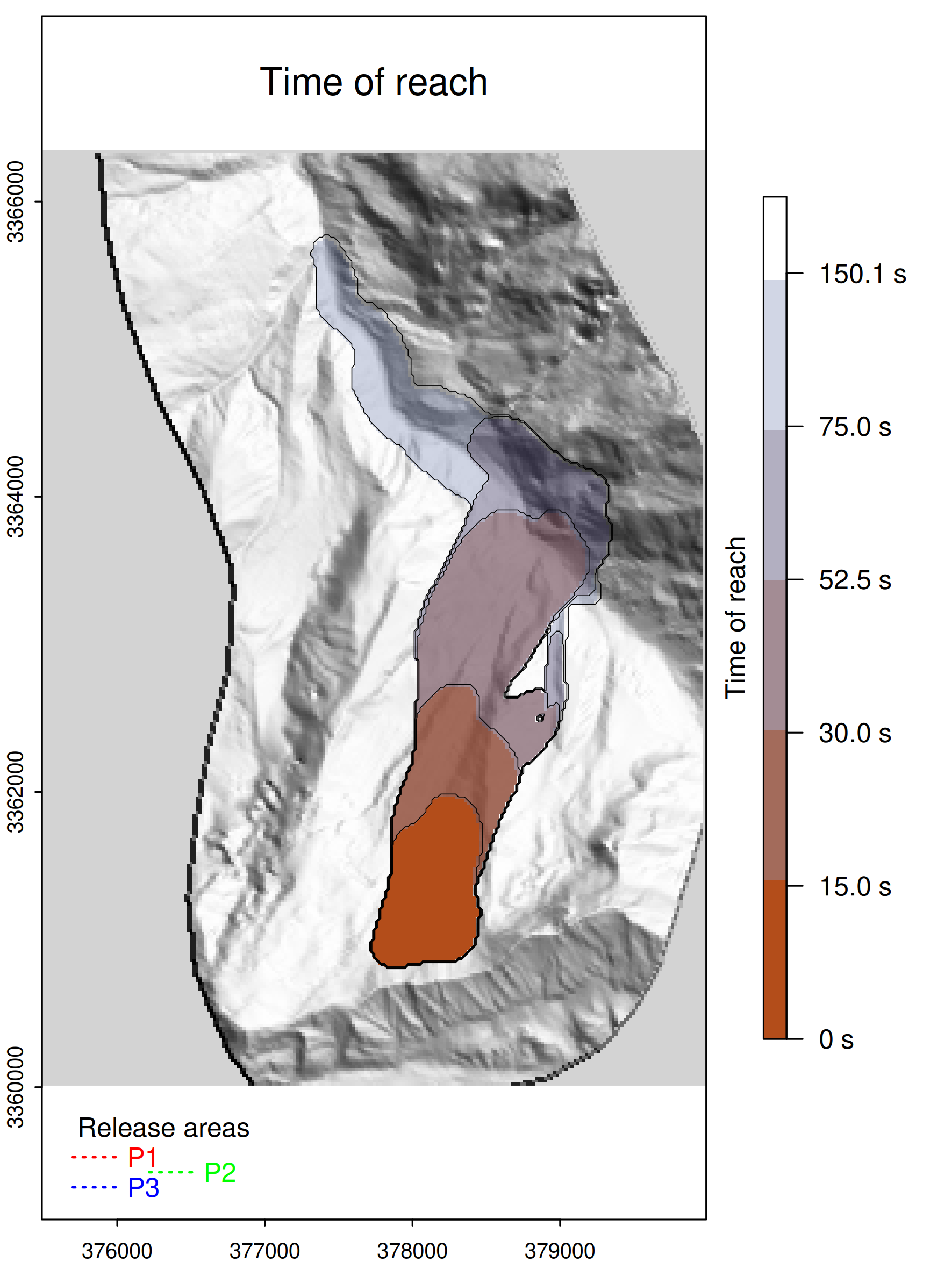}
  \includegraphics[width=6cm]{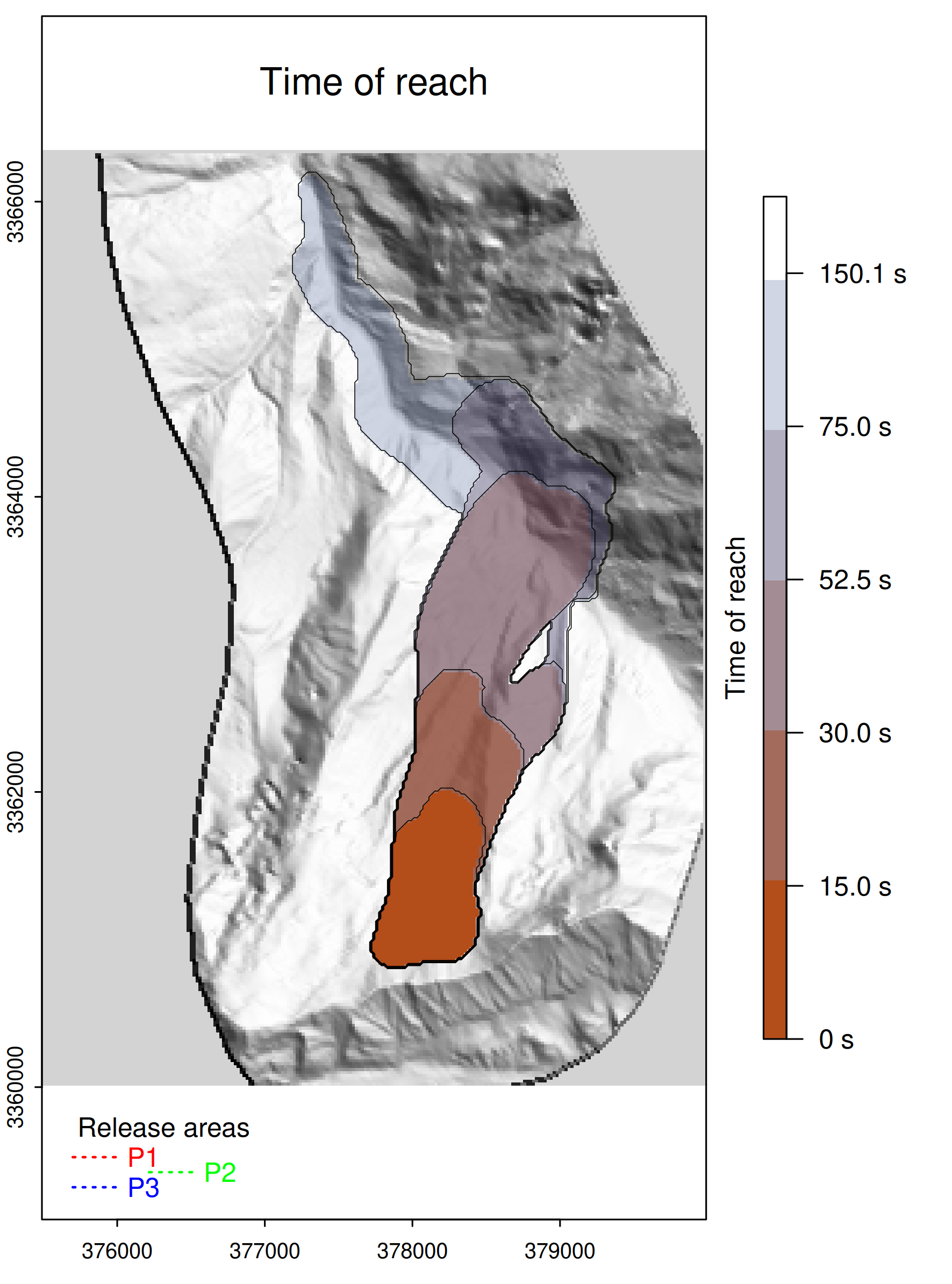}
  \includegraphics[width=6cm]{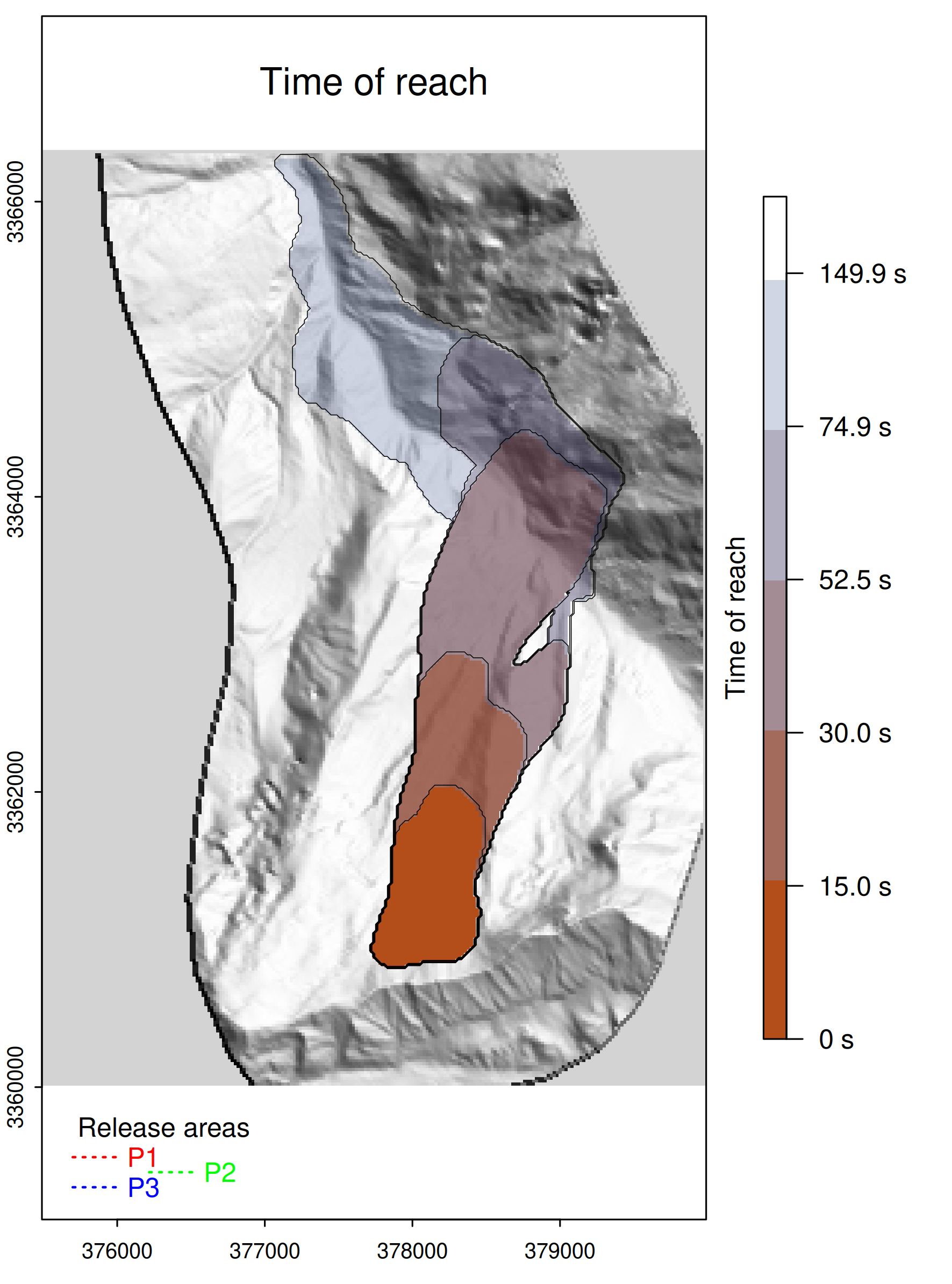}
  \end{center}
  \caption[]{The control of rock friction on the dynamics, spreading and mobility of rock-ice avalanche with ice melting. Decreased rock friction results in changed ice-melt process, and significantly enhances the net flow spreading and mobility. Also displayed are the reach time.}
  \label{Fig-Ch4}
\begin{picture}(0,0)
\put(22, 545){$\delta_r = 25^\circ$}
\put(195,545){$\delta_r = 20^\circ$}
\put(370,545){$\delta_r = 15^\circ$}
\put(117,545){A}
\put(290,545){B}
\put(465,545){C}
\end{picture}
\end{figure}

Rock material in rock-ice avalanche may contain different types of rock particles, or the surface where it slides. Depending on this, the effective rock friction may differ significantly. For this reason, next, we present simulation results for three different (basal) frictions for the rock component. We seek the answer to the question: as the rock friction reduces, what happens to the thermo-mechanics, dynamics and the mobility of the rock-ice avalanche?
\\[3mm]
The results are displayed in Fig. \ref{Fig-Ch4} with the rock frictions $\delta_r = 25, 20, 15^\circ$, respectively ($\alpha_i = 0.5, l_s = 0.1$). As the friction decreases, the rock mass propagates farther downstream relatively easily because of the decreased energy dissipation associated with the friction reduction. Yet, amazingly, the ice-melting process is quite different and is substantially delayed with the decrease in the rock friction as the entire frictional mass slips down economically, resulting in quite different flow dynamics.
\\[3mm]
The results clearly indicate that the reduced rock friction enhances the mobility for the entire duration of the flow, because, it remains rock, equally contributing to the melting-rate due to the unchanged load of the rock in producing the frictional heat. Yet, at the same time, the ice melting produced fluid turns the rock-ice mass into debris flow by largely weakening the mechanical resistance. This further enhances the flow dynamics. With decreased rock friction, there emerges a complex positive interplay between the frictional weakening and the melt-rate weakening resulting in the substantially enhanced net flow mobility. This is in contrast to the thermo-mechanics associated with the reduced ice friction. This is an innovation in the rock-ice avalanche dynamic simulation.
\\[3mm]
The essentially different controls of the reduced ice friction and the reduced rock friction on the state of flow and mobility are great new understanding in rock-ice avalanche dynamics.
Probably the most important fact revealed here is that the ice melting inherently characterizes and explains the mobility of the rock-ice avalanche. Without this, the study of rock-ice avalanche lacks the fundamental essence. All these phenomena are hydro-mechanically consistent. Explaining these seemingly plausible counter-intuitive, complex thermo-mechanical processes was possible with the new unified, multi-phase thermo-mechanical rock-ice avalanche model presented at Section 2.

\section{Summary}

Here, we fundamentally advanced our knowledge by proposing a physically-based, multi-phase thermo-mechanical model for rock-ice avalanches consisting of rock, ice and fluid. The main focus is the development of the general temperature equation for the rock-ice avalanche, the first of this kind in the simulation of mass transport. This explains the comprehensive advection-diffusion of the temperature of the rock-ice mass including the   complex heat exchange processes across the avalanche body, basal heat conduction, the uniquely combined production and loss of heat due to the frictional shear heating and the ice melting, and the enhancement of the temperature associated with the entrainment of the basal material. The new temperature equation is structurally interesting and possesses many important thermo-mechanical and dynamical properties of rock-ice avalanches. It is mathematically consistent and physically-explained. It includes an important composite term, constituting a very special form and is a crucial innovation here. The intensity of fragmentation and melting determines its rate of changes. This special form contains two coupled dynamics in it: the rate of change of the thermal conductivity and the rate of change of temperature of the bulk mixture. It emerged due to the variation of the mixture conductivity. As the constituent volume fractions evolve due to the melting of the ice, the conductivity evolves accordingly. In contrast to the classical diffusion and the advection processes, this composite structure is the only term that involves the variation of both of these quantities. This term uniquely characterizes the distinctive thermo-mechanical processes and their dynamical consequences in the rock-ice avalanche propagation in a tightly coupled way. 
 \\[3mm]
 The model highlights many essential thermo-mechanical aspects of rock-ice avalanches. Thicker avalanches produce more heat than the thinner avalanches. Longitudinal or lateral heat productions, that are included in the new temperature equation, play important role in the temperature evolution. Fast moving avalanches produce higher amount of heat than the slow moving avalanches. The shear heating and the change of temperature of the rock-ice avalanche control the ice melting rate. Conversely, the fast ice melting results in the substantial change in the temperature of the avalanching body. The intensity of the fluid mass production (or ice mass loss) depends on the melting efficiency as described by the newly developed unified effective ice melt rate. The fluid mass production rate is formally derived for multi-phase mixture, and is in general form. Entrainment of the basal material into the propagating body changes the state of temperature of the rock-ice avalanche. The model includes the internal mass and momentum exchanges between the ice and fluid in the rock-ice avalanche, and the mass and momentum production rates due to the entrainment of the basal material through the basal erosion. These mass and momentum productions are fundamentally different processes in the rock-ice avalanche. Yet, the first is an exclusive characteristic property of the rock-ice avalanche. However, both may play important role in the evolution of the temperature in the rock-ice avalanche. If the heat entrainment across the avalanche boundary is substantial, then, the temperature changes are rapid. It also applies to the basal heat conduction. We revealed that there exists a strong coupling between the existing mass and momentum balance equations for the phases (rock, ice and fluid) and the newly proposed general temperature equation for the bulk motion of the rock-ice avalanche, and the unified ice melting rate. We also developed a thermo-mechanically described advection-diffusion-decay-source model for the evolution of temperature of rock-ice avalanche and its exact-analytical solutions. This offers a fundamentally novel, analytical understanding of the complex process of rock-ice avalanche propagation, providing some deep insights into the underlying dynamics with important thermo-mechanical features of the temperature evolution.
 \\[3mm]
Considering the 2021 Chamoli event, the essential functionality of the thermo-mechanical rock-ice avalanche model is illustrated, for the first time, with the simulations based on the multi-phase computational tool r.avaflow. Four different scenarios are considered to study how the rock-ice avalanche dynamics is controlled by the changes in: the ice-melt-efficiency, the fraction of ice, and the ice and rock quality (friction). We analyzed on how these aspects govern the process of melting, flow transformation, mass propagation, spreading and their mobility. Simulation results demonstrate the ice-melt-induced flow mobility, providing the essential computational framework in simulating the purpose-based real rock-ice avalanche events. There appeared a complex counter-play between the frictional weakening and the melt-rate weakening with decreased ice friction resulting in the negligible net flow mobility. However, with decreased rock friction, there emerged a positive interplay between the frictional weakening and the melt-rate weakening resulting in the substantially enhanced net flow mobility. These essentially different controls of the reduced ice and rock frictions on the state of the flow and mobility are great new displays in rock-ice avalanche dynamics. The most important fact revealed here is that the ice melting brings an unprecedented thermo-mechanical play that distinctly characterizes the motion and explains the mobility of the rock-ice avalanche. Yet, these phenomena are hydro-mechanically consistent. Explaining these seemingly plausible, counter-intuitive, complex thermo-mechanical processes was possible with the new model. So, the novel multi-phase thermo-mechanical model for rock-ice avalanches provides a useful method for practitioners and mountain engineers in better solving the problems associated with the hazard  mitigation and planning.

 \section*{Acknowledgments}
 
 Financial support from the German Research Foundation (DFG) through  the research project: ``Landslide mobility with erosion: Proof-of-concept and application - Part I: Modelling, Simulation \& Validation; Project number 522097187" is acknowledged.

\end{document}